\documentclass[12pt]{article}
\pdfoutput=1

\usepackage{cite}
\usepackage{booktabs}
\usepackage[english]{babel}
\usepackage{amsmath,amssymb,amsbsy,amstext, amsthm, simplewick}
\usepackage{hyperref}
\usepackage{graphicx}
\usepackage{amsfonts}
\usepackage{amssymb}
\usepackage[small]{caption}
\usepackage{upgreek}
\usepackage[svgnames,dvipsnames,x11names,table]{xcolor}
\usepackage{multirow} 
\usepackage{geometry}
\usepackage[hang,flushmargin]{footmisc}
\usepackage{bm}
\usepackage{braket}
\usepackage{subcaption}
\usepackage{mathtools}
\usepackage{setspace}
\usepackage{cleveref}
\usepackage{comment}
\usepackage{scalerel}
\usepackage[normalem]{ulem}
\usepackage{slashed}
\usepackage{enumitem}
\usepackage{tikz}
\usetikzlibrary{decorations.markings}
\usetikzlibrary{shapes.misc}

\makeatletter
\g@addto@macro\bfseries{\boldmath}
\makeatother

\hypersetup{
    colorlinks=true,
    linkcolor={red!50!black},
    citecolor={blue!50!black},
    urlcolor={blue!80!black}
}

\usepackage{colortbl}

\makeatletter
\newlength{\apb@width}
\newcommand{\autoparbox}[2][c]{\settowidth{\apb@width}{#2}\parbox[#1]{\apb@width}{#2}}

\makeatother

\definecolor{lightgray}{gray}{0.9}

\usepackage[framemethod=default]{mdframed}
\newmdenv[skipabove=7pt,
skipbelow=7pt,
rightline=false,
leftline=false,
topline=false,
bottomline=false,
backgroundcolor=gray!10,
linecolor=gray,
innerleftmargin=5pt,
innerrightmargin=5pt,
innertopmargin=5pt,
innerbottommargin=5pt,
leftmargin=0cm,
rightmargin=0cm,
linewidth=4pt]{eBox}

\usepackage[most]{tcolorbox}
\tcbset{colback=white, colframe=black,
        highlight math style= {enhanced, %<-- needed for the ?remember? options
            colframe=red,colback=red!10!white,boxsep=0pt}
        }
\definecolor{light-gray}{gray}{0.95}

\crefname{table}{Table}{Tables}
\crefname{equation}{Eq.}{Eqs.}
\crefname{appendix}{App.}{Apps.}
\crefname{section}{Sec.}{Secs.}
\crefname{figure}{Fig.}{Figs.}

\numberwithin{equation}{section}

\def\beq{\begin{equation}}
\def\eeq{\end{equation}}

\def\bea{\begin{eqnarray}}
\def\eea{\end{eqnarray}}

\def\bvar{{\underline{\varphi}}}
\def\vp{\varphi_{+}}
\def\vm{\varphi_{-}}
\def\tvp{{\tilde{\varphi}_{+}}}

\def\d{{\rm d}}
\def\dd{{\rm d}}

\def\beq{\begin{equation}}
\def\eeq{\end{equation}}
\def\bea{\begin{eqnarray}}
\def\eea{\end{eqnarray}}

\def\d{{\rm d}}
\def\dd{{\rm d}}

\def\O{{\cal O}}

\def\d{{\rm d}}

\def\H{{\cal H}}

\def\k{{\vec{\scaleto{k}{7pt}}}}
\def\ksub{{\vec{k}}}
\def\kp{{\!\!\vec{{\scaleto{\,\, k}{7pt}}^{\s\prime}}}}

\def\q{{\vec q}}
\def\p{{\vec p}}

\def\x{{\vec x}}

\def\t{\texttt{t}}

\DeclareRobustCommand{\SkipTocEntry}[4]{}

\newcommand{\s}{\hspace{0.8pt}}

\definecolor{colorTC}{rgb}{.2,.7,.2}

\definecolor{amethyst}{rgb}{0.6, 0.4, 0.8}

\definecolor{acolor}{rgb}{0.4, 0.2, 0.4}

\definecolor{blue3}{RGB}{31, 119, 180}
\definecolor{red3}{RGB}{	214, 39, 40}
\definecolor{orange3}{RGB}{255, 127, 14}
\definecolor{green3}{RGB}{44, 160, 44}

\begin{document}

\begin{titlepage}
\setcounter{page}{1} \baselineskip=15.5pt 
\thispagestyle{empty}
$\quad$
\vskip 70 pt

\begin{center}
{\fontsize{18}{18} \bf Stochastic Inflation at NNLO}
\end{center}

\vskip 20pt
\begin{center}
\noindent
{\fontsize{12}{18}\selectfont  Timothy Cohen$^{\s 1}$, Daniel Green$^{\s 2}$, Akhil Premkumar$^{\s 2}$, and Alexander Ridgway$^{\s 2}$}
\end{center}

\begin{center}
\vskip 4pt
\textit{ $^1${\small Institute for Fundamental Science, University of Oregon, Eugene, OR 97403, USA}
}
\vskip 4pt
\textit{ $^2${\small Department of Physics, University of California at San Diego,  La Jolla, CA 92093, USA}}

\end{center}

%=========================================
\vspace{0.4cm}
 \begin{center}{\bf Abstract}
 \end{center}
\noindent 
Stochastic Inflation is an important framework for understanding the physics of de Sitter space and the phenomenology of inflation.
In the leading approximation, this approach results in a Fokker-Planck equation that calculates the probability distribution for a light scalar field as a function of time.  Despite its successes, the quantum field theoretic origins and the range of validity for this equation have remained elusive, and establishing a formalism to systematically incorporate higher order effects has been an area of active study.  In this paper, we calculate the next-to-next-to-leading order (NNLO) corrections to Stochastic Inflation using Soft de Sitter Effective Theory (SdSET).  In this effective description, Stochastic Inflation manifests as the renormalization group evolution of composite operators.  The leading impact of non-Gaussian quantum fluctuations appears at NNLO, which is presented here for the first time; we derive the coefficient of this term from a two-loop anomalous dimension calculation within SdSET.  We solve the resulting equation to determine the NNLO equilibrium distribution and the low-lying relaxation eigenvalues.  In the process, we must match the UV theory onto SdSET at one-loop order, which provides a non-trivial confirmation that the separation into Wilson-coefficient corrections and contributions to initial conditions persists beyond tree level.  Furthermore, these results illustrate how the naive factorization of time and momentum integrals in SdSET no longer holds in the presence of logarithmic divergences.  It is these effects that ultimately give rise to the renormalization group flow that yields Stochastic Inflation.

\end{titlepage}

\setcounter{page}{2}

\restoregeometry

\begin{spacing}{1.2}
\newpage
\setcounter{tocdepth}{2}
\tableofcontents
\end{spacing}

\setstretch{1.1}
\newpage

\section{Introduction}
The study of quantum fields in de Sitter (dS) space provides insight into the foundations of inflationary cosmology.  In particular, the equal-time in-in correlation functions of light scalar fields form the theoretical underpinnings of the predictions for the observed density fluctuations sourced during inflation~\cite{Maldacena:2002vr,Weinberg:2005vy,Weinberg:2006ac}.  These correlators encode a wealth of information about the inflationary era that could be revealed by measurements of primordial non-Gaussianity~\cite{Meerburg:2019qqi}.  Yet, despite their importance, our understanding of cosmological correlators beyond tree level is quite limited.  For light scalars, explicit loop calculations have revealed the presence of infrared (IR) divergences and unbounded time-dependent ``secular'' growth~\cite{Ford:1984hs,Antoniadis:1985pj,Tsamis:1994ca, Tsamis:1996qm,Tsamis:1997za,Burgess:2010dd,Rajaraman:2010xd,Marolf:2010zp, Marolf:2011sh,Marolf:2012kh,Beneke:2012kn,Akhmedov:2013vka,Anninos:2014lwa, Akhmedov:2017ooy,Hu:2018nxy,Akhmedov:2019cfd}. 

Stochastic Inflation~\cite{Starobinsky:1986fx,Nambu:1987ef,Starobinsky:1994bd} is a framework for treating the IR dynamics of a massless scalar field in a dS background, and has long been suspected to provide the non-perturbative resolution to the IR issues associated with massless fields in dS~\cite{Enqvist:2008kt,Finelli:2008zg,Podolsky:2008qq,Seery:2010kh,Garbrecht:2014dca,Burgess:2015ajz,Gorbenko:2019rza,Baumgart:2019clc,Mirbabayi:2019qtx,Cohen:2020php,Mirbabayi:2020vyt,Baumgart:2020oby}.  The idea is to reframe the problem in terms of the probability distribution for the 
scalar field as a function of time, resulting in a Fokker-Planck equation that depends on the scalar field potential. There are two contributions to the evolution, resulting from quantum noise induced by fluctuations of the field as it crosses the dS horizon and classical drift due to the potential. This framework forms the conceptual basis for slow-roll eternal inflation~\cite{Linde:1986fd,Goncharov:1987ir,Freivogel:2011eg} and can be used to describe the onset of eternal inflation quantitatively~\cite{Creminelli:2008es}. Moreover, these results hint at the physical meaning of the dS entropy~\cite{ArkaniHamed:2007ky,Dubovsky:2008rf,Lewandowski:2013aka}, which remains a significant unsolved problem~\cite{Witten:2001kn}.

Notwithstanding the conceptual and technical appeal of Stochastic Inflation, it is necessarily approximate.  For example, we expect that there are non-Gaussian contributions to the quantum noise that result from the UV interaction, which are not modeled by Stochastic Inflation.  Furthermore, the Fokker-Planck formalism obscures the connection to cosmological correlators, and it is not a prioi obvious how to incorporate higher-order corrections.  It would be ideal if we could understand how the success of Stochastic Inflation relates to other results regarding the IR behavior of fields in dS, such as the freeze-out of superhorizon metric fluctuations which has been shown to all orders in perturbation theory~\cite{Salopek:1990jq,Assassi:2012et,Senatore:2012ya}, or the loop generated anomalous scaling for the time-evolution of massive fields~\cite{Green:2013rd,Green:2020txs}.  One of our goals in exploring the corrections to Stochastic Inflation is to understand how they fit into the broader context of quantum field theory in dS.

A framework that accomplishes this ambitious goal is the Soft de Sitter Effective Theory (SdSET)~\cite{Cohen:2020php}.  By following the standard Effective Field Theory (EFT) playbook, this approach isolates the dynamics that persist in the superhorizon limit, yielding more efficient calculations of loop corrections to long wavelength cosmological correlators.  Taking a real scalar field in dS as the UV description, SdSET describes the IR limit of this model by relying on two degrees of freedom that correspond to the growing and decaying modes which are familiar from the solving the Klein-Gordon equation classically in a dS background.  This representation admits a power counting prescription that systematically expands about the long wavelength limit in terms of a local Lagrangian.  Loop dependence on time and space is manifestly factorized throughout the calculation, allowing an efficient isolation of the time dependent IR divergent logs.  Such logs lead to secular growth for both massive and light fields, and appear in SdSET as contributions to the anomalous dimensions of local operators.  In the case of light fields, an infinite number of operators become degenerate and Starobinsky's model of Stochastic Inflation is equivalent to the leading order (LO) dynamical renormalization group (RG) that governs their mixing as a function of time.\footnote{The dynamical RG flow described here should not be confused with the RG flow that appears in a holographic dual via dS/CFT~\cite{Witten:2001kn,Strominger:2001pn,Mazur:2001aa,Maldacena:2002vr,Maldacena:2011nz}.  The key difference is that our dynamical RG applies directly to the in-in correlators and not to the wavefunction of the universe.}  This implies that corrections to Stochastic Inflation can be computing by simply extending the RG analysis to higher orders. 

Taking the UV description to be massless $\lambda \phi^4$ theory, the endpoint of this RG flow is a non-trivial fixed point where the field values are $\phi  \sim H \lambda^{-1/4}$.  Corrections to this description around this fixed point must account  for this non-perturbative scaling with $\lambda$.  In this paper, we will calculate the evolution of operators to next-to-next-to leading order (NNLO) in this power counting.  At NLO, our results reproduce previous calculations~\cite{Gorbenko:2019rza,Mirbabayi:2020vyt}; as we will show, these contributions can be attributed to field definitions within SdSET.  In contrast, at NNLO we find a universal correction in the form of a two-loop anomalous dimension that introduces the first higher derivative correction to Stochastic Inflation.  In the process, we perform the  full one-loop matching in SdSET, which further elucidates the relationship between the EFT and the UV descriptions.

One novel feature of SdSET is that consistently matching a UV theory onto the EFT requires specifying both Wilson coefficients and (time-independent) initial conditions.  Deriving the RG that yields Stochastic Inflation at NNLO requires performing this matching explicitly at one-loop order.  This provides a highly non-trivial check of the SdSET formalism, and these results can be utilized for a wide variety of correlator calculations.  We will also use this calculation as an opportunity to demonstrate the power of the symmetry preserving ``dynamical dimensional regularization'' technique introduced in~\cite{Cohen:2020php}.

This paper is organized as follows.  We begin with a review of Stochastic Inflation in~\cref{sec:StochInf}, with an emphasis on its origins as a Markovian process, which provides a framework with which we can organize corrections.  Then~\cref{sec:SdSET} reviews the most salient aspects of the SdSET formalism.  The new calculations begin in~\cref{sec:Matching}, where we present the one-loop matching results that are relevant for our applications here.  These are then applied in~\cref{sec:StocInfNNLO}, where we compute the composite operator anomalous dimensions that feed into Stochastic Inflation up to NNLO, and leads to the main result of this work in~\cref{eq:SIatNNLO}.  We then explore the implications of this formula in \cref{sec:Implications}, and finally conclude in~\cref{sec:Conc}.  An appendix on the relevant, but somewhat technical, six-point function matching is provided in~\cref{sec:MatchingSixPt}, and the hard cutoff version of the main calculations are given in~\cref{sec:HardCutoff}.

\vspace{10pt}
\noindent \textbf{Guide for the reader:}  We have attempted to make this paper accessible to a wide ranging audience.  For the casual reader, we suggest ($i$) reading \cref{sec:StochInf} and \cref{sec:SI_sdset}, ($ii$) studying the matching diagrams in \cref{sec:Matching}, and ($iii$) studying the composite operator mixing diagrams in \cref{sec:StocInfNNLO}.  This brings the reader to ($iv$) the nearly final result in \cref{eq:SIatNNLORaw}, where the explicit corrections are given in \cref{eq:Summary}. Finally, ($v$) \cref{sec:Stoc_final_sub} provides the derivation that results in the simplified form of the NNLO formula for Stochastic Inflation given in \cref{eq:SIatNNLO}.  One might also be interested in the NNLO equilibrium probability distribution and relaxation eigenvalues presented in \cref{sec:Implications}, which brings the reader to the conclusions in \cref{sec:Conc}.

\section{Stochastic Inflation}
\label{sec:StochInf}
In a theory of a massless scalar field $\phi$ in a dS background with Hubble constant $H$, the field's value will fluctuate by  $O(H)$
as each momentum mode crosses the dS horizon.  An equivalent point of view is that these stochastic fluctuations are the result of the non-zero temperature within dS.
This effect has a natural interpretation as a random walk, an idea that was made precise by Starobinsky~\cite{Starobinsky:1986fx} and led the formalism known as Stochastic Inflation.  In this section, we discuss this approach, and emphasize the structure of higher order corrections.  

For concreteness, we will assume the canonical example of $\lambda \phi^4$ theory in a dS background, whose UV description in terms of a scalar field $\phi$ is
\beq
S_{\rm UV} = \int \d^4 x\, \sqrt{-g}\,\bigg[-\frac{1}{2}\s g_{\mu\nu}\s \partial^\mu \phi\, \partial^\nu \phi + \frac{1}{2}\s m^2\s \phi^2 + \frac{1}{4!}\s \lambda\s \phi^4\s \bigg] \, ,
\label{eq:Lphi4}
\eeq
where $g_{\mu\nu}$ is the dS metric and $g \equiv \det g_{\mu\nu}$ as usual.  The essential formalism developed here holds for general models.  However, we will not be able to derive corrections to Stochastic Inflation generically, and so we will work with this simple and well studied example when we calculate explicit corrections.

In the process of discussing the general structure of corrections to Stochastic Inflation, we will arrive at a natural interpretation for higher-order corrections in terms of the transition amplitudes for the field $\phi$.  Unfortunately, how to determine the corrections directly is not transparent in this description.  In \cref{sec:SI_sdset} we will show how the corrections discussed in this Section arise from operator mixing, and the remainder of the paper is devoted to deriving these corrections and their implications using SdSET.  

Before moving on, we emphasize that the formalism we develop in this section will rely on the assumption that the late time evolution of $\phi$ can be modeled as a Markovian system (as described in \cref{sec:BLO} below).  This will be justified by the concrete calculation of the dynamical renormalization group using the SdSET that is developed later in this paper.  It ultimately is due to the fact that the dynamics of the SdSET degrees of freedom are governed by a first order equation, which is a consequence of the EFT power counting.

\subsection{Leading Order}
\label{sec:SILO}
The framework of Stochastic Inflation results in a probability distribution $P(\phi,t)$ for the field $\phi$ at a time $t$.  To leading order, $P(\phi,t)$ obeys a Fokker-Planck equation
\beq
\frac{\partial}{\partial t} P(\phi,t)= \frac{H^3}{8 \pi^2} \frac{\partial^2}{\partial \phi^2} P(\phi, t)  + \frac{1}{3 H}  \frac{\partial}{\partial \phi} \big [V'(\phi) P(\phi,t ) \big] \, ,
\label{eq:FPeqStocInf}
\eeq
where the first term captures the stochastic noise from the inherent quantum variance of $\phi$, while the second term is due to the classical drift induced by the potential where $V'(\phi) \equiv \partial V/\partial\phi$.
One interesting application of this equation is to solve for the fixed point that the scalar field would reach if it lived in an eternal dS background.  To find the fixed point, we enforce that $\partial P_{\rm eq}/\partial t=0$ for the equilibrium solution $P_{\rm eq}$, which implies
\beq
\frac{\partial^2}{\partial \phi^2} P_{\rm eq}(\phi) = - \frac{8\pi^2}{3 H^4} \frac{\partial}{\partial \phi} \big [V'(\phi) P_{\rm eq}(\phi) \big] \,.
\eeq
Integrating both sides of this equation twice leads to the solution
\beq
P_{\rm eq}(\phi) =  C e^{- 8 \pi^2 V(\phi)/ 3 H^4}\,.
\label{eq:equilibrium_0}
\eeq
We can use this leading order solution to organize corrections to Stochastic Inflation as a perturbative series in the UV coupling.

\subsection{Beyond Leading Order} \label{sec:BLO}
Having reviewed the leading order formalism and its consequence for pure dS, we now turn to exploring the form we can expect corrections to take.  In order to generalize this Fokker-Planck equation, we return to its origins.  The underlying assumption is that the system is Markovian, in that the time slice of interest is entirely determined by the information contained in the previous step.  In other words, a Markovian system has no ``memory.''  Since this assumption holds for the spectrum of scalar field fluctuations at horizon crossing, the resulting formalism will tell us what kinds of corrections to Stochastic Inflation we can expect.

The Markovian assumption leads directly to the Chapman-Kolmogorov equations, which describe a probability distribution $P(\phi,t+\d t)$ that is fully determined by $P(\phi,t)$:
\beq
\frac{\partial}{\partial t} P(\phi, t) = \int \d \phi'\, \Big[ P(\phi',t)\s W(\phi| \phi') -P(\phi, t) \s W(\phi' | \phi) \Big]\,,
\label{eq:CKeq}
\eeq
where $W(\phi | \phi')$ is a ``transition rate'' in that it sets the rate for transitioning to $\phi$ from another value $\phi'$ in a differential amount of time.  This equation simply expresses that the probability distribution for $\phi$ at $t+\d t$ is fully determined by the weighted sum of the possible transitions that yield $\phi$ minus the sum of all the weighted transitions for $\phi$ to change value.

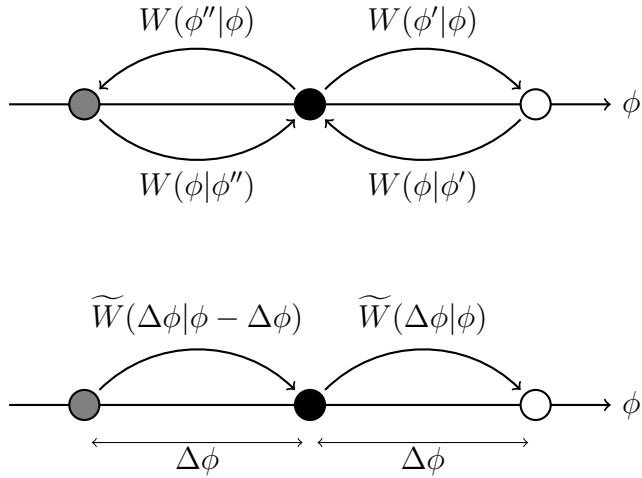
\begin{figure}[t!]
\centering
\begin{tikzpicture}
    % % A helpful grid
    % \draw[step=1cm,gray,very thin] (0,0) grid (12,6);
    % \draw[thick,->] (0,0) -- (12,0) node[anchor=north west] {x axis};
    % \draw[thick,->] (0,0) -- (0,6) node[anchor=south east] {y axis};
    % \foreach \x in {0,1,2,3,4,5,6,7,8,9,10,11}
    %   \draw (\x cm,1pt) -- (\x cm,-1pt) node[anchor=north] {$\x$};
    % \foreach \y in {0,1,2,3,4,5}
    %     \draw (1pt,\y cm) -- (-1pt,\y cm) node[anchor=east] {$\y$};
    
    \begin{scope}[shift={(0,2)}]
        % The states
        \draw[thick,->] (0,3) -- (8,3) node[right] {$\phi$};
        \filldraw[thick,draw=black,fill=gray] (1,3) circle (.2);
        \filldraw[thick,draw=black,fill=black] (4,3) circle (.2);
        \filldraw[thick,draw=black,fill=white] (7,3) circle (.2);
        
        % The transitions
        \draw[thick,->] ({4-.2},{3+.2}) to [out=135,in=45] ({1+.2},{3+.2});
        \draw[thick,->] ({4+.2},{3+.2}) to [out=45,in=135] ({7-.2},{3+.2});
        \draw[thick,->] ({1+.2},{3-.2}) to [out=-45,in=225] ({4-.2},{3-.2});
        \draw[thick,->] ({7-.2},{3-.2}) to [out=225,in=-45] ({4+.2},{3-.2});
        
         % The labels
         \node at (2.5,4.1) {$W(\phi''|\phi)$};
         \node at (5.5,4.1) {$W(\phi'|\phi)$};
         \node at (2.5,1.9) {$W(\phi|\phi'')$};
         \node at (5.5,1.9) {$W(\phi|\phi')$};
    \end{scope}
    
    % The states
    \draw[thick,->] (0,1) -- (8,1) node[right] {$\phi$};
    \filldraw[thick,draw=black,fill=gray] (1,1) circle (.2);
    \filldraw[thick,draw=black,fill=black] (4,1) circle (.2);
    \filldraw[thick,draw=black,fill=white] (7,1) circle (.2);
    
    % The transitions
    \draw[thick,->] ({1+.2},{1+.2}) to [out=45,in=135] ({4-.2},{1+.2});
    \draw[thick,->] ({4+.2},{1+.2}) to [out=45,in=135] ({7-.2},{1+.2});
    
    % The labels
    \node at (2.5,2.2) {$\widetilde{W}(\Delta \phi|\phi-\Delta \phi)$};
    \node at (5.5,2.2) {$\widetilde{W}(\Delta \phi|\phi)$};
    
    \draw[<->] (1.1,0.5) -- (3.9,0.5);
    \draw[<->] (4.1,0.5) -- (6.9,0.5);
    \node at (2.5,0.25) {$\Delta \phi$};
    \node at (5.5,0.25) {$\Delta \phi$};
\end{tikzpicture}
\caption{Visualization of the Kramers-Moyal expansion.  The top panel shows the probability of ``hopping" from $\phi$ to $\phi'$, $W(\phi'|\phi)$ or from $\phi'$ to $\phi$, $W(\phi|\phi')$, or equivalently any other point such as $\phi''$.  On the bottom, we see the process in terms of the probability $\widetilde W(\Delta\phi|\phi)$ to hop from a specific starting point $\phi$ by a distance $\Delta\phi$. }
\label{fig:hopping}
\end{figure}
 
Next, we will reorganize \cref{eq:CKeq} using what is known as the Kramers-Moyal expansion, visualized in~\cref{fig:hopping}. This is effectively the assumption that the transitions are dominated by ``local'' jumps \cite{gardiner2004handbook}.  The first step is to make a substitution of $\Delta \phi = \phi - \phi'$ in the first term and $\Delta \phi = \phi' - \phi$ in the second term.  In other words, when $\phi$ is in the final state, then $\phi' = \phi-\Delta \phi$ and when $\phi'$ is in the final state $\phi' =  \phi+ \Delta \phi$.  This yields 
\beq
\frac{\partial}{\partial t} P(\phi, t) = \int \d \Delta \phi \Big[ P(\phi-\Delta \phi,t)\s W(\phi | \phi- \Delta \phi) -P(\phi, t)\s W(\phi+\Delta\phi | \Delta \phi) \Big]\,,
\label{eq:CKeqIntermediate}
\eeq
where we have included two compensating relative minus signs, one from the different changes of variables for the two terms, and another due to needing to flip the limits of integration after switched the integration variable $\Delta\phi \rightarrow -\Delta\phi$ in the second term.
Since we are assuming the main support of $W$ comes from local jumps, we want to Taylor expand the first term for fixed $\phi$.  To make performing this expansion transparent, it is then useful to redefine $W$ using
\begin{align}
\widetilde{W}(y,x) \equiv W(x+y|x)\,,
\label{eq:defWtilde}
\end{align}
so that \cref{eq:CKeqIntermediate} becomes
\begin{align}
\frac{\partial}{\partial t} P(\phi, t) = \int \d \Delta \phi \Big[ P(\phi-\Delta \phi,t)\s \widetilde{W}(\Delta\phi, \phi- \Delta \phi) -P(\phi, t)\s \widetilde{W}(\Delta\phi, \phi) \Big]\,.
\label{eq:CKeqIntermediate2}
\end{align}
Then Taylor expanding the first term about a fixed value of $\phi$ yields
\begin{align}
 \int \d \Delta \phi\, P(\phi-\Delta \phi,t)\s \widetilde{W}(\Delta\phi, \phi- \Delta \phi) &=  \int \d \Delta\phi \sum_{n=0}^\infty \frac{1}{n!} 
\bigg(\!\! - \!\Delta \phi \frac{\partial}{\partial\phi} \bigg)^n P(\phi,t)\s \widetilde{W}(\Delta\phi,\phi)\notag\\[4pt]
&= \sum_{n=0}^\infty \frac{1}{n!}  \frac{\partial^n }{\partial\phi^n}   \int \d \Delta \phi\, \big(\!-\!\Delta \phi\big)^n \,P(\phi,t)\widetilde{W}(\Delta\phi,\phi)\,,
\end{align}
where in the last line, we used the fact that $\Delta \phi$ is independent of $\phi$ to pull the derivatives outside of the integral.
Plugging this expansion into \cref{eq:CKeqIntermediate2}, we see that the $n=0$ term cancels so that 
\beq
\frac{\partial}{\partial t} P(\phi, t) = \sum_{n=1}^\infty \frac{1}{n!}  \frac{\partial^n }{\partial\phi^n}  \,\Omega_n(\phi)\s P(\phi,t)\,,
\eeq
where 
\beq
\Omega_n(\phi) \equiv  \int \d \Delta \phi\, \big(\!-\!\Delta \phi\big)^n\, W(\phi + \Delta\phi|\Delta \phi)\,,
\label{eq:alphan}
\eeq 
and we have used \cref{eq:defWtilde} to write this expression in terms of the original transition rate $W$.  Note that all terms in this expansion are total derivatives, as required for the conservation of the total probability.  We also see that $\Omega_n(\phi)$ encodes the $\phi$-dependence of the $n^\text{th}$ moment of the distribution.  For the theories of interest here, $\Omega_n(\phi)$ will admit a polynomial expansion in $\phi$: 
\begin{align}
\Omega_n(\phi) = \sum_{m=0}^\infty \frac{1}{m!} \Omega_n^{(m)} \phi^m\,,
\label{eq:alphanm}
\end{align}
for some coefficients $\Omega_n^{(m)}$.

Thus far, all of this discussion was very general.  If we specialize to the case of leading order Stochastic Inflation, we can compare this expanded result with \cref{eq:FPeqStocInf} to identify that the $n=1$ ``drift'' term is proportional to the derivative of the potential, 
\begin{align}
V = \sum_{\ell} \frac{1}{\ell!} c_\ell \phi^\ell\,.
\end{align}
so that matching with \cref{eq:FPeqStocInf} implies
\begin{align}
\Omega_1^{(m)} = \frac{1}{3H} c_{m+1} \,.
\end{align}
Hence, for $n=1$, $m$ tracks the polynomial interactions that could appear in a generic potential.  Moving to the $n=2$ ``noise'' term, we again can compare to \cref{eq:FPeqStocInf} to find 
\begin{align}
\Omega_2^{(0)} = \frac{H^3}{4\pi^2}\,.
\end{align}
In this case, the $m>0$ terms correspond to higher order corrections.

To summarize, if we assume that the UV theory has only a $\lambda \phi^4$ interaction, we conclude that the generalized evolution equation that describes Stochastic Inflation takes the form 
\begin{align}
\frac{\partial}{\partial t} P(\phi, t) =  \sum_{n=2}^\infty \frac{1}{n!}  \frac{\partial^n }{\partial\phi^n} \bigg[\sum_{m=0}^\infty \frac{1}{m!} \Omega_n^{(m)} \phi^m\, P(\phi,t)\bigg] + \frac{1}{3H}  \frac{\partial}{\partial\phi}\bigg [ V'(\phi) P(\phi,t ) \bigg] \,,
\label{eq:BLOFokkerPlMasterEq}
\end{align}
where $V'$ is the $\phi$ derivative of the potential, which includes the matching corrections required to obtain the accuracy of interest.

To develop some intuition for what the $\Omega_n$ corrections are capturing, we can interpret $W(\phi + \Delta\phi|\Delta \phi)$ as the probability distribution of transitions of size $\Delta \phi$.  Then $\Omega_n(\phi)$ is simply the $n^\text{th}$ moment of this distribution; the first and second moments $\Omega_1(\phi)$ and $\Omega_2(\phi)$ are the complete set of inputs for a Gaussian distribution. Furthermore, if $\Omega_2$ has non-trivial $\phi$ dependence, i.e., $\Omega_2^{(m\neq0)}\neq 0$, then the variance of the noise depends on the starting location of the jump.  Finally, if $\Omega_{n=3}(\phi) \neq 0$, then we know our distribution is non-Gaussian. More generally, we interpret the $n>2$ terms in the generalized equation governing Stochastic Inflation in \cref{eq:BLOFokkerPlMasterEq} as encoding contributions from non-Gaussian noise generated by the UV $\lambda \phi^4$ interaction.  Therefore, we conclude that corrections to \cref{eq:FPeqStocInf} are of three types: 
\begin{itemize}
\item Higher order ``noise'' terms captured by $\partial^n/\partial\phi^n$ with $n>2$, see \cref{eq:alphan}.
\item Higher polynomial terms in $\Omega_n$, see \cref{eq:alphanm}.
\item Higher order terms in the potential $V$ via corrections to coefficients and the generation of higher polynomial $\phi$ terms.
\end{itemize}
Next, we will argue for how to relate the expansion in each of these quantities to the expansion in the UV quartic coupling $\lambda$ as it corrects the equilibrium solution in \cref{eq:equilibrium_0}. 

\subsubsection{Organizing Corrections Systematically}
Due to the underlying $\lambda \phi^4$ potential, we expect that the statistics of $\phi$ are neither Gaussian nor independent of the background value of the field.  We therefore expect corrections to the equation for Stochastic Inflation of the form discussed previously.  In this section, we will take the equilibrium solution in \cref{eq:equilibrium_0} and apply it to a UV theory with 
\beq
V(\phi) =\frac{1}{4!} \lambda\s \phi^4 \qquad \Longrightarrow \qquad P_{\rm eq}(\phi) =  C e^{- 8 \pi^2 \lambda \phi^4/ (3\cdot 4! \, H^4)} \equiv C e^{- (\phi/\phi_\text{eq})^4} \,.
\eeq
where $\phi^4_\text{eq} = 9 H^4/(\pi^2 \lambda)$.  This distribution has support over a field range $|\phi | \lesssim \phi_\text{eq}$ such that 
\beq
\big\langle \phi^4 \big\rangle =\frac{1}{4} \Gamma\big[\tfrac{5}{4}\big]\times\phi^4_\text{eq}  =  \frac{9}{4 \pi^2} \Gamma\big[\tfrac{5}{4}\big] \frac{H^4}{\lambda} \, .
\eeq
Therefore, we will organize the possible corrections by assuming the equilibrium scaling
\begin{align}
\phi \simeq \phi_\text{eq} \sim H \lambda^{-1/4}\,.
\end{align}
We will further assume that the corrections are generated as an expansion in perturbation theory, so that
\begin{align}
\Omega_2^{(2m)} \sim \lambda^{m};\qquad \Omega_3^{(2m+1)} \sim \lambda^{m+1};\qquad \Omega_4^{(2m)} \sim \lambda^{m+1};\qquad \text{and}\qquad c_{2\ell} \sim \lambda^{\ell-1}\,,
\end{align}
which we will see agrees with the explicit calculations presented below.  Putting all of this together allows us to determine the order in $\lambda$ for each term that appears in \cref{eq:BLOFokkerPlMasterEq}.  Note that we will assume the $\phi \to -\phi$ UV symmetry is preserved, which explains the absence of many terms. These contributions are as follows.  
\vskip 10pt
\noindent {\bf Leading Order (LO):}  Stochastic Inflation at leading order is given by \cref{eq:FPeqStocInf}:
\beq
\frac{\partial}{\partial t} P(\phi,t)= \frac{H^3}{8 \pi^2} \frac{\partial^2}{\partial \phi^2} P(\phi, t)  + \frac{1}{3 H}  \frac{\partial}{\partial \phi} \left [\frac{1}{3!} \lambda \phi^3 P(\phi,t ) \right] \,,
\eeq
where we have used the known leading order results $\Omega_2^{(0)} = H^3/(4\pi^2)$ and $V'(\phi) = \lambda \phi^3/3!$.
Both terms on the right hand side of this equation are ${O}\big(\lambda^{1/2}\big) \times P(\phi,t)$, which defines what we mean by ``LO.''    

\vskip 10pt
\noindent {\bf Next-to-leading Order (NLO):} Accounting for both $\Omega_n^{(m)}$ and corrections to the potential, we can determine that the next-to-leading corrections to Stochastic Inflation should take the form
\begin{align}
\frac{\partial}{\partial t} P(\phi,t) = O\big(\lambda^{1/2}\big) + \frac{\partial^2}{\partial \phi^2} \Big[ \Omega_2^{(2)} \phi^2 P(\phi, t) \Big] + \frac{1}{3 H}  \frac{\partial}{\partial \phi} \bigg [\frac{1}{5!} c_6  \phi^5 P(\phi,t ) \bigg] \,,
\end{align}
where the correction to the noise term $\Omega_2^{(2)} \sim \lambda$ and the correction to the potential $c_6 \sim \lambda$ will both be determined below, and $\Omega_2^{(1)} = 0$ due to the $\phi \to -\phi$ symmetry.  We see that these NLO terms are $O(\lambda)\times P(\phi,t)$.  These corrections have been previously calculated in~\cite{Gorbenko:2019rza,Mirbabayi:2020vyt}.
%\vskip 10pt

\clearpage
\noindent {\bf Next-to-next-to-leading Order (NNLO):} Following the same logic, we can find the form that the next order terms take: 
\begin{align}
\frac{\partial}{\partial t} P(\phi,t) =\,& O\big(\lambda^{1/2}\big) + O\big(\lambda\big) + \frac{H^3}{8 \pi^2} \frac{\partial^2}{\partial \phi^2} \Big( \Omega_2^{(4)} \phi^4 P(\phi, t) \Big)  \notag\\
&+ \frac{1}{3H}  \frac{\partial}{\partial \phi} \bigg [\frac{1}{7!} c_8 \phi^7 P(\phi, t ) \bigg] +  \frac{\partial^3}{\partial \phi^3}  \Big( \Omega_3^{(1)} \phi  P(\phi, t) \Big)\,,
\end{align}
where $\Omega_2^{(4)} \sim \lambda^2$ and $\Omega_3^{(1)} \sim\lambda$ will be determined by operator mixing in the next section and $c_8 \sim \lambda^3$, and $\Omega_3^{(0)} = 0$ due to the $\phi \to -\phi$ symmetry.  These NNLO terms are $O(\lambda^{3/2})\times P(\phi,t)$.  

Note that in addition to these corrections, we must also include subleading corrections to the parameters that already appear at lower order; these do not change the structure of the equation, but will of course be accounted for as we perform the calculation.  The rest of this paper is devoted to determining these coefficients systematically using the framework of SdSET, with a brief discussion of the physical implications of working with NNLO Stochastic Inflation.

\section{Soft de Sitter Effective Theory}
\label{sec:SdSET}
Stochastic Inflation, and corrections to it, are the consequence of quantum field theory in dS for scalar particles with masses $m^2 \ll H^2$.  This can be seen directly at leading order, where a variety of methods have been used to derive the Fokker-Planck equation~\cite{Starobinsky:1986fx,Nambu:1987ef,Starobinsky:1994bd,Enqvist:2008kt,Finelli:2008zg,Podolsky:2008qq,Seery:2010kh,Garbrecht:2014dca,Burgess:2015ajz,Gorbenko:2019rza,Baumgart:2019clc,Mirbabayi:2019qtx,Cohen:2020php,Mirbabayi:2020vyt,Baumgart:2020oby}.  However, many of these methods become cumbersome beyond leading order and often obscure how corrections arise. 

Soft de Sitter Effective Theory offers a method to compute the equations of Stochastic Inflation systematically to any order.  The key advantage offered by SdSET is that power counting is manifest, thereby making corrections easy to identify. Furthermore, loop integrals are scaleless and regulated by (dynamical) dim reg and thus preserve the power counting that is manifest in the action.  In addition to fixing the values of the SdSET Wilson coefficients, the UV theory sets the initial conditions for the effective theory fields.   We will be specifically interested in understanding the corrections to Stochastic Inflation for the concrete example of $\lambda\phi^4$ theory in dS, where the UV action is given above in~\cref{eq:Lphi4}.

In this section, we will review the machinery of SdSET and how it arises from a given UV theory.  In the subsequent sections, we will use this technology to match SdSET to $\lambda \phi^4$ theory and then use it derive the equations of Stochastic Inflation at NNLO.

\subsection{In-In Correlators}
\label{sec:InIn}
As we will see below, Stochastic Inflation is equivalent to the renormalization group equations that govern how composite operators mix.  One approach to determining the operator mixing is to compute the divergences of in-in correlation functions involving composite operators.  This section is devoted to setting up the relevant framework. We work in the interaction picture, where fields are quantized using the solutions to their quadratic equations of motion.  A free scalar fields in dS can be expressed as a mode expansion:
\beq
\phi(\x,\tau) = \int \frac{\dd^3 k}{(2\s\pi)^3}\s e^{i\s \ksub\cdot \x} \Big(\bar \phi\big(\s \k,\tau\big) a_{\ksub}^\dagger + \bar \phi^*\big(\s \k,\tau\big) a_{-\ksub} \Big) \, ,
\eeq
where $\tau = -1/[aH]$ is the conformal time, and  $a_\ksub^{\dagger}$ and $a_\ksub$ are the canonical creation and annihilation operators respectively that satisfy 
\begin{align}
\Big[a_\ksub^{\dagger}\s,\s a_{\ksub'} \Big] = (2\s \pi)^3\s \delta\big(\k-\kp\big)\,.
\end{align}
In the Bunch-Davies vacuum, one finds the positive frequency modes are given by
\beq\label{eq:mode_m}
\bar \phi\big(\s\k,\tau\big) =-i\s e^{i\s\left(\nu+\frac{1}{2}\right) \frac{\pi}{2}} \frac{\sqrt{\pi}}{2} H(-\tau)^{3 / 2} \text{H}_{\nu}^{(1)}(-k\s \tau) \qquad \text{with}\qquad \nu= \sqrt{\frac{9}{4} -\frac{m^2}{H^2}}\, ,
\eeq
so that $\nu = 3/2$ corresponds to a massless field (see e.g.~\cite{Baumann:2018muz} for review).  The observables of this theory are equal-time in-in correlation functions, which are computed via 
\begin{align}
\langle {\rm in}|  Q(t) |{\rm in} \rangle &= \sum_{N= 0}^{\infty} i^{N} \int_{-\infty}^{t} \d t_{1} \int_{-\infty}^{t_{1}} \d t_{2} \,...\, \int_{-\infty}^{t_{N-1}} \d t_{N} \nonumber \\[4pt] 
&\hspace{60pt} \times\Big\langle\big[H_{{\rm int}}\big(t_{N}\big), \big[H_{{\rm int}}\big(t_{N-1}\big),  \,...\,\big[H_{{\rm int}}\big(t_{1}\big), Q_{\mathrm{int}}(t) \big] \,...\,\big]\big]\Big\rangle  \, , 
\label{eq:commutator}
\end{align}
using 
\beq
H_{\rm int}(t) = \int \d^3 x \sqrt{-g}\s \frac{\lambda}{4!} \phi^4(\x,t) \, .
\label{eq:Hint}
\eeq
We will be interested in multi-field correlators in the long wavelength limit, so that  $Q(t) = \phi\big(\k_1, t\big) \,...\,\phi\big(\k_n,t\big)$. In general, this expression must be generalized to allow for the $i\epsilon$ prescription that projects the initial state onto the interacting vacuum, but this form is the most useful starting point for understanding the superhorizon evolution.

For illustration, we can compute the tree-level four point correlation function as
\begin{align}
\Big\langle \phi\big(\s\k_1\big) \phi\big(\s\k_2\big) \phi\big(\s\k_3\big) \phi\big(\s\k_4\big) \Big\rangle'_{\rm tree} =  i\frac{\lambda}{4!} \int^t \d t_1\, \d^3 x\,  a(t)^3  \Big\langle \Big[ \phi^4(\x,t_1),\phi\big(\s\k_1\big) 
\,...\, \phi\big(\s\k_4\big)\Big] \Big\rangle'\,,\label{eq:tree_tri}
\end{align}
where the $\phi\big(\k\s\big)$ fields are all evaluated at the same time $t$, and we have introduced the notation
\begin{align}
\big\langle \,...\,\rangle = (2\pi)^3 \delta^3\big(\textstyle\sum \k_i\big) \big\langle \,...\,\rangle'\,.
\label{eq:defPrime}    
\end{align}
At tree-level, we can simply use the massless mode functions,\footnote{We will use $\frac{3}{2} - \nu \equiv \alpha \neq 0$ as a regulator when we encounter loops.}
\begin{align}
\label{massless mode functions sub horizon}
\bar \phi\big(\s\k,\tau\big) \,\,\to\,\, \frac{H}{\sqrt{2 k^3} } (1- i k\tau) e^{i k \tau} \, ,
\end{align}
to evaluate this expression 
\begin{align}
&\Big\langle \phi\big(\s\k_1\big) \phi\big(\s\k_2\big) \phi\big(\s\k_3\big) \phi\big(\s\k_4\big) \Big\rangle'_{\rm tree} 
=\lambda\, 2\s {\rm Im}  \int^{\tau} \frac{\d\tau_1}{(-H \tau_1)^4}  \prod_{i=1}^4 \frac{(1-i k_i \tau_1) (1+i k_i \tau)}{2 k_i^3}e^{i k_i (\tau_1-\tau)} \nonumber
\\[5pt]
&=  \frac{\lambda}{8(k_1 k_2 k_3 k_4)^3} \Bigg[ \frac{1}{3}\Bigg(\sum_i k_i^3\Bigg) \bigg(\log \frac{k_t}{[aH]} + \gamma_E + \frac{1}{3}-2\bigg)-\frac{k_1 k_2 k_3 k_4}{k_t} - \frac{1}{9} k_t^3 \nonumber \\[5pt]
&\hspace{112pt} + 2 \sum_{i<j<\ell}  k_i k_j k_\ell +\frac{1}{3} k_t \Bigg(\sum k_i^2 - \sum_{i<j} k_i k_j\Bigg) \Bigg]\,,
\label{eq:fourPtCorUV}
\end{align}
where $k_t =k_1+k_2+k_3+k_4$.  In the final step, we have used $\tau = -1/[aH]$, have kept the $\log [aH]$ and time-independent contributions, and have not included terms that vanish as $[aH] \to \infty$. In deriving this expression, we have expressed the commutator as the imaginary part of the integral, which holds for real fields at first order in the $H_{\rm int}$ expansion, namely
\beq
i \Big( \big\langle H_{\rm int}(t_1) Q(t) \big\rangle -\langle  Q(t) H_{\rm int}(t_1) \rangle \Big) = 2\s {\rm Im} \, \big\langle  Q(t) H_{\rm int}(t_1) \big\rangle \,, 
\eeq
for a real operator $Q(t)$.  

\subsection{Taking the Long Wavelength Limit}
In this section, we review how to determine in-in correlators in the soft limit using SdSET.  The starting point is to decompose the UV fields according to 
\begin{align}
\phi_S\big(\s \k, \t\big) 
&= H \Big( [a(\t)\s H]^{-\alpha} \varphi_+(\k,\t) + [a(\t)\s H]^{-\beta} \varphi_-\big(\s \k,\t\big) \Big) \ . \label{eq:phiS}
\end{align}
where $\phi(\x,\t) = \phi_S(\x,\t) + \Phi_H(\x,\t)$ is split into soft (superhorizon) and hard (subhorizon) modes, and we have introduced a dimensionless time variable $\t \equiv H t$ so that the mass dimension of operators tracks the EFT power counting.  The parameters $\alpha$ and $\beta$ are subject to the constraint $\alpha+\beta =3$.  This is straightforward to derive from the top down, as these parameters are determined by the mass via $\alpha = 3/2 -\nu$ and $\beta = 3/2+ \nu$, where $\nu$ is defined in \cref{eq:mode_m}.

The decomposition of $\phi_S$ into $\vp$ and $\vm$ is exact in the free theory. Taking the limit $k\s\tau \ll 1$ and using $\tau  = -1/[a\s H]$ in \cref{eq:mode_m}, we find
\begin{subequations}
\begin{align}
\varphi_+\big(\x, t\big) &= \int \frac{\dd^3 k}{(2\s\pi)^3} e^{i\s \ksub\cdot \vec x} \s \bar{\varphi}_+\big(\s\k,t\big) \tilde{a}_{\k} \\[8pt]
\varphi_-\big(\x, t\big) &= \int \frac{\dd^3 k}{(2\s\pi)^3}e^{i\s \ksub\cdot \vec x}\s \bar \varphi_-\big(\s\k,t\big) \tilde{b}_{\k} \ ,
\end{align}
\end{subequations}
where
\begin{align}
\bar \varphi_+ = C_\alpha  \frac{1}{\sqrt{2}\s k^{\frac{3}{2}-\alpha}}\,,  \qquad \text{and} \qquad
\bar \varphi_- = D_\beta \frac{1}{\sqrt{2}\s k^{\frac{3}{2}-\beta}} \, ,
\end{align}
and

\begin{align}
C_\alpha &= 2^{1-\alpha} \s \frac{ \Gamma\big(\frac{3}{2} - \alpha\big) }{\sqrt{\pi}}\,,
\qquad \text{and} \qquad
D_\beta =-2^{1-\beta }  \s \frac{\sqrt{\pi }  }{\cos (\pi\s \beta )\s\Gamma \big(\beta -\frac{1}{2}\big)} \ .
\label{eq:Calpha}
\end{align}
The operators $\tilde{a}_{\k}$ and $\tilde{b}_{\k}$ are given in terms of the UV creation and annihilation operators of the form
\begin{align}
\tilde a_\ksub = e^{i\s \delta_\nu} a^{\dagger}_\ksub + e^{-i\s \delta_\nu} a_{-\ksub}\,, \qquad\text{and} \qquad
\tilde b_\ksub =i\Big( e^{-i\s \delta_\nu} a^{\dagger}_\ksub - e^{i\s \delta_\nu} a_{-\ksub}\Big) \, .
\end{align}
From the UV theory, we determine that the operators commute with themselves
\begin{align}
\Big[\tilde a_\ksub^\dag \s,\s \tilde a_\ksub\Big] = \Big[\tilde b_\ksub^\dag \s,\s \tilde b_\ksub\Big]  =0\ .
\label{eq:tildecomm}
\end{align}
Nevertheless, these operators still have non-zero correlation function
\begin{subequations}
\begin{align}
\Big\langle \tilde a_\ksub\, \tilde a_{\ksub'} \Big\rangle &= (2\pi)^3\s \delta\big(\k+\kp\big) \\[5pt]
\Big\langle \tilde b_\ksub\, \tilde b_{\ksub'} \Big\rangle &= (2\pi)^3\s \delta\big(\k+\kp\big) \ ,
\end{align}
\label{eq:CreAnnStocRanVarCondition}%
\end{subequations}
where $\langle .. \rangle \equiv \langle 0 | .. | 0 \rangle $, and $|0\rangle$ is the vacuum that is annihilated by $a_{\vec k}$.  This gives rise to classical statistical power spectra
\begin{subequations}
\begin{align}
\Big\langle \varphi_+\big(\k \s\big)\, \varphi_+\big(\kp\s \big) \Big \rangle &=\frac{C_\alpha^2}{2} \frac{1}{k^{3-2\s\alpha}}  \s (2\s\pi)^3 \s\delta\big( \k+\kp \big)
\label{eq:GaussianPowerSpec}\\[8pt]
\Big\langle \varphi_-\big(\k \s\big)\, \varphi_-\big(\s\kp \big) \Big\rangle &=\frac{D_\beta^2}{2} \frac{1}{k^{3-2\s\beta}}  \s (2\s\pi)^3 \s\delta\big( \k+\kp \big)\,,
\end{align}%
\label{eq:GausianPowerSpecBoth}%
\end{subequations}%
where $C_\alpha$ and $D_\beta$ are defined in \cref{eq:Calpha} above.  Note that in the massless limit $\alpha \to 0$, we reproduce the famous scale invariant power spectrum.

The corrections to this mapping can be systematically accounted for by matching between the UV theory and the EFT, see \cref{sec:Matching}.
The fields $\vp$ and $\vm$ have well defined power counting; they carry operator dimension $\alpha$ and $\beta$ respectively.    
After utilizing field redefinitions, on-shell conditions, and power counting to remove redundant operators, the low energy effective action is given by 
\begin{align}
S_{\pm} = \int \dd^3 x\, \dd\t\s  \Bigg[ -\nu\s\big(\dot{\varphi}_+\s \varphi_-  - \varphi_+\s\dot{\varphi}_-\big) -  \sum_{n\geq 2}^{\infty}\s [a\s H]^{3- n\s\alpha -\beta}\s \frac{c_{n, 1}}{n!}\s \varphi_+^{n}\s \varphi_- \Bigg] \,,
\label{eq:ActionEFT}
\end{align}
where the $c_{n,1}$ are dimensionless Wilson coefficients.  Note that $\t$ carries dimension zero by SdSET power counting, so marginal operators are dimension three.  This explains why we have only included operators with a single factor of $\varphi_-$ since these are the only terms that become marginal in the massless limit ($\alpha \to 0$).
Additionally, we have not included any terms with $\vec\partial$, which start at dimension five and are therefore power suppressed by at least $k^2/[aH]^2$.

In addition to an action, SdSET requires specifying \emph{initial conditions} for the fields $\vp$ and $\vm$ that are acquired from the time evolution prior to horizon crossing.  These initial conditions are random such that, to leading order, $\vp$ behaves as a classical stochastic variable with correlations fixed by matching
\beq
\Big\langle \varphi_{+}\big(\s \k_1\big) \,...\,  \varphi_{+}\big(\s \k_N\big) \Big\rangle_{{\rm IC}^{(n)}} =  \mathcal{K}^{-3\s(N-1) + N\s \alpha} F_{(n)}\big(\{\q_i \} \big) \s (2\s\pi)^3\s \delta\Big(\sum \k_i\Big) \,,
\label{eq:NpointCorr}
\eeq 
where $\mathcal{K}$ is a reference momentum scale,  $F_{(n)}\big(\{\q_i \} \big)$ encode the dependence on the rescaled momenta $\vec{q}_i = \vec{k}_i/\mathcal{K}$, and the $(n)$ subscripts track the order in the $\lambda$ perturbative expansion for each contribution; the two point correlators in \cref{eq:GausianPowerSpecBoth} should be viewed as $\langle\,...\,\rangle_{{\rm IC}^{(0)}}$.
Because $\vp$ is time independent to leading order in the EFT, the initial conditions are determined by matching the time independent terms.

We evaluate time integrals in the EFT using 
\beq
\int^\t_{-\infty} \d\t' \big[a\big(\t'\big)H\big]^{\gamma} = \frac{1}{\gamma} [a(\t)H]\,,
\label{eq:timeInts}
\eeq
where we assume that this holds even when $\gamma < 0$.  This analytic continuation enforces that the contributions from early times vanish, thereby ensuring that power law divergences associated with physics at horizon crossing are automatically absorbed into the initial conditions (in close analogy with how dim reg treats power law divergences).

Cosmological correlators are determined in SdSET using the same in-in formalism as applied to the UV theory, see \cref{sec:InIn}.  For illustration, we can compute the tree-level trispectrum using \cref{eq:commutator} and the canonical commutator
\beq
\Big[\varphi_+ \big(\x,\t\big) \s,\s \varphi_-\big(\x^{\s \prime},\t\big)\Big] =   -  \frac{i}{2\s\nu}\s  \delta\big(\x-\x^{\s\prime}\big)  \, .
\eeq
Performing the time integrals using~\cref{eq:timeInts}, we find 
\begin{align}
 &\Big\langle \varphi_+\big(\k_1\big)\s  \varphi_+\big(\k_2\big)\s \varphi_+\big(\k_3\big)\s \varphi_+\big(\k_4\big) \Big\rangle = \bigg\langle \bigg[\varphi_+(\k_1)  \varphi_+(\k_2) \varphi_+(\k_3) \varphi_+(\k_4)\,,\notag\\[3pt]
 &\hspace{160pt}(-i)\frac{c_{3,1}}{3!} \int \d \t'\, \d^3  x'\, \big[a\big(\t'\big) H\big]^{-2\alpha} \varphi_+^3\s \varphi_-\big(\x\s',\t'\big) \bigg]\bigg\rangle \nonumber \\[7pt]
 &=- \frac{c_{3,1}}{2\nu} \bigg\langle \varphi_+(\k_2) \varphi_+(\k_3) \varphi_+(\k_4)  \int \d^3  x' \varphi_+^3\s\bigg\rangle \int \d\t' \big[a\big(\t'\big) H\big]^{-2\alpha}+{\rm permutations}  \nonumber \\[7pt]
 &= \frac{c_{3,1}}{2\nu} \frac{C_\alpha^6 \sum_i k_i^{3-2\alpha}}{\big(k_1 k_2 k_3 k_4\big)^{3-2\alpha}} \bigg(\frac{[aH]^{-2\alpha}}{2\alpha} \bigg)  \,,
\label{eq:eft_tri}
\end{align}
where $c_{3,1}$ is the Wilson coefficient for the $\varphi_+^3 \varphi_-$ operator, see~\cref{eq:ActionEFT}, and we used \cref{eq:GaussianPowerSpec} to evaluate the field contractions.
In addition, we must include any trispectrum associated with the initial conditions.

\subsection{(Dynamical) Renormalization}

Loops corrections are calculated in the SdSET using dynamical dimensional regularization (dynamical dim reg).  Rather than varying the spacetime dimension, we instead float the dynamical exponents $\alpha$, and evaluate loop integrals by analytic continuation in $\alpha$. Then when we encounter divergences as $\alpha \to 0$, they will be accompanied by log corrections to the time evolution, in exact analogy with conventional dim reg.  To keep the units fixed as we vary $\alpha$, we will introduce the necessary powers of $[aH]$ such that $\vp$ stays dimensionless.  Then we take $\alpha = 0$ at the end of the calculation, so that:
\begin{align}
\phi_S\big(\s \k, \t\big) 
\to H \Big(  \varphi_+\big(\k,\t\big) + [a(\t)\s H]^{-3} \varphi_-\big(\s \k,\t\big) \Big) \, , \label{eq:phiS_0}
\end{align}
so that $\vp$ corresponds to a massless mode.

Since we are working within the EFT, we will typically encounter vanishing scaleless integrals.  Then we can isolate the UV divergence in the usual way by regulating the IR with a dimensionful parameter $K$:
\begin{align}
\langle \O \,...\,\rangle &\propto [aH]^{-2\alpha}\int \frac{\d^3 p}{(2\pi)^3} \frac{1}{p^{3-2\alpha} } \notag \\[5pt]
&\to [aH]^{-2\alpha}\int \frac{\d^3 p}{(2\pi)^3} \frac{1}{(p^2+K^2)^{3/2-\alpha}} = [aH]^{-2\alpha}\frac{1}{8\pi^{3/2}}\frac{\Gamma[-\alpha]}{\Gamma[3/2-\alpha]} K^{2\alpha}\notag\\[5pt]
&\to -\frac{1}{2\pi^2} \left(\frac{1}{2\alpha} + \log \frac{K}{[aH]} -\log 2\right)\,,
\end{align}
where we have taken the limit $\alpha \to 0$ in the third line.
Having isolated these UV divergent contributions, we can use them to determine the (dynamical) RG flow, which resums a series in $\log [aH]$. 

Having regulated the divergence, we can then absorb it into the renormalization of the operator
\beq
\O = Z_\O\s \O_R    \qquad\text{with} \qquad Z_\O -1 \propto   -\frac{1}{2\pi^2} \left(\frac{1}{2\alpha} + \log \frac{K_\star}{[aH]_\star} \right)\,,
\eeq
so that 
\beq
\langle \O_R \,...\,\rangle \propto \frac{1}{2\pi^2} \log \frac{\hspace{2pt}[aH]_\star}{\hspace{-1pt}[aH]} + \log \frac{K}{\hspace{1pt} K_\star}\,,
\eeq
where $[aH]_\star$ and $K_\star$ are energy and momentum scales we have invented to make the logs small, i.e., subtraction points.  By definition the bare operator $\O$ is independent of these arbitrary parameters, and thus
\beq
\frac{\d}{\d \log [aH]_\star}\s \O = 0 \quad\Longleftrightarrow\quad \frac{\d}{\d \log [aH]_\star} \s\O_R  = -\frac{\d\log Z_\O}{\d \log [aH]_\star}\s \O_R \equiv \gamma\s \O_R\,,
\eeq
where the anomalous dimension $\gamma$ is independent of $[aH]$.  We are working with a scheme where $Z_\O$ is diagonal at the scale $[aH]_\star$.
In general, $\gamma \to \gamma_{ij}$ is a matrix that acts on the space of operators, and which encodes both the anomalous scaling and mixing of these operators.  

%%%%%%%%%%%%%%%%%%%%%%%%%%%%%%%%%%%%%%%%%%%%%
\subsection{Matching and Initial Conditions}
%%%%%%%%%%%%%%%%%%%%%%%%%%%%%%%%%%%%%%%%%%%%%
SdSET provides an effective description for the time evolution of scalar modes that have crossed the Hubble horizon.  Their state at the time of horizon crossing cannot be computed within the EFT, and instead has to be provided as an additional input.  This is why SdSET requires matching for both the initial conditions and the EFT Wilson coefficients.  This is not unique to SdSET, but is necessarily part of any EFT description of the post inflationary universe as well, see e.g.~\cite{Baumann:2010tm}.

When defined using dynamical dim reg, SdSET is a so-called continuum EFT~\cite{Georgi:1994qn}.  Concretely, the time integrals include arbitrary early times, even though the EFT does not provide a model of the subhorizon physics, since it relies on the use of the long-wavelength mode functions at all times.  Importantly, as with all continuum EFT, these early time integrals only make scaleless (and therefore vanishing) contributions.  Thus we can integrate over all times, such that the regulated integrals respect the low energy symmetries and the EFT power counting.  Underlying the validity of this procedure is the fact that the subhorizon physics only alters the initial conditions and thus we can fully account for all subhorizon evolution by matching a UV theory onto SdSET.

One can demonstrate how matching separates into Wilson coefficient and initial condition corrections more directly using a hard cutoff to evaluate time and momentum integrals, i.e.,~treating the theory as a Wilsonian EFT~\cite{Georgi:1994qn}.  Let $k_i$ denote the magnitudes of the momenta appearing in the cosmological correlator, and let the cutoff of the momentum integrals $\Lambda$ be much greater than any of the $k_i$.  In addition, we denote the time cutoff by $t_\Lambda$; this is the time when all the EFT modes are in the superhorizon limit, and thus it specifies when we set the initial conditions for the EFT.

Before $t_\Lambda$, a subset of EFT modes are subhorizon, and so one must use the UV theory to describe them.  To match onto the EFT, it is useful to split the full theory time evolution into pieces before and after $t_\Lambda$.  Let $U_I(t,t')$ represent the interaction picture propagator, and $|\Omega\rangle$ be the UV vacuum state.  This decomposition can then be written as 
\begin{align}
&\big\langle\Omega\big|\phi\big(\k_1,t\big)\,...\,\phi\big(\k_n,t\big)\big|\Omega\big\rangle = \big\langle\Omega\big| U_I^\dagger(t,-\infty) \phi_{I}\big(\k_1,t\big)\,...\,\phi_{I}\big(\k_n,t\big)U_I(t,-\infty)\big|\Omega\big\rangle\notag\\[5pt]
&\hspace{29pt}= \big\langle\Omega\big| U_I^\dagger(t_{\Lambda},-\infty) U_I^\dagger(t,t_{\Lambda})  \phi_{I}\big(\k_1,t\big)\,...\,\phi_{I}\big(\k_n,t\big)U_I(t,t_{\Lambda})U_I(t_{\Lambda},-\infty)\big|\Omega\big\rangle\,,
\label{full theory decomposition}
\end{align}
where $\phi_I$ is the interaction picture field. 

One can trivially re-write \cref{full theory decomposition} as an expectation value of $\ket{\psi} = U_I(t_{\Lambda},-\infty)\ket{\Omega}$, the state of the full theory fields at $t_\Lambda$.  One can then integrate out the modes whose wave vector magnitudes satisfy $k > \Lambda$, so that the remaining modes are superhorizon after $t_\Lambda$.  It was shown in~\cite{Cohen:2020php} that, after integrating out these so-called hard modes, the resulting action for the superhorizon modes $\vp$ and $\vm$ is local.  We can then evolve these modes from $t_\Lambda$ to $t$ using the unitary time evolution operators defined within the SdSET itself:
\begin{align}\label{EFT representation of correlator}
\big\langle\Omega\big|\phi\big(\k_1,t\big)\,...\,\phi\big(\k_n,t\big)\big|\Omega\big\rangle&= \notag \\[5pt]
\big\langle\psi_{\rm EFT}(t_\Lambda)\big| U_{I,{\rm EFT}}^\dagger&(t,t_{\Lambda})  \phi_{S}\big(\k_1,t\big)\,...\,\phi_{S}\big(\k_n,t\big)U_{I,{\rm EFT}}(t,t_{\Lambda})\big|\psi_{\rm EFT}(t_\Lambda)\big\rangle\,,
\end{align}
where $U_{I,{\rm EFT}}$ is the interaction picture propagator obtained from \cref{eq:ActionEFT}, and  $\phi_S$ is given by \cref{eq:phiS_0}. The state $|\psi_{\rm EFT}(t_\Lambda) \rangle$ is the EFT state inherited from $|\psi\rangle$ that results from integrating out the hard modes; this state encodes the initial conditions for $\varphi_+$ and $\varphi_-$.

One then fixes the EFT parameters in \cref{eq:ActionEFT} and the initial state $\ket{\psi_{\rm EFT}(t_\Lambda)}$ by matching to full theory correlators.  In practice, the full form of $\ket{\psi_{\rm EFT}(t_\Lambda)}$ is more information than is needed to derive the late time behavior of $n\text{-point}$ correlation functions of $\phi$.  Instead, it is sufficient to determine a finite number of $n$-point functions,
\begin{align}
  \big\langle\psi_{\rm EFT}(t_\Lambda)\big| \vp\big(\k_1\big)\,...\, \vp\big(\k_n\big)\big|\psi_{\rm EFT}(t_\Lambda)\big\rangle \, ,
\end{align}
from matching.  Furthermore, this shows that all of the contributions from $t < t_\Lambda$ are encode in the state $\ket{\psi_{\rm EFT}(t_\Lambda)}$ or, equivalently, in the initial conditions set at $t_\Lambda$.  Finally, since $t_\Lambda$ is an unphysical cutoff parameter, no physical results can depend on it.  In particular, the initial conditions can be identified as the \emph{time-independent} contribution from the UV correlators.  This implies that we do not need to rely on hard cutoff to derive the initial conditions.  In what follows, we will regulate the theory using dynamical dim reg, and will derive the initial conditions by identifying the time-independent contributions to correlators that appear when matching.

To illustrate the matching procedure, it is useful to first consider a simple example.   By expanding \cref{eq:commutator} to leading order in $c_{3,1}$ and setting $\ket{\rm in} = \ket{\psi_{\rm EFT}}$ we find that the tree-level EFT prediction for the four point function of $\phi$ is 
\begin{align}
&\big\langle\Omega\big|\phi\big(\k_1,t\big)\,...\,\phi\big(\k_4,t\big)\big|\Omega\big\rangle = \left(\frac{\tau}{\tau_\Lambda} \right)^{4\alpha}\bigg[ \big\langle\psi_{\rm EFT}(t_\Lambda)\big|\vp\big( \k_1\big) \,...\,\vp\big( \k_4\big) \big|\psi_{\rm EFT}(t_\Lambda)\big\rangle\notag \\[5pt]
&\hspace{6pt} + 2i c_{3,1} \int_{\tau_\Lambda}^\tau \frac{\d\tau_1}{\tau_1^{1+2\alpha}}  \int \d^3 x\,  \big\langle\psi_{\rm EFT}(t_\Lambda)\big|\big[\vp\big( \k_1\big) \,...\,\vp\big( \k_4\big), \vp^3 \vm(\x,\tau_1)\big] \big|\psi_{\rm EFT}(t_\Lambda)\big\rangle \bigg]\,.
\label{four point example}
\end{align}
As in all perturbative SdSET calculations, the time dependence and field contractions factorize at the integrand level.  The EFT prediction is then parameterized by the coupling $c_{3,1}$ and two expectations values of $\ket{\psi_{\rm EFT}}$.  

The expectation values parameterize the subhorizon evolution of the EFT modes before $t_{\Lambda}$.  If the full theory is perturbative, then the subhorizon evolution is approximately Gaussian.  Assuming the UV theory is given by $\lambda\phi^4$, then to $O(\lambda)$ we can write 
\begin{align}
&\big\langle\psi_{\rm EFT}(t_\Lambda)\big|\varphi_{I,+}\big( \k_1\big) \,...\,\varphi_{I,+ }\big(\k_4\big) \big|\psi_{\rm EFT}(t_\Lambda)\big\rangle =\Big(\big\langle\varphi_{I,+}\big( \k_1\big) \varphi_{I,+}\big( \k_2\big)\big\rangle_{{\rm IC}^{(0)}}\big\langle\varphi_{I,+}\big( \k_3\big)\varphi_{I,+}\big( \k_4\big) \big\rangle_{{\rm IC}^{(0)}}\notag\\[5pt]
&+ 2 \big\langle\varphi_{I,+}\big( \k_1\big) \varphi_{I,+}\big( \k_2\big)\big\rangle_{{\rm IC}^{(0)}}\big\langle\varphi_{I,+}\big( \k_3\big)\varphi_{I,+}\big( \k_4\big) \big\rangle_{{\rm IC}^{(1)}} + {\rm perms}\Big) +  \big\langle \varphi_{I,+}\big( \k_1\big) \,...\,\varphi_{I,+}\big( \k_4\big)\big\rangle_{{\rm IC}^{(1)}}\,, \notag\\
\end{align}  
where the final term on the RHS encodes the non-Gaussian contribution to the subhorizon evolution of the modes. There is a similar formula for the expectation value on the second line of \cref{four point example}.  However, since it is already multiplied by $c_{3,1}$, we only need the $\langle ...\rangle_{\text{IC}^{(0)}}$ contribution to the expectation value at this order.

The SdSET two point functions derived from the free theory are given in~\cref{eq:GausianPowerSpecBoth}.  To include the impact of the UV interaction on the EFT, we compute $ \big\langle \varphi_{I,+}\big( \k_1\big)\,...\,\varphi_{I,+}\big( \k_4\big)\big\rangle_{\rm IC}$ and $c_{3,1}$ by matching to the full theory.  The superhorizon evolution between $t_\Lambda$ and $t$ generates the term proportional to $c_{3,1}$, and so we can isolate the initial conditions contribution by evaluating both sides of \cref{four point example} at $t_\Lambda$, giving
\begin{align}
\label{four point initial conditions matching}
\big\langle \varphi_{I,+}\big( \k_1\big) \,...\,\varphi_{I,+}\big( \k_4\big)\big\rangle_{\rm IC} = \big\langle\Omega\big|\phi\big(\k_1,t_\Lambda\big)\,...\,\phi\big(\k_4,t_\Lambda\big)\big|\Omega\big\rangle_{\rm connected}\,,
\end{align}
where the ``connected'' subscript refers to the fact that this does not include the contributions from products of lower point contractions.
The RHS of \cref{four point initial conditions matching} can be computed using \cref{eq:commutator}, where $H_I$ is given by the full theory interaction Hamiltonian and the time integrals extend from $-\infty$ to $t_\Lambda$.  Since the integral's main region of support occurs when the modes are subhorizon, we cannot replace the mode functions with their late time behavior, and instead have to use their full UV form given in \cref{eq:mode_m}.  The UV mode functions simplify for scalars whose mass is much lighter than the Hubble constant and can be approximated by \cref{massless mode functions sub horizon}.

One can then fix $c_{3,1}$ by demanding that \cref{four point example} reproduces the full theory prediction for the correlator in the regime $t > t_\Lambda$.  While the split between subhorizon ($t<t_\Lambda$) and superhorizon ($t > t_\Lambda$) evolution is manifest in the EFT, this split has to be inputted by hand in the full theory, as was done in \cref{full theory decomposition}; this is effectively making a choice of scheme.  Practically, one can decompose the time integrals in \cref{eq:commutator} into regions before and after $t_\Lambda$.  In our four point example, subhorizon contribution is already taken care of by the initial conditions, i.e.,~\cref{four point initial conditions matching}, while the second term in \cref{four point example} must reproduce the superhorizon evolution.  The Wilson coefficient $c_{3,1}$ is fixed by this condition.

Since the split time $t_\Lambda$ is arbitrary from the perspective of the full theory, all full-theory and EFT predictions of the $\phi$ correlators must be independent of it.  This provides an additional check on the matching calculation in the hard cutoff scheme.  Fortunately, when we calculate with dynamical dim reg, the time integrals are manifestly independent of $t_\Lambda$, while still maintaining the split between initial conditions and time-evolution.  In this sense, the hard-cutoff scheme proves the validity of dynamical dim reg, while dynamical dim reg makes it manifest that we can implement this procedure without breaking symmetries.  This matches the more intuitive argument that our treatment of initial conditions and time evolution in SdSET is identical to the continuum EFT approach.

%%%%%%%%%%%%%%%%%%%%%%%%%%%%%%%%%%%%%%%%%%%%%
\subsection{Stochastic Inflation from SdSET} \label{sec:SI_sdset}
%%%%%%%%%%%%%%%%%%%%%%%%%%%%%%%%%%%%%%%%%%%%%%%

From the point of view of our EFT, Stochastic Inflation can be understood as a consequence of operator mixing.  Specifically, for light fields for which $\alpha \to 0$, the composite operators $\vp^n$ are degenerate to leading order (in that they have the same dimension as determined by the EFT power counting).  Assuming the correlations of these fields are only due to the Gaussian contribution given in \cref{eq:GausianPowerSpecBoth}, one encounters a UV divergence from a one-loop contraction
\begin{align}
\big\langle \vp^n(\x) \,...\, \big\rangle &\supset  \big\langle \vp^{n-2}(\x) \,...\, \big\rangle \times  \binom{n}{2}\s \frac{C_\alpha^2}{2} \int \frac{\dd^3 p}{(2\s\pi)^3} \frac{H^{2-2\s\alpha}}{p^{3-2\s\alpha}} \notag \\[7pt]
& \supset  \big\langle \vp^{n-2}(\x) \,...\, \big\rangle \times  \binom{n}{2}\s \frac{C_\alpha^2}{4 \pi^2}\log [aH]\,.
\end{align}
The dynamical RG associated with this operator mixing can be written as
\beq
\frac{\partial}{\partial\t}  \big\langle \vp^n(\x) \,...\, \big\rangle =\frac{n(n-1)}{8\s \pi^2} \s \big\langle \vp^{n-2}(\x) \,...\, \big\rangle - \frac{n}{3} \sum_{m>1}\frac{c_{m,1}}{m!}  \big\langle \vp^{n-1}(\x)\s \vp^m(\x) \,...\, \big\rangle  \, , \label{eq:mixing_leading}
\eeq
where the second term arises from the classical time evolution.  This equation contains the same information as Starobinsky's formulation of Stochastic Inflation.  Specifically, we can use the Fokker-Planck equation given in \cref{eq:FPeqStocInf} to see that these two approaches are equivalent:
\bea
\frac{\partial}{\partial \t}  \big\langle\vp^n\big\rangle &=& \frac{\partial}{\partial \t}  \int \d\vp \vp^n P(\vp,\t) \notag\\[5pt]
&=&  \int \d\vp \vp^n\left( \frac{1}{8 \pi^2} \frac{\partial^2}{\partial \vp^2}  P(\vp, \t)  + \frac{1}{3 }  \frac{\partial}{\partial \vp} \left [\sum_{m>1} \frac{c_{m,1}}{m!} \vp^m P(\vp,\t ) \right] \right)\notag\\[7pt]
&=&  \int \d\vp \left( \frac{n(n-1)}{8\pi^2} \vp^{n-2}  P(\vp, \t)- \frac{n}{3} \vp^{n-1} \sum_{m>1} \frac{c_{m,1}}{m!} \vp^m P(\vp,\t )  \right) \notag\\[5pt]
&=&  \frac{n(n-1)}{8\pi^2}  \big\langle \vp^{n-2} \big\rangle - \frac{n}{3}\sum_{m>1}\frac{c_{m,1}}{m!}  \big\langle \vp^{n-1}  \vp^m \big\rangle \,,
\eea
where in the second line we plugged in \cref{eq:FPeqStocInf}, $H t =  \t$, the $\alpha = 0$ relation $\phi|_{\varphi_- = 0} \to H \varphi_+$, and 
\beq
V'(\phi) \,\,\to\,\, \frac{\partial}{\partial \vm} V(\vp,\vm)\Big|_{\vm=0} = \sum_{m>1} \frac{c_{m,1}}{m!} \vp^m \,.
\eeq
This shows that \cref{eq:mixing_leading} is equivalent to the leading order equation for Stochastic Inflation~\cref{eq:FPeqStocInf}.  Therefore, calculating corrections to Stochastic Inflation has been reduced to the straightforward task of computing the higher order dynamical RG equation using SdSET.  Concretely, we would expect to find mixing between composite operators $\vp^n$ and all possible $\vp^{n'}$ such that 
\begin{align}
    \frac{\partial}{\partial \t}  \big\langle \vp^n \big\rangle =&- \frac{n}{3} \sum_{m>1}^{\text{odd}}\frac{c_{m,1}}{m!}  \big\langle \vp^{n+m-1}\big\rangle 
+  \binom{n}{2} \sum_{m=0}^{\infty}b_m \big\langle \vp^{n+2m-2} \big\rangle \notag\\[5pt]
&-\binom{n}{3} \sum_{m=0}^{\infty}d_m \big\langle \vp^{n+2m-2} \big\rangle
+\binom{n}{4} \sum_{m=0}^{\infty}e_m \big\langle \vp^{n+2m-4} \big\rangle  + ...\ . 
\end{align}
Note the role of the binomial coefficient which will originate from the number of fields inside $\vp^n$ whose contractions are responsible for mixing with a given operator, leading to a single log divergence.  Repeating the above argument, we see that this dynamical RG is equivalent to 
\begin{align}
    \frac{\partial}{\partial \t} P(\vp,\t) &=\frac{1}{3}  \frac{\partial}{\partial\vp} \left [\partial_{\vm}V (\vp,\vm)|_{\varphi_-=0} P(\vp,\t ) \right]+ \frac{\partial^2}{\partial \vp^2} \left[ \sum_{m=0}^\infty \frac{ b_m}{2!} \vp^{2m} P(\vp, \t) \right] \notag\\[5pt] 
&\hspace{14pt} +  \frac{\partial^3}{\partial \vp^3}  \left(\vp \sum_{m=0}^\infty \frac{d_m}{3!} \vp^{2m} P(\vp, \t) \right) +  \frac{\partial^4}{\partial \vp^4}  \left( \sum_{m=0}^\infty \frac{ e_m}{4!} \vp^{2m} P(\vp, \t)  \right) + \ldots \ , \notag\\
\end{align}
where we see that number of derivatives is related to the binomial coefficient of the associated mixing term.  Comparing with \cref{sec:BLO}, we see that the NLO corrections are determined by $c_{5,1}$ and $b_1$ while the NNLO coefficients are $c_{7,1}$, $b_2$ and $d_0$.  Achieving NNLO accuracy requires matching $\lambda \phi^4$ theory onto the SdSET at one loop, the subject of the next section.  Note, however, that we have described Stochastic Inflation in terms of $\vp$ rather than the UV field $\phi$. This distinction will be important because the equations of Stochastic Inflation are not invariant under field redefintions.  We will address these issues in detail in \cref{sec:Stoc_final_sub}.   Finally, we note that this result demonstrates the Markovian assumption which led to~\cref{eq:BLOFokkerPlMasterEq} does in fact hold as a consequence of SdSET power counting.  Specifically, the fact that the $\varphi_-$ dynamics are irrelevant to the evolution of the $\varphi_+$ correlators implies that the evolution of the system is indeed linear, and thus it has no ``memory.''

\section{Matching $\lambda \phi^4$ Onto SdSET at One-Loop}
\label{sec:Matching}
In this section we will show how to match correlators of $\phi$ to correlators of $\vp$ in the SdSET to determine the EFT parameters in terms of UV data.  This then serves as input for any calculation of cosmological correlators for $\lambda \phi^4$ theory in the long wavelength limit, including the corrections to Stochastic Inflation we will discuss in subsequent sections.  Furthermore, by extending this program to one-loop order for the first time, this calculation will serve as a non-trivial check of the SdSET framework.

Unlike conventional EFTs, we match both the couplings of SdSET and the stochastic initial conditions, the former results from studying the time-dependent terms in the EFT while the latter is fixed at the time of horizon crossing. As a result, consistent matching of time-dependence of the UV correlators is non-trivial and requires that the SdSET is a complete representation of the long wavelength dynamics.  In contrast, time-independent contributions to a given correlator can always be absorbed into the initial conditions (up to composite operators as discussed in~\cref{sec:match_alpha}).  As a result, for tree and one-loop matching we will be particularly focused on time-dependent UV contributions.

\subsection{Tree-level Matching and Field Redefinitions}

Tree-level matching in the interacting theory is non-trivial due to the impact on the initial conditions.  We will need to introduce non-Gaussian initial conditions in order to match higher point correlation functions as calculated by the UV theory.  

We can understand many important aspects of matching by Taylor expanding the UV calculation in the long wavelength limit.  As a simple demonstration, we can explore the superhorizon behavior of the operator $\phi$.  At first order in the coupling, we can apply the definition of the in-in correlator given in~\cref{eq:commutator} with $Q(t) = \phi$, which gives
\begin{align}
[\phi]_\lambda\big(\x,\tau\big) &= i\int^\tau \d\tau_1 \Big[H_{\rm int}(\tau_1),  \phi_{\rm in}\big(\x,\tau\big) \Big] \notag\\[7pt]
&=\frac{\lambda}{3!}  \int^\tau \frac{\d\tau_1}{(-H\tau_1)^4} \int \d^3 x_1 \,  \phi_{\rm int}(\x_1,\tau_1)^3 \, i \Big[ \phi_{\rm int}\big(\x_1,\tau_1 \big), \phi_{\rm int}\big(\x,\tau\big)\Big] \ ,
\end{align}
where $\phi_{\rm int}$ are the interaction picture fields.  Because the time integral runs over all times, this includes both the regime where the modes are hard (UV) and the long wavelength limit where the EFT applies.
Nevertheless, if we expand in the long wavelength limit and evaluate the integrals with dynamical dim reg, we will only get contributions from late times.  Since $\phi_{\rm in}$ are the free field operators, we can use the map given in \cref{eq:phiS} to determine the long wavelength behavior of the full theory:\footnote{Note we are not working in the EFT yet, because we have not integrated out the hard modes.  All we are doing here is taking the long wavelength limit.} 
\begin{subequations}
\begin{align}
\phi_{\rm int} &\to H \big([aH]^{-\alpha}\varphi_+ +  [aH]^{-3+ \alpha} \varphi_-\big) \,. 
\end{align}%
\label{eq:trivial_map}%
\end{subequations}%
Then using $\tau = -1/[aH]$ and keeping only terms that survive as $[aH] \to \infty$, one finds
\begin{align}
[\phi]_\lambda\big(\x,\t\big)  &\to \frac{\lambda}{3!} \, H \, \int^\t \d\t_1 \int  \d^3 x_1 \,  [a(\t_1)H]^{-2\alpha} \vp^3\big(\x_1\big) \notag\\[5pt]
& \hskip 40pt \ i \bigg(   \Big[\vm\big(\x_1\big),\vp\big(\x\big)\Big] + \frac{[a(\t_1)H]^{3}}{[a(\t)H]^3} \Big[\vp\big(\x_1\big),\vm\big(\x\big)\Big] \bigg)\notag\\[8pt] 
&=   \frac{\lambda}{3!}  \frac{H}{3} \bigg(- \bigg(\!-\frac{1}{2\alpha}+ \log [a(\t)H] \bigg) + \frac{1}{3} \bigg)  \vp^3\big(\x\big) \,, \label{eq:phi_lambda}
\end{align}
where in the last line we expanded in $\alpha \ll 1$.  Note that the scaling dimension of $\vp$ is still $\alpha$ and thus will provide the necessary distance scale to make the log dimensionless inside of a correlator, as is familiar from conventional dim reg.

Now we turn to exploring the same effect within the EFT direction by calculating the time evolution of $\vp$ using SdSET.  Using \cref{eq:phiS} with $Q=\vp$ and $\H_{\rm int} = c_{3,1} \vp^3 \vm/3!$ we have  
\begin{align}
[\vp]_\lambda (\x,\t) &= \frac{c_{3,1}}{3!}\int^\t \d\t_1 \int \d^3x_1  [aH(\t_1)]^{-2\alpha} \vp^3(\x_1) i \Big[\vm\big(\x_1\big),\vp\big(\x\big)\Big] \notag \\[5pt]
&= -\frac{c_{3,1}}{3!} \frac{1}{3}\bigg(\!-\frac{1}{2\alpha}+ \log [a(\t)H] \bigg) \vp^3(\x) \,, \label{eq:vp_lambda}
\end{align}
where it is trivial to match the tree-level UV interaction to the EFT interaction, such that
\begin{align}
c_{4,0} = c_{3,1} = \,...\, = \lambda + O\big(\lambda^2\big)\,.
\label{eq:treeMatch}
\end{align}
The equality between $c_{n,0}$ and $c_{n-m,m}$ found in matching is also fixed by the reparametrization invariance of SdSET.  We see that the EFT is capturing the first term in the Taylor expansion of the UV theory given in \cref{eq:phi_lambda}, but not the second.  

The origin of the missing term is two-fold.  First, we are only considering correlations of $\vp$ instead of the full UV field, which also includes $\vm$.  This alone would not matter, since $\vm$ is suppressed by $[aH]^{-3}$.  However, in order to organize the interactions within the EFT, we removed the $c_{4,0}\vp^4/4!$ term by a field redefinition.  Specifically, to remove the $c_{n,0} \varphi_+^n/ n!$ operator, we take
\beq\label{eq:vm_field_redef}
\varphi_- \to \varphi_- + \frac{c_{n,0}}{9 (n-1)!} [aH]^{3-(n-1)\alpha} \varphi_+^{n-1}  \ .
\eeq
Therefore, keeping track of the field redefinition implies that we should use  
\beq
\phi_{\rm EFT} \equiv \bvar = H \left( [aH]^{-\alpha} \varphi_+ + [aH]^{-\beta} \varphi_- + \frac{c_{4,0}}{9}\frac{1}{ 3!} [aH]^{-3\alpha} \varphi_+^{3} + \,...\, \right)\,,
\label{eq:FieldMapAfterEFTFieldRedef}
\eeq
with $\alpha \to 0$ and $\beta \to 3$.  Now the quantities on the RHS live purely in the EFT.   As  a result $\varphi_-$ will not contribute to correlation functions of $\bvar$ because they are suppressed by powers of $[aH]^{-3}$.  Combining this with \cref{eq:vp_lambda}, we find
\beq
\big[\,\bvar\,\big]_\lambda =  \vp + \frac{1}{3!} \frac{H}{3} \bigg(c_{3,1}\bigg(\!\frac{1}{2\alpha}- \log [aH] \bigg) + \frac{c_{4,0}}{3} \bigg) \vp^3(\x) \label{eq:EFT_bvar}  \, ,
\eeq
where we dropped terms suppressed by powers of $[aH]$.  Now we see that this matches the UV expression in~\cref{eq:phi_lambda} when we use the tree-level matching relations $c_{3,1} = c_{4,0} = \lambda$ given in \cref{eq:treeMatch}.  

This result also provides the map between SdSET and~Refs.~\cite{Baumgart:2019clc,Baumgart:2020oby}, which derived the soft behavior by explicitly expanding the UV in-in correlator in the superhorizon limit.  The tree-like structure they observe is a consequence of our power counting as only $c_{n,1}$ is marginal and hence interactions only include a single factor of $\vm$.  The nested set of commutators in \cref{eq:commutator} ensures that the marginal operators always have a tree-like structure.  In SdSET, this is manifest from dynamical RG, and all the additional finite terms arise from the field redefinitions.

By similar considerations, we must apply the field redefinition to match the UV potential onto $V(\vp,\vm)$ in the EFT. 
Although $\lambda \varphi_+^4$ has been removed, this procedure introduces higher order terms, such as
\begin{subequations}
\begin{align}
\frac{c_{3,1}}{3!} \varphi_+^3 \varphi_- &\,\,\to\,\, \frac{c_{3,1}\,c_{4,0}}{9 (3!)^2} [aH]^3 \varphi_+^6 \\[8pt]
\frac{c_{2,2} }{4} [aH]^{-3} \varphi_+^2 \varphi_-^2 &\,\,\to\,\, \frac{c_{2,2}\, c_{4,0}}{18 (3!)} \varphi_+^5 \varphi_- \,.
\end{align}
\end{subequations}
Removing the first term will introduce a $\varphi_+^7 \varphi_-$ interaction at order $\lambda^3$ and so on.  As a result, our field redefinition requires that 
\beq
V(\varphi_+,\varphi_-) \supset \frac{1}{3!} \varphi_- \left( c_{3,1} \varphi_+^3 + \frac{c_{2,2} c_{4,0} }{18}\varphi_+^5 + \frac{c_{2,2} c_{3,1} c_{4,0}}{162} \varphi_+^7 +\,...\, \right) \ .
\eeq
Finally, using $c_{2,2} = c_{3,1} = c_{4,0} =\lambda$ from matching, we arrive at
\beq
V(\varphi_+,\varphi_-) \supset \frac{\lambda}{3!} \varphi_- \left( \varphi_+^3 + \frac{\lambda}{18}\varphi_+^5 + \frac{\lambda^2}{162} \varphi_+^7 +\,...\, \right) \, . \label{eq:Vvpvm}
\eeq
or $c_{5,1} = \frac{\lambda^2}{18} \frac{5!}{3!}$ and $c_{7,1} = \frac{\lambda^3}{162} \frac{7!}{(3!)}$. 

The correction to $c_{5,1}$ is equivalent to the NLO corrections to the effective potential calculated in Refs.~\cite{Gorbenko:2019rza} and~\cite{Mirbabayi:2020vyt} using complementary techniques.  While these two references approach this problem from different perspectives, the wavefunction of the universe and the dS static patch respectively, both effectively integrate out the decaying mode $\vm$ which leads to an additional term in the potential. Instead, when $\vm$ is included, our corrections arise from insuring $\vm$ does not mix with $\vp$ at higher orders in perturbation theory. As the dimensions of $\vp$ and $\vm$ are well separate for $\alpha \to 0$, removing mixing can always be achieved by such a field redefinition.  Additionally, it is easy to determine the correction to $c_{2n+1,1} \propto \lambda^{n}$ by repeated application of \cref{eq:vm_field_redef}.  Most importantly, we do not integrate out $\vm$ to ensure we have a local action, rather than an open EFT for the growing mode alone~\cite{Burgess:2015ajz}. 

What we have accomplished thus far is to determine the correct basis of operators to match the EFT and UV descriptions.  We have ensured that the superhorizon limit of $\phi$ and $\bvar$ agree as operators at higher orders in $\lambda$.  However, in order to match the correlators of the UV theory, which include the subhorizon evolution, we will need to determine the stochastic initial conditions beyond the Gaussian limit.  

In order to correctly match the four-point function, we write the EFT trispectrum to order $\lambda$ as\footnote{For convenience, we take $C_\alpha \to 1$ for the light fields and have dropped the additional constants that arise from expanding $C_\alpha = 1 + \alpha(\gamma_E - 2+\log 2) +{O}\big(\alpha^2\big)$.  While this choice has no impact on the physics  (since the correct constants will appear in the initial conditions by matching), it will simplify the algebra significantly.}
\begin{align}
\Big\langle \bvar\big(\k_1\big) \bvar\big(\k_2\big) \bvar\big(\k_3\big) \bvar\big(\k_4\big) \Big\rangle_C' &=\Big\langle \big[\,\bvar\,\big]_\lambda \big(\k_1\big)\vp\big(\k_2\big) \vp\big(\k_3\big) \vp\big(\k_4\big) \Big\rangle'_C  + {\rm permutations} \nonumber \\[5pt]
&\hspace{14pt}+\Big\langle \bvar\big(\k_1\big) \bvar\big(\k_2\big) \bvar\big(\k_3\big) \bvar\big(\k_4\big) \Big\rangle'_{{\rm IC}^{(1)}} \notag \\[7pt]
&= \frac{\lambda \, H^4}{8(k_1 k_2 k_3 k_4)^3} \sum_i  \frac{k_i^3}{3} \left(c_{3,1}\bigg(\!\frac{1}{2\alpha}+\log \frac{k_i}{[aH]} \bigg) + \frac{c_{4,0}}{3} \right)\nonumber  \\[5pt]
&\hspace{14pt}+\Big\langle \bvar\big(\k_1\big)  \bvar\big(\k_2\big) \bvar\big(\k_3\big) \bvar\big(\k_4\big) \Big\rangle'_{{\rm IC}^{(1)}} \, , 
\end{align}
where $c_{3,1} = c_{4,0} = \lambda$ as before, and the subscript ``$C$'' denotes that this is only the connected contributions.  Matching this EFT expression to the UV result in \cref{eq:fourPtCorUV} fixes the non-Gaussian contribution to the initial conditions:
\begin{align}
\Big\langle \bvar\big(\k_1\big) \bvar\big(\k_2\big) \bvar\big(\k_3\big) \bvar\big(\k_4\big) \Big\rangle'_{{\rm IC}^{(1)}}  &= \frac{\lambda \, H^4}{8(k_1 k_2 k_3 k_4)^3} \Bigg[ \frac{\sum_i k_i^3\big(-\frac{1}{2\alpha} + \gamma_E-2+ \log\frac{k_t}{k_i}\big) }{3}   \notag \\[5pt]
-\frac{k_1 k_2 k_3 k_4}{k_t}&- \frac{1}{9} k_t^3 + 2 \sum_{h<i<j} k_h k_i k_j +\frac{1}{3} k_t \Big(\sum k_i^2 - \sum_{i<j} k_i k_j\Big) \Bigg]\,, \label{eq:tri_IC}
\end{align}
where $k_t =k_1+k_2+k_3+k_4$. Most significantly, all the time-dependence of the full UV trispectrum is already captured by the EFT and, as expected, the initial conditions are only required for matching the time-independence contributions.  This result is extended to the six-point function in \cref{sec:MatchingSixPt} as expected from general arguments.  

\vskip 10pt
\noindent \textbf{Matching Derivative Operators:}  Before moving on the loop-level matching, let us briefly comment on the case where the UV theory is itself an effective theory.  Specifically, we are only considering the case of a $\lambda \phi^4$ interaction in the UV, while in principle there could be a variety of higher derivative (irrelevant) interactions as well. The first such operator we can write down is $(\nabla_\mu \phi \nabla^\mu \phi)^2 / M^4$.  The full tree level trispectrum was calculated in~Ref.~\cite{Creminelli:2011mw} and is given by
\begin{align}
\hspace{-1pt}\Big\langle\phi\big(\k_{1}\big) \phi\big(\k_{2}\big) \phi\big(\k_{3}\big) \phi\big(\k_{4}\big)\Big\rangle_{\nabla^4} &=(2 \pi)^{3} \delta\left(\sum_{i} \vec{k}_{i}\right) \frac{1}{M^{4}} \frac{H^{8}}{\prod_{i} 2 k_{i}^{3}} \notag\\[4pt]
&\hspace{18pt}\times\Bigg[ -\frac{144 k_{1}^{2} k_{2}^{2} k_{3}^{2} k_{4}^{2}}{k_{t}^{5}}-4\Bigg(\frac{12 k_{1} k_{2} k_{3} k_{4}}{k_{t}^{5}}+\frac{3 \prod_{i<j<l} k_{i} k_{j} k_{l}}{k_{t}^{4}}  \notag\\[4pt]
&\hspace{40pt}+\frac{\prod_{i<j} k_{i} k_{j}}{k_{t}^{3}}+\frac{1}{k_{t}}\Bigg)\left(\left(\vec{k}_{1} \cdot \vec{k}_{2}\right)\left(\vec{k}_{3} \cdot \vec{k}_{4}\right)+2 \text { perms}\right)  \notag\\[4pt]
&\hspace{40pt}+\left(\vec{k}_{1} \cdot \vec{k}_{2}\right)\left(\frac{4 k_{3}^{2} k_{4}^{2}}{k_{t}^{3}}+\frac{12\left(k_{1}+k_{2}\right) k_{3}^{2} k_{4}^{2}}{k_{t}^{4}}+\frac{48 k_{1} k_{2} k_{3}^{2} k_{4}^{2}}{k_{t}^{5}}\right)  \notag\\[4pt]
&\hspace{40pt} +  \text{5 perms} \Bigg]\,.
\end{align}
What we notice right away is that this gives us a completely time-independent result.  It can therefore only be absorbed into the initial conditions of $\varphi_+$:
\bea
\Big\langle\varphi_+\big(\vec{k}_{1}\big) \vp\big(\vec{k}_{2}\big) \vp\big(\vec{k}_{3}\big) \vp\big(\vec{k}_{4}\big)\Big\rangle_{\rm IC}&\supset & H^{-4} \Big\langle\phi\big(\k_{1}\big) \phi\big(\k_{2}\big) \phi\big(\k_{3}\big) \phi\big(\k_{4}\big)\Big\rangle_{\nabla^4}\,.
\eea
This result illustrates a broader feature of physics in dS: higher derivatives decouple at long wavelength and thus contribute, at most, time-independent correlation functions.  This property is made manifest in SdSET and must hold in matching.  As a result, a variety of possible UV theories only impact the initial conditions and not the long wavelength dynamics. In this sense, our choice to focus $\lambda \phi^4$ is not missing more complicated superhorizon evolution. Instead, higher derivative terms are trivially matched in SdSET and thus do not further illuminate the structure of the EFT.

\subsection{One-loop Matching}
\label{sec:match_1loop}
When calculating loops in the UV theory, rather than the EFT, our EFT regulator (dynamical dim reg) is not effective.\footnote{In a forthcoming paper~\cite{Green:2021}, it will be shown that when these integrals are transformed to Mellin space, then it is indeed possible to implement dynamical dim reg in the full theory.}  Loop calculations in the UV are not scaleless and therefore we will need to regulate them differently than the EFT approach.  This difference will be absorbed into the matching calculating.  In~\cref{sec:HardCutoff}, we match using a hard cutoff in both theories for an example with consistent regulators in both and find identical results, up to scheme dependent coefficients.  To avoid the usual challenges of working with a hard cutoff, we will use dimensional regularization in the UV theory via
\beq
\Big\langle\phi\big(\k, \tau\big)\, \phi\big(\!-\!\k, \tau^{\prime}\big)\Big\rangle=\frac{\pi}{4} H^{d-1}(-\tau)^{\frac{d}{2}}\left(-\tau^{\prime}\right)^{\frac{d}{2}} \text{H}_{\nu}\big(\!-\! k \tau\big) \text{H}_{\nu}^{\star}\big(\!-\! k \tau^{\prime}\big)\,,
\eeq
where $\nu = \sqrt{d^2/4 - m^2 /H^2}$.  To simplify calculations, we can fix $\nu = 3/2$ for any dimension, and then can regulate integrals that appear via an analytic continuation in $d$.  We note that while this will regulate the one-loop divergences that appear in this section, dim reg alone is not sufficient in general, as we will see below.  

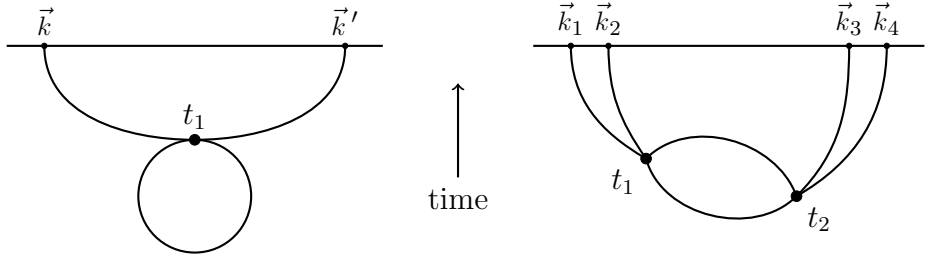
\begin{figure}[t!]
\centering
\begin{tikzpicture}
    % % A helpful grid
    % \draw[step=1cm,gray,very thin] (0,0) grid (12,6);
    % \draw[thick,->] (0,0) -- (12,0) node[anchor=north west] {x axis};
    % \draw[thick,->] (0,0) -- (0,6) node[anchor=south east] {y axis};
    % \foreach \x in {0,1,2,3,4,5,6,7,8,9,10,11}
    %   \draw (\x cm,1pt) -- (\x cm,-1pt) node[anchor=north] {$\x$};
    % \foreach \y in {0,1,2,3,4,5}
    %     \draw (1pt,\y cm) -- (-1pt,\y cm) node[anchor=east] {$\y$};
    
    % The power spectrum    
    \draw[thick] (0,4) -- (5,4);
    \draw[thick] (0.5,4) to [out=-90,in=180] (2.5,2.75) to [out=0,in=-90] (4.5,4);
    \draw[thick] (2.5,2.75-0.75) circle [radius=0.75];
    
    % The labels
    \filldraw (0.5,4) circle (1pt) node[above] {$\k$};
    \filldraw (4.5,4) circle (1pt) node[above] {$\k\,'$};
    \filldraw (2.5,2.75) circle (2pt) node[above] {$t_1$};
    
    % The tri spectrum    
    \draw[thick] (7,4) -- (12.2,4);
    \draw[thick] (7.5,4) to [out=-90,in=150] (8.5,2.5);
    \draw[thick] (8,4) to [out=-90,in=125] (8.5,2.5);
    \draw[thick] (10.5,2) to [out=45,in=-90] (11.2,4);
    \draw[thick] (10.5,2) to [out=30,in=-90] (11.7,4);
    \draw[thick] (8.5,2.5) to [out=-75,in=225] (10.5,2);
    \draw[thick] (8.5,2.5) to [out=45,in=110] (10.5,2);
    
    % The labels
    \filldraw (7.5,4) circle (1pt) node[above] {$\k_1$};
    \filldraw (8,4) circle (1pt) node[above] {$\k_2$};
    \filldraw (11.2,4) circle (1pt) node[above] {$\k_3$};
    \filldraw (11.7,4) circle (1pt) node[above] {$\k_4$};
    \filldraw (8.5,2.5) circle (2pt) node[anchor=north east] {$t_1$};
    \filldraw (10.5,2) circle (2pt) node[anchor=north west] {$t_2$};
    
    % The arrow of time
    \draw[->,thick] (6,2.25) -- (6,3.5);
    \node[below] at (6,2.25) {time};
\end{tikzpicture}
\caption{Diagrams for the one-loop matching, as computed in the UV theory. The horizontal line indicates a surface on constant conformal time $\tau_0$ on which our in-in correlators are evaluated.  {\it Left:} One-loop power spectrum. {\it Right:} One-loop trispectrum. } 
\label{fig:oneloopmatch}
\end{figure}

We will begin with the one-loop power spectrum of the growing mode, illustrated in the left side of \cref{fig:oneloopmatch}.  In the UV theory, a standard in-in calculations gives us
\begin{align}
\Big\langle \phi\big(\s\k\,\big) \phi\big(\s\kp\s\big) \Big\rangle'_{(1)} &= \frac{\lambda }{4 k^3}  \frac{H^2}{3} \left( \frac{1}{\epsilon} + \log \frac{2 k}{[aH]}+\gamma_E - 2  \right) \bigg[[aH]^{-\epsilon} \int \frac{\d^d p}{(2\pi)^d} \frac{1}{2 p^{3}} \bigg]\,.
\end{align}
Regulating the IR by substituting $p^2 \to p^2 + K^2$ in the denominator, we get 
\begin{align}
\Big\langle \phi\big(\s\k\,\big) \phi\big(\s\kp\s\big) \Big \rangle'_{(1)} &= \frac{\lambda }{8 \pi^2 k^3} \frac{H^2}{3} \bigg(\frac{1}{\epsilon}  + \log \frac{2 k}{[aH]}+\gamma_E - 2  \bigg) \bigg[\frac{1}{\epsilon} - \log\frac{K}{[aH]} +\frac{1}{2} \big(\log 4\pi -\gamma_E\big) \bigg]\,.\notag\\
\label{eq:UVTwoPt1LCorr}
\end{align}
This is the complete one-loop power spectrum of the UV theory.

In SdSET, the one-loop power spectrum is given by several terms 
\begin{align}
\Big\langle \bvar\big(\s\k\,\big) \bvar\big(\s\kp\s\big) \Big\rangle'_{(1)} &= H^2 [aH]^{-2 \alpha} \Big\langle \vp\big(\s\k\,\big) \vp\big(\s\kp\s\big) \Big \rangle'_{(1)}+H^2 [aH]^{-2 \alpha} \Big\langle \vp\big(\s\k\,\big) \vp\big(\s\kp\s\big) \Big \rangle'_{\delta\alpha^{(1)}}\nonumber  \\[5pt]
&\hspace{13pt} + 2 \frac{c_{4,0}}{9}\frac{1}{ 3!} H^2 [aH]^{-4 \alpha}\Big\langle \vp\big(\s\k\,\big) \vp^3\big(\s\kp\s\big) \Big\rangle'_{(0)} + \Big\langle \bvar\big(\k\big) \bvar\big(\kp\big) \Big\rangle'_{{\rm  IC}^{(1)}}  \, ,
\label{eq:EFTTwoPt1LCorr}
\end{align}
 where $c_{4,0} \sim \lambda$, and $\bvar$ is given in \cref{eq:FieldMapAfterEFTFieldRedef}. The subscript $(n)$ labels the order in $\lambda$ in which correlator is calculated and $\delta\alpha^{(m)}$ is the contribution to the correlator from a shift in the value of $\alpha$ at order $\lambda^m$. 
The first term is the one-loop power spectrum in the EFT,
\begin{align}
[aH]^{-2 \alpha}\Big\langle \vp\big(\k\big) \vp\big(\kp\big) \Big\rangle'_{(1)} &= -[aH]^{-4 \alpha} \frac{1}{2\nu} \bigg(-\frac{1}{2 \alpha}\bigg) \frac{\lambda}{4 k^{3-2\alpha}} \int \frac{\d^3 p}{(2\pi)^3} \frac{1}{p^{3-2\alpha}} \notag  \\[7pt]
&\to -\frac{1}{6 \pi^2}  \frac{\lambda}{4 k^{3}} \bigg(-\frac{1}{2 \alpha} - \log \frac{k}{[aH]} \bigg)  \bigg( - \frac{1}{ 2\alpha} - \log \frac{K}{[aH]} \bigg) \, .\notag\\
\end{align}
We must also allow for the possibility that, due to matching, we will need to correct the value of $\alpha$ from its free-field value.  Substituting  $\alpha \to \alpha +\delta \alpha^{(1)}$ in \cref{eq:GaussianPowerSpec} and expanding to linear order, we have
\beq
H^2 [aH]^{-2 \alpha} \Big\langle \vp\big(\s\k\,\big) \vp\big(\s\kp\s\big) \Big \rangle'_{\delta\alpha^{(1)}} =\Big\langle \bvar\big(\k\big) \bvar\big(\kp\big)  \Big\rangle_{(0)}\bigg(1 + 2 \delta \alpha^{(1)} \log \frac{k}{[aH]}\bigg )\,,
\eeq
where have absorbed the $[aH]^{-2\delta \alpha}$ into the definition of this term to derive the dimensionless argument of the log.  From the EFT point of view, $\delta \alpha^{(1)}$ is an unknown constant of order $\lambda$ to be fixed by matching.  Finally, we have the contribution to the $\bvar$ power spectrum from the field redefinition:
\bea
 2 \frac{c_{4,0}}{9}\frac{1}{ 3!} H^2 [aH]^{-4 \alpha}\Big\langle \vp\big(\s\k\,\big) \vp^3\big(\s\kp\s\big) \Big\rangle'_{(0)}\to \frac{c_{4,0}}{9}  \frac{1}{2 k^3} \frac{1}{2\pi^2}  \bigg(-\frac{1}{2\alpha} -\log \frac{K}{[aH]} +\log 2\bigg) \, .
\eea
We emphasize that since the coefficient of $\vp^3$ in the definition of $\bvar$ is fixed by matching the superhorizon six-point function, there is no additional freedom within this term that can be used to match the 1-loop power spectrum.

The final term $\delta \big\langle \bvar\big(\k\big) \bvar\big(\kp\big) \big\rangle'_{\rm  IC}$ is time independent and is determined by matching the time-independent part of the UV calculation.  Combining these results and using $c_{4,0} \to \lambda$, we find
\begin{align}
H^2 \Big\langle \bvar\big(\k\big) \bvar\big(\kp\big) \Big\rangle'_{(1)} =&   \frac{H^2}{6 \pi^2}  \frac{\lambda}{4 k^{3}} \bigg(\frac{1}{-2 \alpha} + \log \frac{[a H]}{k}  + \frac{2}{3} \bigg)  \bigg( \frac{1}{- 2\alpha} + \log \frac{[aH]}{K} \bigg)  \notag \\
&+\delta \alpha^{(1)} \frac{H^2}{k^3}\log \frac{k}{a}+ \frac{\lambda}{9}  \frac{1}{2 k^3} \frac{1}{2\pi^2}  \log 2 + \Big\langle \bvar\big(\k\big) \bvar\big(\kp\big) \Big\rangle'_{{\rm  IC}^{(1)}} \ . \label{eq:EFT_comb}
\end{align}
Comparing this result to \cref{eq:UVTwoPt1LCorr}, we see that the time-dependence of the UV and EFT agree after matching the single log coefficient with
\beq
\delta \alpha^{(1)} =  \frac{\lambda}{8\pi^2} \frac{1}{3} \bigg( \gamma_E - \frac{7}{3}+ \log 2  - \frac{1}{2} \big(\log 4\pi -\gamma_E\big)\bigg) \, .
\label{eq:deltaAlpha}
\eeq
This is a non-trivial result as we only have a single free parameter to match the full time-dependence on the UV result. All the time-independent contributions can be matched by adjusting the initial conditions, IC$^{(1)}$, which in this case is just a renormalization of the amplitude of the two point function, namely a shift in $C_\alpha$.

\subsubsection*{One-loop Trispectrum}
Now we move on to the one-loop trispectrum, as illustrated in the right side of \cref{fig:oneloopmatch}.  For our purposes here, it suffices to simply match the UV divergent terms which result in time dependent $\log[aH]$ factors.\footnote{We remind the reader that the ``UV divergences'' that we are isolating here take their origins from the IR divergences of the full theory that are resummed by Stochastic Inflation.  The expansions being done to simplify the integrals that appear on the UV theory side of this matching calculation amount to decomposing the correlator via the method of regions~\cite{Beneke:1997zp, Smirnov:2002pj}.}  This will result in a correction to the $c_{3,1}$ Wilson coefficient in the EFT.  While the initial conditions also receive corrections from matching the trispectrum, these would contribute to Stochastic Inflation beyond NNLO, and so we will not compute them here.

For convenience, we break the UV calculation into two terms
\begin{subequations}
\begin{align}
{\cal I}_4 &=\Big\langle \phi^{(2)}\big(\k_1\big) \phi^{(1)}\big(\k_2\big)\phi^{(2)}\big(\k_3\big) \phi^{(1)}\big(\k_4\big) \Big\rangle +{\rm permutations}\\[5pt]
{\cal K}_4 &=\Big\langle \phi^{(3)}\big(\k_1\big) \phi^{(1)}\big(\k_2\big)\phi^{(1)}\big(\k_3\big) \phi^{(1)}\big(\k_4\big) \Big\rangle +{\rm permutations}\,,
\end{align}
\end{subequations}
where
\begin{subequations}
\begin{align} 
\phi^{(2)}(\k,t) &= i\int^t \d t_1  \Big[ H_{\rm int}(t_1) , \phi\big(\k,t\big)\Big] \\[5pt]
 \phi^{(3)}(\k,t)  &= - \bigg[\int^{t_1} \d t_2 H_{\rm int}(t_2) ,\bigg[\int^t \d t_1 H_{\rm int}(t_1) , \phi(\k,t)\bigg]\bigg]\,,
\end{align}
\end{subequations}
where $H_{\rm int}$ is given in \cref{eq:Hint}.  In what follows, we will show that ${\cal I}_4$ is UV finite, while ${\cal K}_4$ is UV divergent.  Then we will match this divergent contribution between the UV theory and the EFT to derive the one-loop correction to $c_{3,1}$.

We begin by evaluating ${\cal I}_4$.  Expanding in the limit $p \gg k_i$, we have
\begin{align}
{\cal I}_4 = \frac{1}{k_2^3 k_4^3}  &\int^\tau \frac{\d\tau_1}{(-\tau_1)^4} \int \frac{\d^3 p}{(2\pi)^3} G\big(\k_1, \tau, \tau_1\big)  \frac{(1-i p \tau_1)^2 }{p^3} e^{i 2 p \tau_1}  \nonumber \\[5pt]
& \times\int^\tau \frac{ \d\tau_2}{ (-\tau_2)^4} \int \frac{\d^3 p}{(2\pi)^3}  G\big(\k_3, \tau, \tau_2\big) \frac{(1+i p  \tau_2)^2}{p^3} e^{-i 2 p \tau_2}\notag\\[5pt]
&  + {\rm permutations} \,.
\end{align} 
The integrals over $\tau_1$ and $\tau_2$ are independent and we can evaluate them as usual.  Expanding in $k_i \tau_i \ll 1$, we have 
\begin{align}
  \int^\tau\frac{ \d\tau_1}{(-\tau_1)^4} G\big(\k_1, \tau, \tau_1\big)(1-i p \tau_1)^2 e^{i 2 p \tau_1} &\to  -\frac{1}{3} \int^\tau\frac{ \d\tau_1}{(-\tau_1)^4} \big(\tau^3-\tau_1^3\big)(1-i p \tau_1)^2 e^{i 2 p \tau_1}\notag \\[5pt]
  &= \frac{(-i)}{3} \left( \log p \tau - \frac{11}{12} \right)\,.
\end{align}
Putting this together yields
\beq
{\cal I}_4 \simeq \frac{1}{k_2^3 k_4^3}  \int \frac{\d^3 p}{(2\pi)^3}  \frac{1}{p^6}  \left( \log p \tau - \frac{11}{12} \right)^2\,.
\eeq
This integral will not lead to any UV divergences.  Furthermore, this result is exact in $p\tau$ since we have only expanded in $k / p$.  Therefore all the subleading terms will be more UV convergent, and correspond to higher power corrections in the EFT. This shows that all we need to include when matching $c_{3,1}$ is the correction due to ${\cal K}_4$.

The second term, ${\cal K}_4$, gives rise to more interesting UV behavior.  Again expanding for large loop momentum $p \gg k_i$, we have
\begin{align}
{\cal K}_4 &\simeq \frac{H^4}{32 (k_2 k_3 k_4)^3} \int \frac{\d^d p}{(2\pi)^{d}}   \int^\tau \frac{\d\tau_1}{(-\tau_1)^{d+1}}  G\big(\k_1,\tau,\tau_1\big) \notag\\[5pt]
&\hspace{140pt} \times  2{\rm Im} \int^{\tau_1}  \frac{\d\tau_2}{(-\tau_2)^{4}}   \frac{1}{p^6}(1+i p\tau_1)^2 (1-i p \tau_2)^2 e^{-i 2 p (\tau_1-\tau_2)}
\notag\\
&\hspace{14pt} +{\rm permutations}  \, .
\end{align}
We have expressed the commutator acting on the loop momenta in  $\phi^{(3)}$ in terms of the imaginary part to simplify the calculation.  In this form, the time integrals can be evaluated exactly, giving 
\begin{align}\label{eq:K4}
{\cal K}_4 &\simeq \frac{H^4 \lambda^2}{16(k_2 k_3 k_4)^3}  \int \frac{\d^d p}{(2\pi)^d} \frac{1}{p^3} \bigg[ \frac{10}{81} - \frac{1}{27} \gamma_E(2+ 3 \gamma_E) - \frac{5}{36} \pi^2  \notag\\[5pt]
&\hspace{14pt}+ \frac{1}{9} \left(\log \frac{2p}{[aH]}\right)^2 + (1+ 3\gamma_E) \log \frac{2p}{[aH]} + \frac{4}{9} \log \frac{k}{[aH]}
\bigg] + {\rm permutations}\notag\\[8pt]
&=\frac{H^4 \lambda^2}{8 (k_2 k_3 k_4)^3} \frac{1}{4\pi^2} \left(\frac{2}{\epsilon} +  \log\frac{k}{[aH]} +\log \frac{K^2}{4\pi} -\gamma_E  \right) \\
&\hspace{3cm}\times \left[ \frac{10}{81} - \frac{1}{27} \gamma_E(2+ 3 \gamma_E+ 6 
\log 2) - \frac{5}{36} \pi^2 +\frac{2 \log 2 + 3 \log^2 2}{27}\right] \notag \\[5pt]
&\hspace{5cm}+ {O}\Big((\log[aH])^2\Big) + {\rm permutations}  \notag \,.
\end{align}
In the last line, we are only writing the terms that are linear in $\log [aH]$ since this is what determines the contribution to the RG.

We can see the appearance of higher powers of log from the perturbative EFT contribution to the one-loop trispectrum 
\begin{align}
\Big\langle \vp\big(\k_1\big) \,...\, \vp\big(\k_4\big)\Big\rangle' &\supset \frac{C_\alpha^6\sum_{i} k_i^3 }{8 (k_1 k_2 k_3 k_4)^{3-2\alpha}} \frac{1}{9} \notag\\[5pt]
&\hspace{16pt}\times\int^{\t} \d \t_1 \int^{\t_1} \d \t_2 \big([a(\t_1)H] [a(\t_2)H]\big)^{-2\alpha} \int \frac{\d^3 p}{(2\pi)^3} \frac{C_\alpha^2}{2 p^{3-2\alpha}} \notag \\[8pt]
& \to \frac{\sum_{i} k_i^3 }{8 (k_1 k_2 k_3 k_4)^{3-2\alpha}} \frac{1}{9}\, \frac{1}{2} \left(-\frac{1}{2\alpha} + \log [aH] \right)^2 \notag\\[5pt]
& \hspace{16pt}\times \frac{1}{2\pi^2}\left(-\frac{1}{2\alpha} -\log \frac{K}{[aH]} +\log 2\right)\,,
\end{align}
where the compensating dimensionful factors that appear inside the $\log [aH]$ term come from expanding the prefactor in the small $\alpha$ limit.
One can check that this term matches the coefficient of the log$^3$ divergence of the UV calculation.

In order to match the linear log term, we need to keep track of the EFT field redefinition to order $\lambda^2$.  Specifically, we need 
\beq
 \bvar = H \left( [aH]^{-\alpha} \varphi_+ + [aH]^{-\beta} \varphi_- + \frac{\lambda}{9}\frac{1}{ 3!} [aH]^{-3\alpha} \varphi_+^{3}  +  \frac{\lambda^2}{81} \frac{1}{ 3!} [aH]^{- 5\alpha}  \varphi_+^{5}\right) \label{eq:bvar5} \ .
\eeq
to remove the $\vp^6$ operator from the EFT potential.  This ${O}\big(\lambda^2\big)$ term contributes to the trispectrum at one loop:
\bea
\Big\langle \bvar\big(\k_1\big) \bvar\big(\k_2\big)\bvar\big(\k_3\big) \bvar\big(\k_4\big)\Big\rangle' &\supset& \frac{\sum_i k_i^3 }{(k_1 k_2 k_3 k_4)^3}  \frac{\lambda^2}{81} \frac{5!}{3! 2} \int \frac{\d^3 p}{(2\pi)^3} \frac{1}{p^{3-2\alpha}} \,,
\eea
which matches the leading UV term in \cref{eq:K4}, namely the log term proportional to a factor of $10/81$.  After matching this term, we see a fairly complicated expression remains for the linear log.  This can be absorbed into the $c_{3,1}$ Wilson coefficient in the potential using \cref{eq:eft_tri}, such that 
\beq 
c_{3,1}\to \lambda - \frac{\lambda^2}{2 \pi^2} \bigg( \frac{1}{9} \gamma_E(2+ 3 \gamma_E+6 \log 2) + \frac{5}{12} \pi^2 - \frac{2 \log 2 + 3 \log^2 2}{9} \bigg)\,.
\label{eq:shiftc31}
\eeq
This is the one-loop matching correction to an EFT Wilson coefficient, in analogy with \cref{eq:deltaAlpha} above. A priori, one might think that we have to consider the divergence from the mixing of the operators $\vp$ and $\vp^3$.  In later sections, we will see that such a mixing is equivalent to a shift in the potential of the form in \cref{eq:shiftc31}, and therefore these two interpretations of the logarithmic growth are related to each other by a field redefinition.

\subsection{Initial Conditions for Composite Operators}\label{sec:match_alpha}
The above matching procedure is sufficient to regulate the correlation function of $\phi$ and match $\vp$ correlators at separated points.  Composite operators are defined when some of these operators are at coincident points.  In Fourier space, this involves a convolution integral which can produce divergences that require renormalizing the composite operator itself.  

Composite operators can be defined in this way purely within the EFT.  Since we are regulating loops in the EFT with dynamical dim reg, this implies that we will need to know initial conditions for general $\alpha$.  Fortunately, we will be interested in limits where the momenta are hierarchical (UV divergences of the momentum integrals), which simplifies the matching considerably.

We will start by taking the tree-level initial conditions for the trispectrum, and tying two of the legs together to form a $\vp^2[\x = 0]$ composite operator:   
\beq
\Big\langle \vp^2[0] \vp\big(\s\k_1\big) \vp\big(\s\k_2\big) \Big\rangle'_{\rm IC}  = \int \frac{\d^3 p}{(2\pi)^3} \Big\langle \vp\big(\s\p\,\big) \vp\big(-\p- \k_1-\k_2\big) \vp\big(\k_1\big) \vp\big(\k_2\big) \Big\rangle'_{\rm IC} \, ,
\label{eq:vp2IC}
\eeq
where we are interested in determining the integrand of the right hand side in the limit $p \gg k_i$. For $\alpha=0$, we determined the initial conditions exactly in \cref{eq:tri_IC}.  In order to regulate the UV divergence that comes from tying the two legs together, we want to evaluate this for general $\alpha$, while isolating the term of interest, which is proportional to $P_+(k_1) P_+(k_2)$, where $P_+$ is defined by
\beq
\Big\langle \vp\big(\k\s\big) \vp\big(\kp\big) \Big\rangle = P_+(k) (2\pi)^3 \delta\big(\k+\kp\big) \,,
\label{eq:Pplus}
\eeq
so that $\big(P_+(k)\big)_\text{tree} = (2k^3)^{-1}$.  Since $k \ll p$, the initial conditions will arise at the horizon crossing of the modes carrying momentum $p$ when the $k_i$-modes are superhorizon.  As discussed in \cref{sec:match_1loop}, we therefore match using massive mode functions for only the $k_i$ fields, where we can also Taylor expand in $k\tau \ll 1$.  As a result, the initial conditions are given by
\begin{align}
&\Big\langle \vp\big(\s\p\,\big) \vp\big(-\p-\k_1 -\k_2\big) \vp\big(\k_1\big) \vp\big(\k_2\big) \Big\rangle'_{\text{IC}^{(1)}}  \simeq \lambda P_+(k_1) P_+(k_2) \notag \\[5pt]
&\hspace{70pt} \times \lim_{\tau_0 \to 0} 2\s {\rm Im} \int^{\tau_0} \frac{\d\tau}{(-H \tau)^4} [aH]^{-2 \alpha} \frac{1}{4 p^2 } (1-i p \tau)^2 (1+i p \tau_0)^2 e^{i 2 p (\tau-\tau_0)} \,,
\end{align}
where $[aH]^{-2 \alpha} = (-\tau)^{2\alpha}$. The RHS of this expression is the calculation in the UV theory with the appropriate choice of masses for the mode functions.  We are implicitly evaluating the correlation function at a time $\tau_0$ and extracting the $\tau_0$-independent piece in the $\tau_0 \to 0$ limit.  Evaluating the integral, we find
\begin{align}
\Big\langle \vp\big(\s\p\,\big) \vp\big(\p-\k_1 -\k_2\big) \vp\big(\k_1\big) \vp\big(\k_2\big) \Big\rangle'_{\rm IC^{(1)}} &\simeq \nonumber \\[5pt]
\lambda P_+(k_1) P_+(k_2)  &\frac{(1+2\alpha) \Gamma[-1+ 2\alpha]\sin\big(\frac{\pi}{2}(1-2\alpha)\big)}{2^{2\alpha} p^{3+2\alpha}}  \label{eq:tri_1L_IC}\ .
\end{align}
Corrections to this result are suppressed by powers of $k_i / p$ which will not contribute to the one-loop divergences that we will use to determine the RG for composite operator mixing below in \cref{sec:oneLoopOpMix}.

For two-loop divergences, one must determine the initial conditions (see \cref{sec:TwoLoop} below): 
\beq
  \Big\langle \vp^3[0] \vp\big(\k\s\big) \Big\rangle'_{\rm IC^{(1)}} =  \int \frac{\d^3 p_1}{(2\pi)^3} \int \frac{\d^3 p_2}{(2\pi)^3}\Big\langle \vp{}_{\alpha_1}\big(\p_1\big) \vp{}_{\alpha_2} \big(\p_2\big) \vp{}_{\alpha_3}\big(-\k-\p_1-\p_2\big) \vp\big(\k\s\big) \Big\rangle'_{\rm IC^{(1)}}\,,\notag\\
\eeq
where we will take the limit $p_i \gg k$, such that $-\vec{k}-\p_1-\p_2 \simeq -\p_1-\p_2$.  We calculate this contribution for general $\alpha_i$ by matching to the full theory:
\begin{align}
\Big\langle \phi_{\alpha_1}\big(\p_1\big) \phi_{\alpha_2} \big(\p_2\big) \phi_{\alpha_3}\big(\p_3\big) \phi\big(\k\s\big) \Big\rangle'_{(1)} & = 2\s {\rm Im} \bigg[ u_{\nu_1}^*\big(\p_1,\tau\big)u_{\nu_2}^*\big(\p_2,\tau\big) u_{\nu_3}^*\big(\p_3,\tau\big) u_{3/2}^*\big(\k,\tau\big) \notag \\[5pt]
& \hspace{14pt} \times\int^\tau \frac{\d\tau_1}{(-\tau_1)^4} u_{\nu_1}\big(\p_1,\tau_1\big) u_{\nu_2}\big(\p_2,\tau_1\big) u_{\nu_3}\big(\p_3,\tau_1\big) u_{3/2}\big(\k,\tau_1\big) \bigg]\,,\notag\\
\end{align}
where 
\beq
u_\nu\big(\k, \tau\big) = -i\s e^{i\big(\nu+\frac{1}{2}\big) \frac{\pi}{2}} \frac{\sqrt{\pi}}{2} H(-\tau)^{3 / 2} \text{H}_{\nu}^{(1)}(-k \tau)\,,
\eeq
is the positive frequency mode for a field with $\alpha = \frac{3}{2} - \nu$.
Again, since the mode functions at horizon crossing behave as if they are effectively massless, this contribution can be determine using massless mode function of $p\tau_1$, but we must use the massive mode functions for $p \tau$.  The result of integrating over $\tau_1$ is
\beq
\Big\langle \vp\big(\p_1\big) \vp\big(\p_2\big) \vp\big(\p_3\big) \vp\big(\k\s\big) \Big\rangle_{\rm IC^{(1)}}' =
\lambda\, \frac{\prod_{i=1}^3 C_{\alpha_i}\s p^{\alpha_i}}{12 \big(p_1 p_2 p_3\big)^3 } \bigg(\sum_i \kappa_i\s p_i^3  - p_1 p_2 p_3 +\sum_{i\neq j} p_i^2 p_j  \bigg)\,, \label{eq:tri_2L_IC}
\eeq
where the correlators were matched at time $\tau$,
\beq
\kappa_i =  \frac{1}{3} \left(-\frac{1}{2\alpha} + \gamma_E-2+ \log\frac{p_t}{p_i}\right)\ ,
\label{eq:kappai}
\eeq
and $p_t = p_1+p_2+p_3$.

 \begin{figure}[t!]
\centering
\begin{tikzpicture}
    % % A helpful grid
    % \draw[step=1cm,gray,very thin] (-1,0) grid (12,6);
    % \draw[thick,->] (0,0) -- (12,0) node[anchor=north west] {x axis};
    % \draw[thick,->] (0,0) -- (0,6) node[anchor=south east] {y axis};
    % \foreach \x in {0,1,2,3,4,5,6,7,8,9,10,11}
    %   \draw (\x cm,1pt) -- (\x cm,-1pt) node[anchor=north] {$\x$};
    % \foreach \y in {0,1,2,3,4,5}
    %     \draw (1pt,\y cm) -- (-1pt,\y cm) node[anchor=east] {$\y$};
    
    % The 'tree level correlator' with contraction
    \draw[thick] (-1.5,4) -- (4.5,4);
    \draw[thick] (0,2.5) -- (0,4);
    \draw[thick] (-1,4) -- (0,2.5) -- (1,4);
    \filldraw (0,2.5) circle (2pt) node[below] {$t_1$};
    \draw[thick] (3,2) -- (4,4);
    \draw[thick] (2,4) -- (3,2) -- (3,4);
    \filldraw (3,2) circle (2pt) node[below] {$t_2$};
    \draw[thick] (0,2.5) -- (3,2);
    
    % The labels
    \filldraw (-1,4) circle (1pt) node[above] {$\k_1$};
    \filldraw (0,4) circle (1pt) node[above] {$\k_2$};
    \filldraw (1,4) circle (1pt) node[above] {$\p$};
    \filldraw (2,4) circle (1pt) node[above] {$-\p$};
    \filldraw (3,4) circle (1pt) node[above] {$\k_3$};
    \filldraw (4,4) circle (1pt) node[above] {$\k_4$};
    
    % The arrow of time
    \draw[->,thick] (-2,2.25) -- (-2,3.5);
    \node[below] at (-2,2.25) {time};
\end{tikzpicture}
\caption{Tree level pentaspectrum}
\label{fig:pentaMainText}
\end{figure}
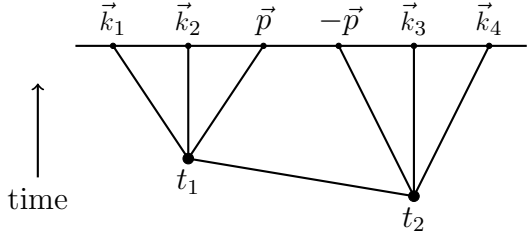

%%%%%%%%%%%%%%%%%%%%%%%%%%%%%%%%%%%%%%
Finally, we must also determined the tree-level six-point initial conditions, illustrated in \cref{fig:pentaMainText}.  We are specifically interested in the correlator
\beq
\Big\langle \vp\big(\s\p\,\big) \vp\big(-\p-\k_t\big) \vp\big(\k_1\big)\, ...\, \vp\big(\k_4\big) \Big \rangle\,,
\eeq
in the limit $p \gg k_i$, where $\k_t\equiv \sum_{i=1}^4 \k_i$.  As in the case of the trispectrum, we will isolate the term
\beq
\Big\langle \vp\big(\s\p\,\big) \vp\big(-\p-\k_t\big) \vp\big(\k_1\big)\,...\,\vp\big(\k_4\big) \Big\rangle_{\rm IC^{(2)}} \supset \lambda^2 \Gamma_{2,4}(p) P_+(k_1)\,...\,P_+(k_4)\,,
\eeq
such that
\begin{align}
\Gamma_{2,4}(p) &=  \int^\tau \frac{\d\tau_1}{(-\tau_1)^{4-2\alpha}} \int^\tau \frac{\d\tau_2}{(-\tau_2)^{4-2\alpha}} \Big\langle  \phi^2(p,\tau_1)\phi^2(p,\tau) \phi^2(p,\tau_2) \Big\rangle \notag\\[5pt]
&\hspace{14pt}- 2\s {\rm Re} \int^\tau \frac{\d\tau_1}{(-\tau_1)^{4-2\alpha}} \int^{\tau_1} \frac{\d\tau_2}{(-\tau_2)^{4-2\alpha}}\Big\langle  \phi^2(p,\tau) \phi^2(p,\tau_1) \phi^2(p,\tau_2) \Big\rangle\,.
\end{align}
By direct calculation we find that 
\begin{align}
\Gamma_{2,4}(p)  = \frac{1}{216 p^{3+4\alpha}}\bigg[& 16 + 4 \gamma_E (-11 + 3 \gamma_E) + 3\pi^2 +4 (-11 + 6\gamma_E+3 \log 2)\log 2 \notag \\[4pt]
&+ {O}\bigg(\log\frac{p}{[aH]}\bigg) \bigg]\, .
\label{eq:penta_1L_IC}
\end{align}

\section{Composite Operator Mixing}
\label{sec:StocInfNNLO}
From our above discussion, we argued that corrections to Stochastic Inflation are uniquely determined by the correlation functions of composite operators; computing those that are relevant to correcting Stochastic Inflation up to NNLO is the topic of this section.  We will start by setting up the problem of calculating composite operator renormalization.  We are interested in operators of the form $\vp^n(\x)$.  These are well-defined purely within the EFT, so we must be able to discuss their correlation functions and renormalizations given only the EFT data.  Of course, the crucial information is the initial conditions, which therefore depend on first matching to the UV. 

In general, given the EFT field operator $\vp(\x,\t)$, we can always define a composite operator
\beq
\vp^n(\x) = \prod_{i=1}^n \int \frac{\d^3 p_i}{(2\pi)^3} e^{ - i \p_i \cdot \x} \vp(\p_i)  \, .
\eeq
In a free theory, we find the structure:
\begin{align}
\Big\langle \vp^n\big(\x\big) \vp\big(\k_1\big) \,...\, \vp\big(\k_n\big)\Big\rangle &= n! \prod_{i=1}^n \int \frac{\d^3 p_i}{(2\pi)^3} e^{ - i \p_i \cdot \x} (2\pi)^3 \delta(\k_i + \p_i) P_+(k_i) \notag\\[7pt]
&= n!\s e^{i \sum \vec{k}_i\cdot \x} P_+(k_1) \, ...\, P_+(k_n)\,,
\end{align}
where $P_+$ is defined in~\cref{eq:Pplus}.
Taking the Fourier transform,
\beq
\vp^n\big(\k\s\big) = \int \d^3 x\, e^{i \k\cdot \x} \vp^n\big(\x\big)\,,
\eeq
we have
\beq
\Big\langle \vp^n\big(\k\s\big) \vp\big(\k_1\big) \,...\, \vp\big(\k_n\big)\Big\rangle = n!\s P_+(k_1) \,...\, P_+(k_n)\s \delta\Big(\k + \sum \k_i\Big)\,.
\eeq
For simplicity, it will often be easiest to use $\vp^n(\x=0) \equiv \vp^n[0]$ to avoid the extra $\delta$-function.

Given this definition, we see that we should be able to define all such correlators of composite operators as simply integrals over the correlators of $\vp$.  For example,
\beq
\Big\langle \vp^n\big[\x=0\big] \vp\big(\k_1\big) \,...\, \vp\big(\k_m\big)\Big\rangle = \prod_{i=1}^n \int \frac{\d^3 p_i}{(2\pi)^3} \Big\langle \vp\big(\p_1\big) \,...\, \vp\big(\p_n\big) \vp\big(\k_1\big) \,...\, \vp\big(\k_m\big) \Big\rangle\,.
\eeq
A very important feature of this formula is that the integrand on the RHS is free of divergences, in the sense that we should have already regulated and renormalized the expression.  As a result, all of the renormalization of the composite operator itself (and hence the matrix of anomalous dimensions) has to be associated with integrals over $\p_i$ as opposed to the loop integrals that appear in the calculation of $\big\langle \vp\big(\p_1\big) \,...\, \vp\big(\p_n\big) \vp\big(\k_1\big) \,...\, \vp\big(\k_m\big) \big\rangle$ itself.

\subsection{One-loop Corrections}
\label{sec:oneLoopOpMix}
\subsubsection*{Trispectrum}
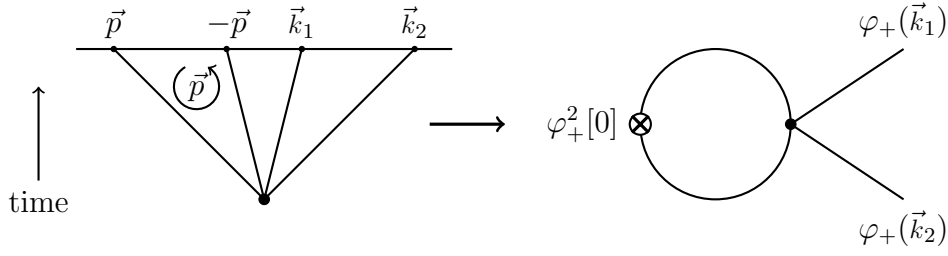
\begin{figure}[t!]
\centering
\begin{tikzpicture}
    % % A helpful grid
    % \draw[step=1cm,gray,very thin] (-1,0) grid (12,6);
    % \draw[thick,->] (0,0) -- (12,0) node[anchor=north west] {x axis};
    % \draw[thick,->] (0,0) -- (0,6) node[anchor=south east] {y axis};
    % \foreach \x in {0,1,2,3,4,5,6,7,8,9,10,11}
    %   \draw (\x cm,1pt) -- (\x cm,-1pt) node[anchor=north] {$\x$};
    % \foreach \y in {0,1,2,3,4,5}
    %     \draw (1pt,\y cm) -- (-1pt,\y cm) node[anchor=east] {$\y$};
    
    % The 'tree level correlator' with contraction
    \draw[thick] (-0.5,4) -- (4.5,4);
    \draw[thick] (0,4) -- (2,2) -- (4,4);
    \draw[thick] (1.5,4) -- (2,2) -- (2.5,4);
    \filldraw (2,2) circle (2pt);
    
    % The labels
    \filldraw (0,4) circle (1pt) node[above] {$\p$};
    \filldraw (1.5,4) circle (1pt) node[above] {$-\p$};
    \filldraw (2.5,4) circle (1pt) node[above] {$\k_1$};
    \filldraw (4,4) circle (1pt) node[above] {$\k_2$};
    \draw[->,thick] ({1.1 + 0.3*cos(120)},{3.5 + 0.3*sin(120)}) arc[radius = 3mm, start angle=120, end angle=420];
    \node at (1.1,3.5) {$\p$};
    % \node[above] at (0,2.8) {$\int d^3 p$};
    
    % The loop diagram
    \draw[thick] (8,3) circle [radius=1];
    \draw[thick] (9,3) -- (10.5,4);
    \draw[thick] (9,3) -- (10.5,2);
    \filldraw (9,3) circle (2pt);
    
    % The labels
    \node[above] at (10.5,4) {$\vp(\k_1)$};
    \node[below] at (10.5,2) {$\vp(\k_2)$};
    \filldraw[thick,draw=black,fill=white] (7,3) circle (4pt) node[left] {$\vp^2[0]$ \ };
    \node[cross out, draw=black, very thick, scale=0.7] at (7,3) {};
    
    % The arrow between them
    \draw[->, very thick] (4.2,3) -- (5.2,3);
    
    % The arrow of time
    \draw[->,thick] (-1,2.25) -- (-1,3.5);
    \node[below] at (-1,2.25) {time};
\end{tikzpicture}
\caption{One loop correction to $\varphi_+^2$ that looks like an anomalous dimension ($\varphi_+^2 \to \varphi_+^2$).  We start from the tree level trispectrum and integrate over two of the fields to form the composite operator $\vp^2$.  \textit{Left:} The Witten diagram with a boundary at future infinity.  \textit{Right:} The Feynman diagram with the same momentum flow.}
\label{fig:gamma22_massive}
\end{figure}

We will start by computing $b_1$, which is determined from the four-point function via 
\bea
\Big\langle \vp^2[0] \vp\big(\k_1\big) \vp\big(\k_2\big)\Big\rangle = \int \frac{\d^3 p}{(2\pi)^3} \Big\langle \vp\big(\s\p\,\big) \vp\big(-\p- \k_1-\k_2\big) \vp\big(\k_1\big) \vp\big(\k_2\big) \Big\rangle'\,.
\eea
The relationship between this loop contribution and the tree-level trispectrum is illustrated in \cref{fig:gamma22_massive}.  The contribution to the four-point function from the time evolution already yields a $\log [aH]$.  Therefore, any anomalous scaling (which should also only be a single log) must arise from the initial conditions.  The relevant contribution was calculated in \cref{eq:tri_1L_IC}.

Performing the integration over $p$ using dynamical dim reg, we find
\begin{align}
\Big\langle \vp^2[0] \vp\big(\k_1\big) \vp\big(\k_2\big)\Big\rangle  &= \lambda\s P_+(k_1) P_+(k_2) \nonumber \\[4pt]
&\hspace{14pt}\times \frac{(1+2\alpha) \Gamma[-1+ 2\alpha]\sin\big(\frac{\pi}{2}(1-2\alpha)\big)}{2^{2\alpha} } \frac{K^{-2\alpha} \pi^{3/2} \Gamma[\alpha]}{\Gamma\big[\frac{3}{2} +\alpha\big]} \, .
\end{align}
Expanding for $\alpha \to 0$ we have
\begin{align}
\Big\langle \vp^2[0] \vp\big(\k_1\big) \vp\big(\k_2\big)\Big\rangle  &=  \lambda\s P_+(k_1) P_+(k_2)  \bigg( \frac{1}{48 \pi^2 \alpha^2} +\frac{ (4 - 3 \gamma_E -3 \log K) }{72 \pi^2 \alpha} + {\rm finite} \bigg) \notag \\
&\to \frac{\lambda}{36 \pi^2 }\s P_+(k_1) P_+(k_2) (4 - 3 \gamma_E)  \bigg( \frac{1}{2\alpha} - \log \frac{[aH]}{k_i} \bigg) +\ldots \ .
\end{align}
We can repeat this calculation with $\vp^n[0]$ to determine the one-loop anomalous dimension for all $n$.  Keeping track of combinatorics, one finds
\begin{align}
\Big\langle \vp^n[0] \vp\big(\k_1\big)\,...\, \vp\big(\k_{n}\big)\Big\rangle &\supset  \frac{\lambda}{36 \pi^2 } \binom{n}{2}  n!\s P_+(k_1)\,...\,P_+(k_{n}) \notag\\[4pt]
&\hspace{14pt}\times\sum_i  (4 - 3 \gamma_E)  \bigg( \frac{1}{2\alpha} - \log \frac{[aH]}{k_i} +{\cal O}\left(\frac{1}{\alpha^2}\right)+ {\rm finite} \bigg)\,. \label{eq:b1_dim_reg}
\end{align}
This NLO $\log[aH]$ dependence can be resummed by including the following correction in the dynamical RG:
\beq
b_1 = -\frac{ \lambda }{36 \pi^2 } (4 - 3 \gamma_E)\, .
\eeq

\subsubsection*{Six-point}

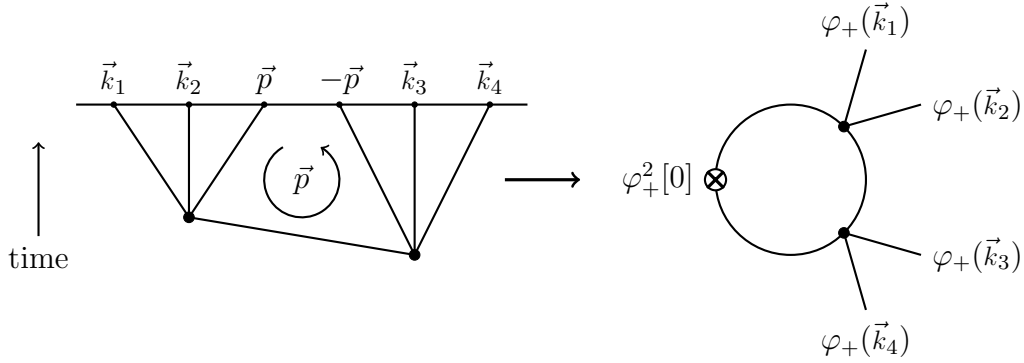
\begin{figure}[t!]
\centering
\begin{tikzpicture}
    % A helpful grid
    % \draw[step=1cm,gray,very thin] (-1,0) grid (12,6);
    % \draw[thick,->] (0,0) -- (12,0) node[anchor=north west] {x axis};
    % \draw[thick,->] (0,0) -- (0,6) node[anchor=south east] {y axis};
    % \foreach \x in {0,1,2,3,4,5,6,7,8,9,10,11}
    %   \draw (\x cm,1pt) -- (\x cm,-1pt) node[anchor=north] {$\x$};
    % \foreach \y in {0,1,2,3,4,5}
    %     \draw (1pt,\y cm) -- (-1pt,\y cm) node[anchor=east] {$\y$};
    
    % The 'tree level correlator' with contraction
    \draw[thick] (-1.5,4) -- (4.5,4);
    \draw[thick] (0,2.5) -- (0,4);
    \draw[thick] (-1,4) -- (0,2.5) -- (1,4);
    \filldraw (0,2.5) circle (2pt);
    \draw[thick] (3,2) -- (4,4);
    \draw[thick] (2,4) -- (3,2) -- (3,4);
    \filldraw (3,2) circle (2pt);
    \draw[thick] (0,2.5) -- (3,2);
    \draw[->,thick] ({1.5 + 0.5*cos(120)},{3 + 0.5*sin(120)}) arc[radius = 5mm, start angle=120, end angle=420];
    
    % The labels
    \filldraw (-1,4) circle (1pt) node[above] {$\k_1$};
    \filldraw (0,4) circle (1pt) node[above] {$\k_2$};
    \filldraw (1,4) circle (1pt) node[above] {$\p$};
    \filldraw (2,4) circle (1pt) node[above] {$-\p$};
    \filldraw (3,4) circle (1pt) node[above] {$\k_3$};
    \filldraw (4,4) circle (1pt) node[above] {$\k_4$};
    % \node[above] at (1.5,4.5) {$\int d^3 p$};
    \node at (1.5,3) {$\p$};
    
    % The loop diagram
    \draw[thick] (8,3) circle [radius=1];
    \filldraw ({8 + cos(45)}, {3 + sin(45)}) circle (2pt);
    \filldraw ({8 + cos(-45)}, {3 + sin(-45)}) circle (2pt);
    \draw[thick] ({8 + cos(45)}, {3 + sin(45)}) -- ({8 + 2*cos(60)}, {3 + 2*sin(60)}) node[above] {$\vp(\k_1)$};
    \draw[thick] ({8 + cos(45)}, {3 + sin(45)}) -- ({8 + 2*cos(30)}, {3 + 2*sin(30)}) node[right] {$\vp(\k_2)$};
    \draw[thick] ({8 + cos(-45)}, {3 + sin(-45)}) -- ({8 + 2*cos(-30)}, {3 + 2*sin(-30)}) node[right] {$\vp(\k_3)$};
    \draw[thick] ({8 + cos(-45)}, {3 + sin(-45)}) -- ({8 + 2*cos(-60)}, {3 + 2*sin(-60)}) node[below] {$\vp(\k_4)$};
    % \draw[thick] (9,3) -- (10.5,2);
    
    % The labels
    % \node[above] at (10.5,4) {$\vp(\k_1)$};
    % \node[below] at (10.5,2) {$\vp(\k_2)$};
    \filldraw[thick,draw=black,fill=white] (7,3) circle (4pt) node[left] {$\vp^2[0]$ \ };
    \node[cross out, draw=black, very thick, scale=0.7] at (7,3) {};
    
    % The arrow between them
    \draw[->, very thick] (4.2,3) -- (5.2,3);
    
    % The arrow of time
    \draw[->,thick] (-2,2.25) -- (-2,3.5);
    \node[below] at (-2,2.25) {time};
\end{tikzpicture}
\caption{Diagram of contribution one loop contribution to $\Gamma_{2,4}$ ($\varphi_+^2 \to \varphi_+^4$).  We start from the tree level pentaspectrum (6 points function) and integrate over two of the fields to form the composite operator $\vp^2$.  \textit{Left:} The Witten diagram with a boundary at future infinity.  \textit{Right:} The Feynman diagram with the same momentum flow.}
\label{fig:gamma24_massive}
\end{figure}

Our next task is to compute $b_2$, which we determine by evaluating the six point function, 
\begin{align}
\hspace{-8pt}\Big\langle \vp^2[0] \vp\big(\k_1\big) \,...\, \vp\big(\k_4\big)\Big\rangle = \int\! \frac{\d^3 p}{(2\pi)^3} \bigg\langle \vp\big(\s\p\,\big) \vp\bigg(\!\!-\p-\sum_{i=1}^4 \vec{k}_i\bigg) \vp\big(\k_1\big) \,...\, \vp\big(\k_4\big) \bigg\rangle' \,,
\end{align}
with $p \gg k_i$.  The relationship between this one-loop contribution and the tree-level six-point function is illustrated in \cref{fig:gamma24_massive}. As before, only the initial conditions arising from 
\beq
\Big\langle \vp\big(\s\p\,\big) \vp\big(-\p-\k_t\big) \vp\big(\k_1\big) \,...\, \vp\big(\k_4\big) \Big\rangle' \supset \lambda^2\s \Gamma_{2,4}(p) P_+(k_1)\,...\,P_+(k_4)\,,
\eeq
will contribute to the operator mixing, where $\Gamma_{2,4}(p)$ was calculated in \cref{eq:penta_1L_IC}.
Performing the momentum integration using dynamical dim reg, we find
\begin{align}
 \lambda^2 \int \frac{\d^3 p}{(2\pi)^3}\Gamma_{2,4}(p) 
& = \left( \frac{1}{8 \pi^2 \alpha}  - \frac{1}{2\pi^2} + \,...\, \right)\left[ b_{2,4}  + { O}\left(\frac{1}{\alpha}\right)  \right]+ {\rm finite}\,,
\end{align}
where   
\beq
b_{2,4} =  \frac{\lambda^2}{216 }\Big[ 16 + 4 \gamma_E (3 \gamma_E-11 ) + 3\pi^2 +4\log 2 ( 6\gamma_E+3 \log 2-11) \Big] \ .
\eeq
Now we restore the factor $( k_1 k_2 k_3 k_4)^\alpha [a H]^{-4 \alpha}$ associated with $P(k_i)$ for general $\alpha$ and take the limit $\alpha \to 0$ to get
\begin{align}
\Big\langle \vp^2[0] \vp\big(\k_1\big) \,...\, \vp\big(\k_4\big)\Big\rangle &\supset b_{2,4} P_+(k_1)\,...\,P_+(k_4)\left[ \frac{1}{8 \pi^2 \alpha}  + \frac{1}{2 \pi^2} \sum_{i=1}^4 \frac{1}{4} \log \frac{k_i}{[aH]} \right]\,.
\end{align}
Repeating this calculation for $\vp^n$ we have
\begin{align}
\Big\langle \vp^n[0] \vp\big(\k_1\big)\,...\, \vp\big(\k_{n+2}\big)\Big\rangle &\supset  b_{2,4}\s \binom{n}{2}  (n+2)!\s P_+(k_1)\,...\,P_+(k_{n+2}) \notag \\[4pt]
&\hspace{22pt}\times\left[ \frac{1}{8 \pi^2 \alpha}  + \frac{1}{2 \pi^2} \sum_{i=1}^4 \frac{1}{4} \log \frac{k_i}{[aH]} \right]\,.
\end{align}
This NLO $\log[aH]$ dependence can be resummed by including the following correction in the dynamical RG:
\beq
b_2 =\frac{1}{2 \pi^2} \frac{\lambda^2}{216 }\Big[ 16 + 4 \gamma_E (3 \gamma_E-11 ) + 3\pi^2 +4\log 2\, ( 6\gamma_E+3 \log 2-11) \Big]\,.
\label{eq:b2Final}
\eeq

\subsection{Two-loop Corrections}
\label{sec:TwoLoop}
\begin{figure}[t!]
\centering
\begin{tikzpicture}
    % % A helpful grid
    % \draw[step=1cm,gray,very thin] (-1,0) grid (12,6);
    % \draw[thick,->] (0,0) -- (12,0) node[anchor=north west] {x axis};
    % \draw[thick,->] (0,0) -- (0,6) node[anchor=south east] {y axis};
    % \foreach \x in {0,1,2,3,4,5,6,7,8,9,10,11}
    %   \draw (\x cm,1pt) -- (\x cm,-1pt) node[anchor=north] {$\x$};
    % \foreach \y in {0,1,2,3,4,5}
    %     \draw (1pt,\y cm) -- (-1pt,\y cm) node[anchor=east] {$\y$};
    
    % The 'tree level correlator' with contraction
    \draw[thick] (-0.5,4) -- (4.5,4);
    \draw[thick] (0,4) -- (2,2) -- (4,4);
    \draw[thick] (1.5,4) -- (2,2) -- (2.5,4);
    \filldraw (2,2) circle (2pt);
    
    % The labels
    \filldraw (0,4) circle (1pt) node[above] {$\p_1$};
    \filldraw (1.5,4) circle (1pt) node[above] {$\p_2$};
    \filldraw (2.5,4) circle (1pt);
    \node[above] at (2.7,4) {$-\p_1-\p_2$};
    \filldraw (4,4) circle (1pt) node[above] {$\k$};
    % \draw[->,thick] ({1.1 + 0.3*cos(120)},{3.5 + 0.3*sin(120)}) arc[radius = 3mm, start angle=120, end angle=420];
    % \node at (1.1,3.5) {$\p$};
    % \node[above] at (0,2.8) {$\int d^3 p$};
    
    % The loop diagram
    \draw[thick] (8,3) circle [radius=1];
    \draw[thick] (7,3) -- (10,3);
    \draw[thick] (9,3) -- (10.5,3) node[right] {$\vp(\k)$};
    \filldraw (9,3) circle (2pt);
    
    % The labels
    \filldraw[thick,draw=black,fill=white] (7,3) circle (4pt) node[left] {$\vp^3[0]$ \ };
    \node[cross out, draw=black, very thick, scale=0.7] at (7,3) {};
    \node[above] at (8,4) {$\p_1$};
    \node[below] at (8,3) {$\p_2$};
    \node[below] at (8,2) {$-\p_1-\p_2$};
    
    % The arrow between them
    \draw[->, very thick] (4.2,3) -- (5.2,3);
    
    % The arrow of time
    \draw[->,thick] (-1,2.25) -- (-1,3.5);
    \node[below] at (-1,2.25) {time};
\end{tikzpicture}
\caption{Diagram of contribution two loop contribution to the mixing $\varphi_+^3 \to \varphi_+$.  We start from the tree level trispectrum and integrate over {\it three} of the fields to form the composite operator $\vp^3$.  We first drew this as a Witten diagram with a boundary at future infinity.  At the bottom we have shown the Feynman diagram with the same momentum flow. }
\label{fig:twoloopmix}
\end{figure}
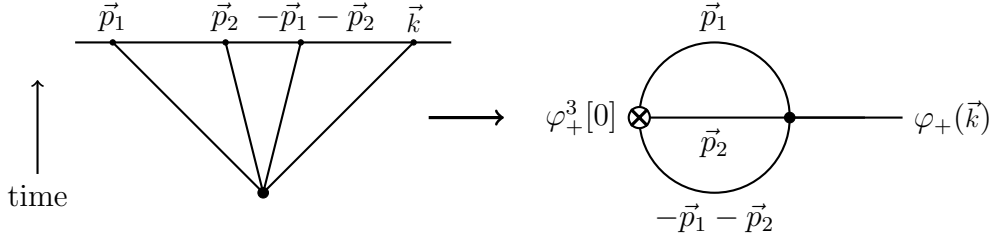

Next, we move to the calculation of the two-loop anomalous dimension that generates the NNLO non-Gaussian noise term for Stochastic Inflation.  This represents a novel contribution which is calculated here for the first time. In particular, our goal is to calculate: 
\beq
\Big\langle \vp^3[0] \vp\big(\k\s\big)\Big\rangle = \int \frac{\d^3 p_1\, \d^3 p_2\, \d^3 p_3}{(2 \pi)^9} \Big\langle \vp\big(\p_1\big) \vp\big(\p_2\big) \vp\big(\p_3\big) \vp\big(\k\s\big) \Big\rangle\,,
\label{eq:twoLoopStart}
\eeq
where $p_{1,2,3} \gg k$.  The relationship between this two-loop contribution and the tree-level trispectrum is illustrated in \cref{fig:twoloopmix}. The correlation function on the RHS was calculated in \cref{eq:tri_2L_IC} for general $\alpha_i$ (for each $p_i$) such that we can regulate the integral with dynamical dim reg.  Making the above substitution
\begin{align}
\Big\langle \vp^3[0] \vp\big(\k\s\big)\Big\rangle
\supset P_+(k)\int \frac{\d^3 p_1\, \d^3 p_2}{(2\pi)^6} &\lambda \frac{\prod_{i=1}^3 C_{\alpha_i} p^{\alpha_i}}{12 (p_1 p_2 p_3)^3 } \left(\kappa \sum_i p_i^3  - p_1 p_2 p_3 +\sum_{i\neq j} p_i^2 p_j  \right)\,,
\end{align}
where $\kappa$ is fixed by the full calculation for the reasons described above, see~\cref{eq:kappai}.  We note, that this result will require us to calculate several integrals of the form
\beq
I_3 =\int \frac{\d^3 p_1\, \d^3 p_2}{(2\pi)^6}  \frac{1}{(p_1^2)^a (p_2^2)^b ((\p_1+\p_2)^2)^c}\,,
\eeq
where $a$, $b$ and $c$ are half integers when $\alpha = 0$. We can evaluate this integral as follows:
\begin{align}
I_3&= \frac{\Gamma[b+c]}{\Gamma[b] \Gamma[c]} \int_0^1 \d x \int \frac{\d^3 p_1}{(2 \pi)^3} \frac{1}{p_1^{2a}} \int \frac{\d^3 \bar p_2}{(2\pi)^3}\frac{x^{b-1} (1-x)^{c-1}}{(\bar p_2^2 + x(1-x) p_1^2)^{b+c}} \notag\\[5pt]
&=  \frac{\Gamma[b+c-\tfrac{3}{2}]}{\Gamma[b] \Gamma[c]}  \frac{1}{(4 \pi)^{3 / 2}}\int  \frac{\d^3 p_1}{(2 \pi)^3} \frac{1}{p_1^{2a+2b+2c-3}} \int \d x \frac{x^{b-1} (1-x)^{c-1}}{(x(1-x))^{b+c-3/2}}  \notag\\[5pt]
&=\frac{\Gamma[b+c-\tfrac{3}{2}]}{\Gamma[b] \Gamma[c]} \frac{\Gamma[\tfrac{3}{2}-b] \Gamma[\tfrac{3}{2} -c]}{\Gamma[3-b-c]}  \frac{1}{(4 \pi)^{3 }} \frac{\Gamma[a+b+c-3]}{\Gamma[a+b+c-\tfrac{3}{2}]} K^{6-2a-2b-2c}\,,
\label{eq:2loophelper}
\end{align}
where we introduced an IR regulator $K$ as we did for our 1-loop divergence.  This IR regulator is only needed when $a+b+c\simeq 3$ (the integral vanishes by dynamical dim reg otherwise).  One should also notice that enforcing $a+b+c \to 3$ restores the invariance under permutations of $a$, $b$ and $c$.

We always have at least one log from $\Gamma[a+b+c -3]$.  Therefore, the only contributions that are not $\log^2$ (or higher) are those where all the other $\Gamma$ functions are finite.  Up to permutations, there are three relevant cases ($i$) $a=3/2$, $b=3/2$, $c=0$, ($ii$) $a=3/2$, $b=1$, $c=1/2$, and (iii) $a=b=c=1$.  Only ($iii$) is a single log as expected from above.  Isolating just the single log term, for example using $a=1-\alpha/2$ with $b=c=1$, we get
\beq
\Big\langle \vp^3[0] \vp\big(\k\s\big)\Big\rangle_{(1)}  =- P_+(k)\s \frac{\lambda}{12}  \frac{1}{16\pi^2} \bigg( -\frac{1}{\alpha} + \log \frac{[aH]}{k} + \log \frac{k}{K} + \,...\, \bigg)  + { O}\bigg(\frac{1}{\alpha^2} \bigg)\,.
\label{eq:2loopResult}
\eeq
Repeating this calculation for $\vp^n$ we have
\begin{align}
\Big\langle \vp^n[0] \vp\big(\k_1\big) \,...\, \vp\big(\k_{n-2}\big)\Big\rangle_{(1)} &\supset - \frac{\lambda}{192 \pi^2 } \binom{n}{3}  (n-2)! P_+(k_1)\,...\, P_+(k_{n-2})\notag\\[3pt]
&\hspace{70pt}\times\sum_i \bigg( -\frac{1}{\alpha} - \sum_i \log \frac{k_i}{[aH]}\bigg)\,.
\end{align}
This NNLO $\log[aH]$ dependence can be resummed by including the following correction in the dynamical RG:
\beq
d_0 = \frac{\lambda}{192 \pi^2 } \, .
\label{eq:d0}
\eeq

\subsection{Stochastic Inflation at NNLO}\label{sec:Stoc_final_sub}

Now we have computed all the necessary pieces.
To summarize our results, we have found that at NNLO,
the Fokker-Planck equation for Stochastic Inflation becomes 
\begin{align}
\frac{\partial}{\partial \t} P(\vp,\t)
    &= \frac{1}{3}  \frac{\partial}{\partial\vp} \bigg [\partial_{\vm}V (\vp,\vm)\Big|_{\vm=0} P(\vp,\t ) \bigg] \nonumber \\[5pt]
    &\hspace{13pt}+\frac{1}{2} \frac{\partial^2}{\partial \vp^2} \Big[ (b_0+ b_1 \vp^{2} + b_2 \vp^4)P(\vp, \t) \Big]  +\frac{1}{3!}  \frac{\partial^3}{\partial \vp^3}  \Big(d_0\s \vp   P(\vp, \t) \Big)  \,,
\label{eq:SIatNNLORaw}
\end{align}
where
\begin{subequations}
\begin{align}
V(\varphi_+,\varphi_-) &= \frac{\lambda}{3!} \varphi_- \left( \varphi_+^3 + \frac{\lambda}{18}\varphi_+^5 + \frac{\lambda^2}{162} \varphi_+^7 +\,...\, \right)\\[5pt]
b_0 &= \frac{1}{4\pi^2}\\[5pt]
b_1 &= -\frac{ \lambda }{36 \pi^2 } (4 - 3 \gamma_E) \\[5pt]
b_2 &=\frac{1}{2 \pi^2} \frac{\lambda^2}{216 }\Big[ 16 + 4 \gamma_E (3 \gamma_E-11 ) + 3\pi^2 +4\log 2\, ( 6\gamma_E+3 \log 2-11) \Big]  \\[5pt]
d_0 &=\frac{\lambda}{192 \pi^2 } \,.
\end{align}%
\label{eq:Summary}%
\end{subequations}%

The rest of this section is devoted to expressing this result in a particularly simple basis.  As we have emphasized above, the coefficient $b_1$ and $b_2$ are basis dependent, in the sense that they can be changed by taking a field redefinition.  For the purposes of solving for the NNLO equilibrium distribution, we will find it useful to first perform a field redefinition that moves their effects into the potential.  For contrast, we note that the coefficient $d_0$ is basis independent, and furthermore field redefinitions of this $(\partial/\partial\vp)^3$ term will only induce higher order (NNNLO) terms that we will neglect.

Concretely, we want to redefine the field $\tvp = f(\vp)$ such that $\tilde b_1 = \tilde b_2= 0$.  Using $P = (\d \tvp/\d\vp) \tilde P$ under such a field redefinition, we see that 
\begin{align}
\frac{1}{2}\frac{\partial^2}{\partial \vp^2} &\Big[ \big(b_0+ b_1 \vp^{2} + b_2 \vp^4 \big)P(\vp, \t) \Big] \,\to\, \notag\\[7pt]
&\frac{1}{2} \frac{\partial^2}{\partial  \tvp^2}\Bigg[ \bigg(\frac{\d\tvp}{\d\vp}\bigg)^2\big(b_0+ b_1 \vp^{2} + b_2 \vp^4\big)P(\tvp, \t) \Bigg] \notag\\[7pt]
-& \frac{1}{2} \frac{\partial}{\partial \tvp}\Bigg[ \frac{\d^2\tvp}{\d\vp^2}\frac{\d\tvp}{\d\vp}\big(b_0+ b_1 \vp^{2} + b_2 \vp^4\big)P(\tvp, \t) \Bigg]\,.
\end{align}
Next, in order to set $\tilde{b}_1 = \tilde{b}_2 = 0$, we define $\tvp$ so that 
\begin{align}
\left(\frac{\d\tvp}{\d\vp}\right)^2\big(b_0+ b_1 \vp^{2} + b_2 \vp^4\big) = b_0\,,
\end{align}
We can integrate this equation to determine
\beq
\tvp = \vp - \frac{b_1}{6 b_0} \vp^3 + \frac{3 b_1^2 - 4 b_0 b_2}{4 b_0^2} \vp^5 \quad \to \quad \vp \simeq \tvp + \frac{b_1}{6 b_0} \tvp^3\,.
\eeq
The remaining term is then determined to be
\begin{align}
\frac{\d^2\tvp}{\d\vp^2}\frac{\d\tvp}{\d\vp}\big(b_0+ b_1 \vp^{2} + b_2 \vp^4\big) &= -\frac{1}{2} \frac{b_0}{b_0 + b_1 \vp^2 +b_2 \vp^4} \big(2 b_1 \vp+ 4 b_2 \vp^3\big)\notag\\[7pt]
&\simeq - b_1 \tvp- 
\left(2 b_2 - \frac{5}{6} \frac{b_1^2}{b_0} \right) \tvp^3 \, .
\end{align}
This basis change also impacts the effective potential that appears in the Fokker-Planck equation:  
\begin{align}
V'_{\rm eff}(\tvp) &= \partial_{\vm}V (\vp,\vm)\big|_{\vm=0} +\frac{3}{4} \frac{b_0}{b_0 + b_1 \vp^2 +b_2 \vp^4} \big(2 b_1 \vp+ 4 b_2 \vp^3\big) \notag\\[5pt]
&\to \frac{3}{2} b_1 \tvp + \bigg(\frac{c_{3,1}}{3!}+ 3 b_2- \frac{5}{4}\frac{b_1^2}{b_0}\bigg) \tvp^3 +\,...\,\,.
\end{align}
The first term, $\tfrac{3}{2} b_1 \tvp$, simply provides an ${O}(\lambda)$ correction to $\alpha$, i.e., the quadratic term in the potential $c_{1,1}$ can always be removed by redefining $\alpha$, just as we did above when matching the 1-loop matching power spectrum, see \cref{eq:deltaAlpha}.  The new contribution to the second term, $b_2 = {O}(\lambda^2)$, is simply a correction to the definition $c_{3,1}$, again just as was computed above when matching to the 1-loop trispectrum, see \cref{eq:shiftc31}.  The same is true for higher powers of $\vp$ that we have dropped. In short, we see that $b_1$ and $b_2$ simply shift the definition of the couplings within SdSET at higher order in $\lambda$ but do not introduce any new terms. As a result, we may simply define 
\begin{subequations}
\begin{align}
\lambda_{\rm eff} &= \lambda + 18 b_2 +  3! \, \delta c_{3,1} \\[5pt]
\delta c_{3,1} &= - \frac{\lambda^2}{2 \pi^2} \bigg( \frac{1}{9} \gamma_E(2+ 3 \gamma_E+6 \log 2) + \frac{5}{12} \pi^2 - \frac{2 \log 2 + 3 \log^2 2}{9} \bigg)\,,
\end{align}
\end{subequations}
where $\delta c_{3,1}$ is the correction from one-loop matching in \cref{eq:shiftc31}.  For the other contributions to Stochastic Inflation, $\lambda_{\rm eff} = \lambda$ is sufficient to achieve NNLO accuracy.  In this sense, the coefficients of the NLO and NNLO corrections to the potential are independent of redefinitions of $\lambda$.  This same argument explains the scheme-independence of $\beta$-functions up to two loops.

Putting this all together and relabeling $\tilde{\varphi}_+ \to \vp$, we arrive at a canonical form for the NNLO equation: 

\begin{center}
\begin{tcolorbox}[colback=light-gray]
\begin{minipage}{\textwidth}
\begin{subequations}
\begin{align}
\hspace{-11pt}\frac{\partial}{\partial \t} P(\vp,\t)
    &= \frac{1}{3}  \frac{\partial}{\partial\vp} \big [V'_{\rm eff}(\vp) P(\vp,\t ) \big] 
    +\frac{1}{8\pi^2}\frac{\partial^2}{\partial \vp^2} P(\vp, \t)
    + \frac{\lambda_{\rm eff}}{1152\pi^2}  \frac{\partial^3}{\partial \vp^3}  \big(\vp   P(\vp, \t) \big)
    \label{eq:FPatNNLO}\\[10pt]
   &  \hspace{55pt} V'_{\rm eff} = \frac{\lambda_{\rm eff}}{3!} \bigg( \varphi_+^3 + \frac{\lambda_{\rm eff}}{18}\varphi_+^5 + \frac{\lambda_{\rm eff}^2}{162} \varphi_+^7 +\,...\, \bigg)\,.
    \label{eq:VpEff}
\end{align}%
\label{eq:SIatNNLO}%
\end{subequations}%
\end{minipage}
\end{tcolorbox}
\end{center}

\noindent Presumably this freedom to put these equations into such a canonical form can be recast in terms of a covariant description in field space~\cite{Kitamoto:2018dek,Pinol:2020cdp}, and multi-field generalizations thereof. 

\section{Implications}
\label{sec:Implications}
Having derived novel corrections to the equations that govern the Markovian evolution of the probability distribution for $\vp$, we now turn to solving them.  In particular, this section will explore the physical implications of these new terms by calculating the NNLO equilibrium distribution for $\vp$, assuming a static dS background.  Then we will extend this to account for the dynamics of the system as it relaxes to the equilibrium state.  To this end, we will set up the formalism to calculate the ``relaxation eigenvalues,'' and will numerically solve for these quantities to ${O}(\lambda^{3/2})$.

\subsection{Equilibrium Probability Distribution}

Starting with the canonical form of the NNLO equation governing Stochastic Inflation, we can  understand the impact on the equilibrium probability distribution $P_{\rm eq}(\vp)$, which by definition satisfies $\frac{\partial}{\partial\t} P_{\rm eq}(\vp) = 0$.  We can solve the problem non-perturbatively in the absence of the higher derivative term proportional to $d_0$. Therefore, our strategy will be to find the solution in terms of general $V'_{\rm eff}$, and then include the correction from $d_0$ as a perturbation.

The equilibrium contribution from $V_{\rm eff}$ can be determined non-perturbatively following \cref{sec:SILO}.  The equilibrium solution satisfies
\beq
\frac{\partial^2}{\partial \vp^2} P^{V}_{\rm eq}(\vp) = - \frac{8\pi^2}{3} \frac{\partial}{\partial \vp} V_{\rm eff}'(\vp) P^{V}_{\rm eq}(\vp) \, .
\eeq
where $V'_{\rm eff}(\vp) \equiv \partial_{\vm}V_{\rm eff} (\vp,\vm)|_{\vm=0}$ and $P^V_{\rm eq}$ is defined as being a solution to this equation.\footnote{Note that $P^V_{\rm eq}(\vp)$ includes any higher order correction in $V'_{\rm eff}$ but not does not including higher derivative terms that arise at NNLO and beyond.  In this sense $P^V_{\rm eq}$ is not to be confused with the LO solution, as it contains some (but not all) contributions at every order.}
Integrating twice gives the solution 
\beq\label{eq:PV}
P^{V}_{\rm eq}(\vp)=  C e^{- 8 \pi^2 V_{\rm eff}(\vp)/ 3} \ ,
\eeq
where we defined
\beq
V_{\rm eff}(\vp) \equiv \int^{\vp} \d\tilde{\varphi}_+ V'_{\rm eff}(\tilde{\varphi}_+) \ .
\label{eq:Veff}
\eeq
Note that $V_{\rm eff}$ is only a function of $\vp$ and should not be confused with $V(\vp,\vm)$ in \cref{eq:Vvpvm}.  
Since this solution holds for any $V_{\rm eff}$, it gives the answer at both LO and NLO provided we include the NLO contributions to $V_{\rm eff}$.

At NNLO, in addition to the correction to $V_{\rm eff}$, we must include $d_0 = \lambda_{\rm eff} / (192 \pi^2)$ which alters the equation for the equilibrium solution 
\beq
\frac{d_0}{3!}\frac{\partial^3}{\partial \vp^3}\big( \vp P_{\rm eq} (\vp)\big)+  \frac{\partial^2}{\partial \vp^2} P_{\rm eq}(\vp) = - \frac{8\pi^2}{3} \frac{\partial}{\partial \vp} V_{\rm eff}'(\vp) P_{\rm eq}(\vp) \, .
\eeq
We can integrate this equation once to get
\beq
\frac{1}{P_{\rm eq}(\vp)} \left( \frac{d_0}{3!}\frac{\partial^2}{\partial \vp^2} \vp P_{\rm eq}(\vp)+  \frac{\partial}{\partial \vp} P_{\rm eq}(\vp)\right)= - \frac{8 \pi^2}{3 } V_{\rm eff}'(\phi) \,.
\label{eq:SolvePeqIntermediate}
\eeq
This can be solved using separation of variables, so we will make the ansatz 
\begin{align}
P_{\rm eq}(\vp) = P_{\rm eq}^V(\vp) Q(\vp)\,,
\end{align}
where $P_{\rm eq}^V$ is given in \cref{eq:PV}, including the NNLO corrections to $V_{\rm eff}$. Plugging this ansatz into \cref{eq:SolvePeqIntermediate} gives
\beq
\frac{d_0}{3!}\frac{1}{Q (\vp) P_{\rm eq}^V(\vp)} \frac{\partial^2}{\partial \vp^2} \big(\vp P_{\rm eq}^V(\vp) Q(\vp)\big)  + \frac{1}{Q(\vp)}\frac{\partial}{\partial \vp} Q(\vp) = 0\, .
\eeq
We can solve this equation perturbatively in $Q$.  The zeroth order term is $Q =$ constant.  At the next order, clearly $Q'$ is order $d_0$ so we can neglect derivatives of $Q$ in the first terms such that 
\begin{align}
 \frac{1}{Q(\vp)}\frac{\partial}{\partial \vp} Q(\vp) &= - \frac{d_0}{ 3! P_{\rm eq}^V(\vp)} \frac{\partial^2}{\partial \vp^2} \big(\vp P_{\rm eq}^V(\vp)\big) \notag \\[7pt]
 &= -\frac{d_0}{3!} \left( -  \frac{16\pi^2}{3} V_{\rm eff}'(\phi) + \vp \left(\frac{8 \pi^2}{3} V_{\rm eff}'(\phi) \right)^2 -\frac{8 \pi^2}{3} \vp V_{\rm eff}''(\phi)  \right)\,,
 \end{align}
such that the solution becomes
\beq
\log Q = \frac{d_0}{3!}\Bigg[ \frac{16 \pi^2}{3} V_{\rm eff}(\phi) - \int \d\vp \left(\vp \left(\frac{8 \pi^2}{3} V_{\rm eff}'(\phi) \right)^2 -\frac{8 \pi^2}{3} \vp V_{\rm eff}''(\phi)  \right)\Bigg] \,.
\eeq
Finally, we use $d_0 = \lambda_{\rm eff} / (192 \pi^2)$ and $V_{\rm eff} \simeq \lambda_{\rm eff} \vp^4/4!$, which consistently captures effects up to NNLO accuracy.  We then evaluate the integrals and simplify the expression to find 
\beq
Q(\vp) = \exp\Bigg[ \frac{\lambda_{\rm eff}^2 \vp^4}{1152} \bigg(\frac{5}{9} -\frac{2\pi^2}{81}  \lambda_{\rm eff} \vp^4 \bigg) \Bigg] \, .
\eeq
Using the fact that $\lambda_{\rm eff} \vp^4 = {O}(1)$ for the LO equilibrium solution, we see that both terms in $Q$ are ${O}(\lambda_{\rm eff})$, as expected for NNLO accuracy.  Recall the NLO and one additional contribution at NNLO are encoded in $V_{\rm eff}(\phi_+)$ and are included in $P^V_{\rm eq}(\phi)$.  Combining these terms and writing $P_{\rm eq} = C P_{\rm LO}(\vp) P_{\rm NLO}(\vp) P_{\rm NNLO}(\vp)$, we have 
\begin{subequations}
\begin{align}
P_{\rm LO} &= \exp\left(- \frac{\pi^2}{9} \lambda_{\rm eff} \vp^4 \right)\\[4pt] 
P_{\rm NLO} &= \exp\left(- \frac{\pi^2}{243} \lambda_{\rm eff}^2 \vp^6 \right) \\[4pt]
P_{\rm NNLO} &=  \exp\left( \frac{5}{10368} \lambda_{\rm eff}^2 \vp^4 - \frac{17\pi^2}{46656} \lambda_{\rm eff}^3 \vp^8 \right)\,.
\end{align}
\end{subequations}
In the regime where $\log P_{\rm LO} = O(1)$, we have $\log P_{\rm NLO} = O\big(\lambda_{\rm eff}^{1/2}\big)$ and $\log P_{\rm NNLO} = O(\lambda_{\rm eff})$.

\subsection{Relaxation Eigenvalues}
In this section, we will explore the implications for the time dependence of $P(\vp,\t)$.  This can be characterized by computing the so-called ``relaxation eigenvalues'' as we explain below.  In the previous section, we could calculate the analytic NNLO equilibrium probability distribution where we only had to treat the $(\partial/\partial\vp)^3$ term perturbatively. Here, we must resort to a numerical evaluation of the perturbative expansion, which requires that we treat {\it all} higher order corrections as perturbations.

To begin, we return to the full NNLO equation that governs Stochastic Inflation given in \cref{eq:FPatNNLO}, and rewrite it as a Euclidean Schr\"{o}dinger equation~\cite{Starobinsky:1994bd,Baumgart:2019clc}:
\beq
\frac{\partial}{\partial \t} P(\vp,\t)
= \frac{1}{3}  \frac{\partial}{\partial\vp} \big [V'_{\rm eff}(\vp) P(\vp,\t ) \big] 
+\frac{1}{8\pi^2} \frac{\partial^2}{\partial \vp^2} P(\vp, \t) \,,
\eeq
where $V'_{\rm eff}(\vp) \equiv \partial_{\vm}V_{\rm eff} (\vp,\vm)|_{\vm=0}$.
This equation can be solved using separation of variables 
\beq
P(\vp,\t) =
\exp \bigg[-\frac{4 \pi^2}{3} V_{\rm eff}(\vp)  \bigg]
\sum_{n=0}^{\infty} \Phi_n(\vp) e^{-\Lambda_n \t}\,,
\label{eq:EuclideanSWEsoln}
\eeq
where $V_{\rm eff}(\vp)$ is defined in \cref{eq:Veff}, we have assumed $t_0=0$, and $\Phi_n$ are the eigenfunctions of~\cite{Starobinsky:1994bd,Baumgart:2019clc}
\beq
\frac{\partial^2}{\partial \vp^2} \Phi_n(\vp) - U(\vp) \Phi_n(\vp)
= -8 \pi^2 \Lambda_n \Phi_n(\vp)\,,
\label{eq:EuclideanSWE}
\eeq
where $\Lambda_n$ are the non-negative ``relaxation eigenvalues,'' and the Schr\"{o}dinger potential is
\beq
U(\vp) =
\left( \frac{4 \pi^2}{3} V'_{\rm eff}(\vp) \right)^2 - \frac{4 \pi^2}{3} \frac{\partial}{\partial \vp} V'_{\rm eff}(\vp)\,.
\label{eq:EuclideanPotential}
\eeq
The lowest eigenvalue is zero with the eigenfunction $\Phi_0(\vp) \propto \exp \left[-\frac{4 \pi^2}{3} V_{\rm eff}(\vp) \right]$. Since all other $\Lambda_n$ are positive, at late times the distribution $P(\vp,t)$ relaxes to the fixed point 
\beq
    P(\vp,\t) = N \exp \bigg[-\frac{8 \pi^2}{3} V_{\rm eff}(\vp) \bigg] \,.
\eeq
This reproduces the static result above in~\cref{eq:PV}. However, for the numerical evaluations we take $V'_{\rm eff}(\vp) = \frac{\lambda_{\rm eff}}{3!} \vp^3$, and treat the additional correction to the potential as perturbations.  In addition, for the remainder of this section, we will drop subscript `eff' on the coupling, $\lambda_{\rm eff} \to \lambda$, for brevity.

To explore the effect of the NNLO correction we need to determine the eigenvalues $\Lambda_n$ of \cref{eq:FPatNNLO} for $n \geq 1$. We can find the eigenvalues and eigenfunctions of \cref{eq:FPatNNLO} as perturbative expansions in powers of $\sqrt{\lambda}$,
\begin{subequations}
\begin{align}
    \Lambda_n &= \lambda^{1/2} \Lambda_n^{(0)} + \lambda\s \Lambda_n^{(1)} + \lambda^{3/2} \Lambda_n^{(2)} + \,...\, \\[0.75em]
    \Phi_n &= \Phi_n^{(0)} + \lambda^{1/2} \Phi_n^{(1)} + \lambda \Phi_n^{(2)} + \,...\, \\[0.75em]
    U &= \lambda^{1/2} U^{(0)} + \lambda\s U^{(1)} + \lambda^{3/2} U^{(2)} + \,...\,\,,
\end{align}
\end{subequations}
where $\Lambda_n^{(0)}$ and $\Phi_n^{(0)}$ are the solutions of \cref{eq:EuclideanSWE} with Schr\"odinger potential $U = U^{(0)}$. This potential is obtained from \cref{eq:EuclideanPotential} by setting $V'_{\rm eff}(\vp) = \frac{\lambda}{3!} \vp^3$, the leading term in \cref{eq:VpEff}. Explicitly,
\beq
    \frac{\partial^2}{\partial \vp^2} \Phi_n^{(0)} (\vp)
        - \left[ \frac{4 \pi^4}{81} \lambda^2 \vp^6  - \frac{2 \pi^2}{3} \lambda \vp^2 \right] \Phi_n^{(0)} (\vp)
    = -8 \pi^2 \lambda^{\frac{1}{2}} \Lambda_n^{(0)} \Phi_n^{(0)} (\vp)
    \label{eq:EuclideanSWElowest}
\eeq
Since $\vp \sim \lambda^{-1/4}$ we see that the Schr\"odinger potential, the term in square brackets, indeed scales as $O\big(\lambda^{1/2}\big)$.
This equation can be solved numerically by the shooting method\cite{Markkanen:2019kpv}. Higher order terms in \cref{eq:VpEff} as well as the NNLO term $\partial_{\vp}^3 (\vp P)$ can be treated as perturbative corrections to \cref{eq:EuclideanSWElowest}. That is, if we substitute \cref{eq:EuclideanSWEsoln} into \cref{eq:FPatNNLO} and keep terms up to $O\big(\lambda^{3/2}\big)$, the eigenfunctions $\Phi_n$ must solve
\begin{align}
 \hspace{-9pt}   \frac{\partial^2}{\partial \vp^2} \Phi_n
        - \Big[ \lambda^{1/2} U^{(0)} + \lambda\s U^{(1)} + \lambda^{3/2} U^{(2)} \Big] \Phi_n
        = -8 \pi^2 \Big[ \lambda^{1/2} \Lambda_n^{(0)} + \lambda\s \Lambda_n^{(1)} + \lambda^{3/2} \Lambda_n^{(2)} \Big] \Phi_n\,,
    \label{eq:EuclideanSWEperturbative}
\end{align}
where 
\begin{subequations}
\begin{align}
    \lambda^{1/2} U^{(0)} &= \frac{4 \pi^4}{81} \lambda^2 \vp^6  - \frac{2 \pi^2}{3} \lambda \vp^2 \\[0.75em]
    \lambda\s U^{(1)} &= \frac{4 \pi^4}{729} \lambda^3 \vp^8 - \frac{5 \pi^2}{81} \lambda^2 \vp^4 \\[0.75em]
    \lambda^{3/2} U^{(2)}
        &= \frac{79 \pi^4}{104976} \lambda^4 \vp^{10} - \frac{53 \pi^2}{5832} \lambda^3 \vp^6 - \frac{5}{1728} \lambda^2 \vp^2 
        + \left( \frac{\pi^2}{7776} \lambda^3 \vp^7 - \frac{5}{1728} \lambda^2 \vp^3 \right) \partial_{\vp}\notag \\
        &\qquad\qquad + \left(-\frac{1}{1728} \lambda^2 \vp^4 + \frac{\lambda}{384\pi^2} \right) \partial_{\vp}^2
        + \frac{\lambda}{1152 \pi^2} \vp \partial_{\vp}^3 \, .
\end{align}
\end{subequations}
Finally, the perturbative corrections to the eigenvalues are computed numerically using
\begin{subequations}
\begin{align}
    8 \pi^2 \Lambda_n^{(1)} &= \big\langle \Phi_n^{(0)} \big| U^{(1)} \big| \Phi_n^{(0)} \big \rangle \\[0.75em]
    8 \pi^2 \Lambda_n^{(2)} &=
        \big\langle \Phi_n^{(0)} \big| U^{(2)} \big| \Phi_n^{(0)}  \big\rangle
        + \frac{\lambda^{1/2}}{8\pi^2}
            \sum_{k \neq n}
            \frac{
                \big|\big\langle \Phi_k^{(0)} \big| U^{(1)} \big| \Phi_n^{(0)}  \big\rangle\big|^2
            }{
                \Lambda_n^{(0)} - \Lambda_k^{(0)}
            } \,.
\end{align}
\end{subequations}
The first few relaxation eigenvalues are (recall that $\lambda = \lambda_\text{eff}$ here)

{
\begin{center}
\renewcommand{\arraystretch}{2}
\setlength{\tabcolsep}{2em}
\setlength{\arrayrulewidth}{1.2pt}
\begin{tabular}{l|l}
    \multicolumn{1}{c|}{$n$} & \multicolumn{1}{c}{$\Lambda_n$} \\ \hline
    $1$ & $0.03630 \, \lambda^{1/2} + 0.00076 \, \lambda + 0.00049 \, \lambda^{3/2}$ \\ \hline
    $2$ & $0.11814 \, \lambda^{1/2} + 0.00338 \, \lambda + 0.00138\, \lambda^{3/2}$ \\ \hline
    $3$ & $0.21910 \, \lambda^{1/2} + 0.00795 \, \lambda + 0.00316 \, \lambda^{3/2}$ 
\end{tabular}
\end{center}
}
The contribution to the eigenvalues at ${O}\big(\lambda^{3/2}\big)$ includes both corrections to the equations of Stochastic Inflation at NNLO as well as perturbative corrections to the eigenvalues from the LO and NLO equations.  Working with the full NNLO equations was crucial to obtaining the detailed numeric values. We note that the  NNLO contribution to $V_{\rm eff}$ in \cref{eq:VpEff} dominates; the small numerical coefficient of $d_0 \simeq 9 \times 10^{-5} \lambda$ suppresses its impact on these eigenvalues.  As the form of $V_{\rm eff}$ is determined by SdSET field redefinitions to all orders, it may prove useful in future studies to simply include higher order corrections to the potential as an approximation. We expect this 
minor impact of $d_0$ is due to the fact that we are assuming the UV theory is 
$\lambda \phi^4$ such that the non-Gaussian noise and corrections to the potential are determined by the same parameter.  In contrast, if one were to consider inflationary models with primordial non-Gaussianity, these two effects are controlled by independent parameters, such that the non-Gaussian contribution to the noise could become important~\cite{Meerburg:2019qqi}.  In either case, this investigation is only possible because we have framework in which all corrections to Stochastic Inflation, including contributions from non-Gaussian noise, can be systematically computed.

\section{Conclusions}
\label{sec:Conc}
Understanding the nature of quantum dS space is one of the most basic conceptual problems in cosmology~\cite{Witten:2001kn}.  Stochastic Inflation~\cite{Starobinsky:1986fx,Nambu:1987ef,Starobinsky:1994bd} informs much of the physical intuition for how we think about accelerating cosmologies, particularly as we approach the eternally inflating regime that is dominated by quantum fluctuations~\cite{Linde:1986fd,Goncharov:1987ir, Creminelli:2008es,Freivogel:2011eg}.  Yet,  Stochastic Inflation is itself an approximation whose regime of validity, and corrections thereof, should follow from a more basic starting point.  Ultimately, a complete description should include dynamical gravity, although the simpler case of quantum field theory in a fixed dS background studied here already provides a non-trivial challenge.  

In this paper, we demonstrated precisely how corrections to Stochastic Inflation arise from quantum field theory in dS, namely as a natural consequence of dynamical renormalization group flow within the EFT that emerges in the superhorizon limit.  By working with SdSET, the origin of the stochastic description is a direct consequence of EFT power counting, which also explains why this effect is only relevant for light (massless) scalars.  (This same power counting scheme also explains the all orders conservation of the adiabatic mode~\cite{Cohen:2020php}.)  By matching $\lambda \phi^4$ theory onto SdSET up to one loop, we could then calculate the log enhanced corrections to the mixing of EFT operators up to two loops.  This allowed us to derive the corrections to the equations of Stochastic Inflation at NNLO accuracy.  These results include the first higher derivative correction to the framework, which is the leading signature of the non-Gaussian contribution to the noise as modes cross the horizon.

This work extends derivations of Stochastic Inflation from quantum field theory in dS at LO~\cite{Burgess:2015ajz,Baumgart:2019clc,Cohen:2020php,Baumgart:2020oby} and NLO~\cite{Gorbenko:2019rza,Mirbabayi:2019qtx,Mirbabayi:2020vyt} to NNLO.  Yet, even at NLO, we showed that the ``universal" correction to the effective potential follows from a field redefinition and can be extended to all orders.  This result agrees with~Refs.~\cite{Gorbenko:2019rza,Mirbabayi:2019qtx,Mirbabayi:2020vyt}, which arrive at this NLO correction by (effectively) integrating-out the decaying mode at tree-level.  Furthermore, the first appearance of non-Gaussianity in the stochastic noise appears at NNLO and requires a genuine two-loop calculation.  Higher-loop calculations of inflationary correlators are notoriously difficult, but they are made manageable by working with SdSET, which facilitates the use of the symmetry preserving dynamical dimensional regularization.  Most importantly, SdSET reduces the problem of calculating any corrections to Stochastic Inflation to the determination of the matrix of anomalous dimensions.  Rather than being a mysterious feature of dS space, we now see that the derivation of Stochastic Inflation and corrections thereof is  conceptually and technically similar to calculating the scaling dimensions of operators at the Wilson-Fisher fixed point in $d=4-\epsilon$ dimensions.  Finally, there is an intriguing connection that could be made with a thermodynamic interpretation of the equilibrium probability distributions, $P_\text{LO} \sim \exp(-\beta E)$, where the inverse temperature is $\beta = 2\pi/H$.  It would be interesting to understand the meaning of the NNLO corrections derived here from this point of view.

Phenomenologically, Stochastic Inflation is an important tool for understanding the predicted non-Gaussianity in multi-field inflation~\cite{Salopek:1990jq,Wands:2000dp, Seery:2005gb,Wands:2010af,Tada:2016pmk,Achucarro:2016fby}, where superhorizon evolution can give rise to non-trivial correlations.  Previous work has included the non-Gaussian contributions for the non-linear superhorizon evolution, but thus far the effects of non-Gaussian noise has been missing.  It will be interesting to explore models where both effects are simultaneously important.  For example, one might hope this techniques would elucidate the physics of the small mass regime of quasi-single field inflation~\cite{Chen:2009zp}, which is known to produce large logs.

Conceptually, Stochastic Inflation serves as the basis for much of our understanding of slow-roll eternal inflation.  This description requires coupling a light scalar field to gravity, yet much of the structure is determined by the quantum noise in the Fokker-Planck equation.  Specifically, the regime of slow roll eternal inflation is the limit where the potential becomes flat and the quantum noise dominates the time evolution until inflation ends. While understanding this regime is often considered a conceptual problem, it may have important consequences for cosmological solutions to hierarchy problems, such as~\cite{Graham:2015cka, Geller:2018xvz, Kartvelishvili:2020thd, Giudice:2021viw}.

Finally, underlying our results on Stochastic Inflation is a demonstration that SdSET is a consistent description of dS quantum field theory at loop level.  Calculations in a wide variety of cosmological settings are beset with challenges stemming from the underlying time evolution and lack of consistent regulator.  The successful implementation of SdSET as an organizing principle for calculating quantum correlators in dS offers hope that more of these cosmological problems may be organized and simplified when described with the right degrees of freedom.  The emergence of Stochastic Inflation as a simple consequence of EFT power counting is a non-trivial example of these principles in action.

\paragraph{Acknowledgements}
We are grateful to Daniel Baumann, Matthew Baumgart, Raphael Flauger, Tom Hartman, Mehrdad Mirbabayi, Gui Pimentel, Eva Silverstein, and Raman Sundrum for helpful discussions. T.\,C.~is supported by the US~Department of Energy, under grant no.~DE-SC0011640.  D.\,G., A.\,P.~and A.\,R. were supported by the US~Department of Energy under grant no.~DE-SC0019035.

% -----------------------------------------------------------------------------------------------------------------------------------------
\appendix

%%%%%%%%%%%%%%%%%
\section*{Appendices}
\section{Matching the Six-point Function}
\label{sec:MatchingSixPt}
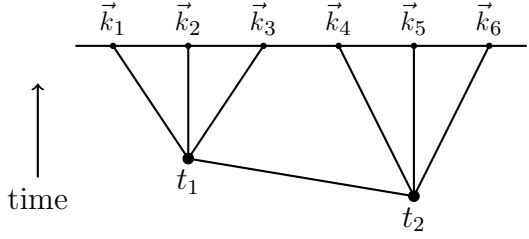
\begin{figure}[t!]
\centering
\begin{tikzpicture}
    % % A helpful grid
    % \draw[step=1cm,gray,very thin] (-1,0) grid (12,6);
    % \draw[thick,->] (0,0) -- (12,0) node[anchor=north west] {x axis};
    % \draw[thick,->] (0,0) -- (0,6) node[anchor=south east] {y axis};
    % \foreach \x in {0,1,2,3,4,5,6,7,8,9,10,11}
    %   \draw (\x cm,1pt) -- (\x cm,-1pt) node[anchor=north] {$\x$};
    % \foreach \y in {0,1,2,3,4,5}
    %     \draw (1pt,\y cm) -- (-1pt,\y cm) node[anchor=east] {$\y$};
    
    % The 'tree level correlator' with contraction
    \draw[thick] (-1.5,4) -- (4.5,4);
    \draw[thick] (0,2.5) -- (0,4);
    \draw[thick] (-1,4) -- (0,2.5) -- (1,4);
    \filldraw (0,2.5) circle (2pt) node[below] {$t_1$};
    \draw[thick] (3,2) -- (4,4);
    \draw[thick] (2,4) -- (3,2) -- (3,4);
    \filldraw (3,2) circle (2pt) node[below] {$t_2$};
    \draw[thick] (0,2.5) -- (3,2);
    
    % The labels
    \filldraw (-1,4) circle (1pt) node[above] {$\k_1$};
    \filldraw (0,4) circle (1pt) node[above] {$\k_2$};
    \filldraw (1,4) circle (1pt) node[above] {$\k_3$};
    \filldraw (2,4) circle (1pt) node[above] {$\k_4$};
    \filldraw (3,4) circle (1pt) node[above] {$\k_5$};
    \filldraw (4,4) circle (1pt) node[above] {$\k_6$};
    
    % The arrow of time
    \draw[->,thick] (-2,2.25) -- (-2,3.5);
    \node[below] at (-2,2.25) {time};
\end{tikzpicture}
\caption{The tree level in-in six-point function.}
\label{fig:penta}
\end{figure}

In this appendix, we provide some details for matching the tree-level six-point function, as illustrated in \cref{fig:penta}.  This serves as an input to the one-loop corrections, see~\cref{fig:gamma24_massive}, and also provides a non-trivial check on the matching the EFT operator $\bvar$ to the UV field $\phi$.  

Assuming the UV interaction is $\lambda \phi^4$, the six-point function first arises at second order in perturbation theory.  Using the commutator form of the in-in correlator, see \cref{eq:commutator}, we can write the full six-point function as
\beq
 \Big\langle  \phi\big(\k_1\big)  \phi\big(\k_2\big)\phi\big(\k_3\big)\phi\big(\k_4\big)\phi\big(\k_5\big)\phi\big(\k_6\big)  \Big\rangle_{\rm tree} = A_6 +B_6 \,,
\eeq
where 
\begin{align}
{A}_6 =  \bigg\langle  \left(i\int \d t_1  \Big[H_{\rm int}(t_1), \phi\big(\k_1\big)\Big] \right) \, & \left(i \int  \d t_2  \Big[H_{\rm int}(t_2), \phi\big(\k_2\big)\Big] \right) \,   \phi\big(\k_3\big)\phi\big(\k_4\big)\phi\big(\k_5\big)\phi\big(\k_6\big)\bigg\rangle\notag\\[3pt]
&\hspace{120pt}+ {\rm permutations}\,,
\end{align}
and 
\begin{align}
B_6 = \bigg\langle \phi\big(\k_1\big)\phi\big(\k_2\big)\phi\big(\k_4\big) \phi\big(\k_4\big)\phi\big(\k_5\big)& i^2 \int \d t_1 \int^{t_1} \d t_2 \Big[H_{\rm int}(t_2),\Big[H_{\rm int}(t_1),\phi\big(\k_6\big) \Big]\Big] \bigg\rangle\notag\\[3pt]
&\hspace{106pt}+ {\rm permutations}\, .
\end{align}
Our goal is to match this expression onto the EFT, so we need to take the limit where all of the fields are superhorizon.  The additional contributions that arise from the subhorizon region, $k_i \tau_j ={O}(1)$, will be absorbed into the initial conditions, which we do not need to calculate explicitly for our purposes in this work.

To match the superhorizon behavior, we can again expand the operator using \cref{eq:phi_lambda} to find
\begin{align}
[\phi]_\lambda &= i\int^t \d t_1  \Big[H_{\rm int}(t_1), \phi\big(\k_1\big)\Big] =\frac{\lambda}{3!}  \frac{1}{3} \bigg(- \bigg(\!-\frac{1}{2\alpha}+ \log [aH] \bigg) + \frac{1}{3} \bigg)  \vp^3(\x) \,. 
\end{align}
It is easy to see that the superhorizon contribution to $A_6$ is determined by $[\phi]_\lambda$,
\begin{align}
{A}_6 &=  \bigg\langle  [\phi]_\lambda(\k_1) \, [\phi]_\lambda\big(\k_2\big) \,   \phi\big(\k_3\big)\phi\big(\k_4\big)\phi\big(\k_5\big)\phi\big(\k_6 \big)\bigg\rangle\notag+ {\rm permutations}\, .
\end{align}
This shows that by matching the trispectrum with \cref{eq:EFT_bvar}, we can also match $A_6$.

Clearly we cannot determine $B_6$ using $[\phi]_\lambda$.  The most straightforward way to determine the superhorizon contribution is to write
\begin{align}
B_6= \frac{\lambda^2}{3!2}\bigg\langle \phi\big(\k_1\big)\phi\big(\k_2\big)\phi\big(\k_4\big) \phi\big(\k_4\big)\phi\big(\k_5\big)& \int^{t}_{-\infty} \d t_1\, a^3(t_1) G\big(\k_6,t_1, t\big) \int \d^3 x_1 \phi^2(\x_1, t_1) \notag\\[4pt]
& \hskip -30pt \times\int^{t_1}_{-\infty} \d t_2\, a^3(t_2) G\big(\k_{123},t_2, t_1\big) \int \d^3 x_2 \phi^3(\x_2,t_3) \bigg\rangle  \, .
\end{align}
where
\begin{align}
G\big(\k, t',t\big)= i \Big\langle \Big[\phi\big(\k,t'\big),\phi\big(\kp,t\big)\Big]\Big\rangle' \simeq \frac{H}{3} \Big([a(t')H]^{-3} -[a(t)H]^{-3}\Big)\,.
\end{align}
Integrating this expression using dynamical dim reg, we find the superhorizon contribution is
\begin{align}
B_6  &=\lambda^2 \frac{H^6 \, \sum_i k_i^3}{2^5 (k_1 k_2 k_3 k_4 k_5 k_6)^3}  \frac{5!}{(3!)^2} \int^\t \d \t_1 \frac{[a(\t_1)H]^{-2\alpha}}{3}\left(-1 + \frac{[a(\t_1)H]^3}{[a(\t)H]^3}\right)\notag\\[5pt]
&\hskip 130pt\times \int^{\t_1} 
\d \t_2 \frac{[a(\t_2)H]^{-2\alpha}}{3}\left(-1 + \frac{[a(\t_2)H]^3}{[a(\t_1)H]^3}\right)\notag\\[5pt]
&=  \frac{ \lambda^2\, H^6\, \sum_i k_i^3}{2^5 (k_1 k_2 k_3 k_4 k_5 k_6)^3} \bigg[ \frac{10}{9} \left(\frac{1}{2\alpha} - \log[aH] +\frac{1}{3} \right)\left(\frac{1}{4\alpha}  -\log[aH] +\frac{1}{3} \right) +\frac{10}{81} \bigg]\,.
\label{eq:B6UV}
\end{align}

Next, we would like to see how this formula arises in SdSET.  First, we calculate the contribution to the six-point function at second order in $c_{3,1}$.  We take the same commutator structure as $B_6$, where one $[\varphi_+,\varphi_-]$ acts on the external line and the other on an internal commutator.  The result is
\bea
B_{3,1} &=& c_{3,1}^2 \frac{5!}{(3!) 2} \frac{H^6}{2^5 (k_1 .. k_6)^{3-2\alpha}} \sum_i k_i^{3-2\alpha} \int^{\t} \d \t_1 \frac{1}{3}  [a(\t_1)H]^{2\alpha}  \int^{\t_1} \d \t_2 \frac{1}{3} [a(\t_2)H]^{2\alpha} \notag \\[5pt]
&=&c_{3,1}^2 \frac{10}{9} \sum_i k_i^{3-2\alpha}  \frac{C_\alpha^{10}}{(k_1 .. k_6)^{3-2\alpha}} \frac{[aH]^{4 \alpha}}{8\alpha^2}\,,
\eea
where the additional factors of $1/3$ are from the commutator $i[\varphi_+ ,\varphi_- ] =\delta\big(\x+\x'\big)/3$ when $\alpha  =0$.  In addition, we have the contribution from the time evolution of $\bvar$ at order $c_{3,1}$ from \cref{eq:FieldMapAfterEFTFieldRedef} and \cref{eq:EFT_bvar}, which yields
\begin{align}
\bvar \supset  \frac{c_{4,0}}{9}\frac{H}{ 3!} [aH]^{-3\alpha} \varphi_+^{3}  &\to \frac{c_{4,0}}{3}\frac{1}{ 3!} [aH]^{-3\alpha} \varphi_+^{2}[\vp]_\lambda \notag\\[5pt]
&\to
 \frac{H}{3!^2} \frac{c_{4,0} c_{3,1}}{9}\bigg(\!\frac{1}{2\alpha}- \log [aH]  \bigg) \vp^5(\x)   \, .
\end{align}
This contribution is in addition, to the $\vp^5$ term in $\bvar$ in \cref{eq:bvar5}, that we determined from our field redefinition, 
\beq
\bvar \supset \frac{\lambda^2}{81} \frac{1}{ 3!} [aH]^{- 5\alpha}  \varphi_+^{5} \, .
\eeq
Combining these two terms in $\bvar$, we get the contribution to the six-point function: 
\beq
B_{\bvar} =     \frac{ \lambda^2\, H^6\, \sum_i k_i^3}{2^5 (k_1 k_2 k_3 k_4 k_5 k_6)^3} \left[ \frac{10}{27} \bigg(\!\frac{1}{2\alpha}- \log [aH]  \bigg)+ \frac{20}{81}\right]\,.
\eeq
Finally, from the field redefinition we found in \cref{eq:Vvpvm}, we also have a correction to the effective potential via $c_{5,1} =\frac{\lambda^2}{18} \frac{5!}{3!}$.  This contributions to the six-point function at linear order in $c_{5,1}$:
\begin{align}
B_{5,1} &= \frac{c_{5,1}}{2\nu} \frac{[aH]^{-4 \alpha}}{4\alpha} \frac{H^6}{2^5 k_1^{3-2\alpha} ..k_6^{3-2\alpha}} \sum_i k_i^{3-2\alpha}\notag\\[5pt]
&\to
 \lambda^2 \frac{10 }{27}  \frac{H^6 \sum_i k_i^{3-2\alpha}}{2^5 k_1^{3-2\alpha} ..k_6^{3-2\alpha}}  \left(\frac{1}{4\alpha} -\log [aH] \right)\,.
\end{align} 
Combining these terms and using $c_{3,1} = c_{4,0} = \lambda$, we match the UV six-point function
\beq
B_6 = B_{3,1}+B_{\bvar}+B_{5,1} \, .
\eeq
Note that there is some ambiguity in the constant term due to scheme dependence associated with regulating our (divergent) time integrals.  Although our expression matches the constant as well, in some other schemes, the initial conditions may play a role in matching.  On the other hand, all powers of $\log aH$ must match in any scheme, as we find here.

%%%%%%%%%%%%%%%%%%%%%%%%
%%%%%%%%%%%%%%%%%%%%%%%%

\section{Hard Cutoff Calculations}
\label{sec:HardCutoff}
In the main text, we used dynamical dim reg for the EFT loop calculations.  Loops in the UV calculations were, in some cases, regulated with dim reg rather than dynamical dim reg.  These regulators offer some technical advantages but one might worry about using different regulators in the matching calculation.  We can therefore gain further conceptual insight by redoing these calculations with a hard cutoff. This regulator can be easily implemented in both the UV and EFT and also makes the origin of divergences more transparent.  Furthermore, we can work directly with the massless mode functions, thereby avoiding the complications of working with massive modes.  In this appendix we will repeat the calculations from the main text using a hard cutoff, reproducing all the above results up to differences in scheme dependent coefficients.

\subsection{Matching}
In this section, we compute the matching for $\lambda \phi^4$ onto the SdSET up to one-loop order.  We will use a hard cutoff for both momentum and time integrals.  Specifically, we regulate the momentum integral with a UV cutoff $\Lambda = [aH]$ and an IR cutoff $K$.  For time integrals, noting that the UV region of integration does not contribute due to our $i\epsilon$ prescription, we simply regulate the IR with a cutoff $\t_\star$, which corresponds to the time of horizon crossing for a mode $k$.

\subsubsection*{Power Spectrum}
The one-loop power spectrum in the UV theory is given by a standard in-in calculation, 
\bea
\Big\langle \phi\big(\k\s\big) \phi\big(\kp\big) \Big\rangle'_{(1)} &=& \frac{\lambda}{4 k^3} \frac{H^2}{3} \left( \log \frac{2 k}{[aH]}+\gamma_E - 2  \right) \int \frac{\d^3p}{(2\pi)^3} \frac{1}{p^3} \,,
\eea
where the primed correlator is defined in \cref{eq:defPrime} above.  The resulting power spectrum is 
\beq
\Big\langle \phi\big(\k\s\big) \phi\big(\kp\big) \Big\rangle'_{(1)}  =\frac{\lambda}{4 k^3} \frac{H^2}{3} \left( \log \frac{2 k}{[aH]}+\gamma_E - 2  \right) \frac{1}{2 \pi^2}\log \frac{[aH]}{K}\,.
\label{eq:UVOneLoop2PtFn}
\eeq  
We can calculate the one-loop power spectrum in the EFT using 
\begin{align}
\Big\langle \bvar\big(\s\k\,\big) \bvar\big(\s\kp\s\big) \Big\rangle'_{(1)} &= H^2 \Big\langle \vp\big(\s\k\,\big) \vp\big(\s\kp\s\big) \Big \rangle'_{(1)}+H^2 \Big\langle \vp\big(\s\k\,\big) \vp\big(\s\kp\s\big) \Big \rangle'_{\delta\alpha^{(1)}}\nonumber  \\[5pt]
&\hspace{13pt} + 2 \frac{c_{4,0}}{9}\frac{1}{ 3!} H^2 \Big\langle \vp\big(\s\k\,\big) \vp^3\big(\s\kp\s\big) \Big\rangle'_{(0)} + \Big\langle \bvar\big(\k\big) \bvar\big(\kp\big) \Big\rangle'_{{\rm  IC}^{(1)}}  \, ,
\end{align}
where we have set $\alpha =0$ for the UV mode functions.  The second term allows for the possibility that $\alpha = \delta \alpha$ is generated by matching and the third term is generated by performing the EFT field redefinition given in \cref{eq:FieldMapAfterEFTFieldRedef}. Using the leading order matching relation for the Wilson coefficient $c_{4,0}= c_{3,1} = \lambda + O(\lambda^2)$, we find
\begin{align}
H^2 \Big\langle \vp\big(\k\s\big) \vp\big(\kp\big) \Big\rangle'_{(1)} &= - \frac{H^2}{3} \log \frac{[aH]}{k}\frac{\lambda}{4 k^{3}} \int \frac{\d^3 p}{(2\pi)^3} \frac{1}{p^{3}} \notag\\[5pt]
&= - \frac{H^2}{3}\log \frac{[aH]}{k} \frac{\lambda}{4 k^{3}} \frac{1}{2 \pi^2} \log \frac{[aH]}{K}\,,
\end{align}
where we evaluated the time integral with a hard cutoff at the time of horizon crossing, $[aH]_{\star} = k$,
\beq
\int_{\t_\star}^\t \d\t' = \log [aH] - \log [aH]_{\star} = \log \frac{[aH]}{k}\,.
\eeq
As with the main text, the contribution from a shift in $\alpha$ is given by
\beq
H^2 \Big\langle \vp\big(\s\k\,\big) \vp\big(\s\kp\s\big) \Big \rangle'_{\delta\alpha^{(1)}} =\Big\langle \bvar\big(\k\big) \bvar\big(\kp\big)  \Big\rangle_{(0)}\bigg(1 + 2 \delta \alpha \log \frac{k}{[aH]}\bigg )\,,
\eeq
and from the field redefinition is
\bea
 2\times \frac{\lambda}{9}\frac{H^2}{ 3!} \Big\langle \vp\big(\k\big) [\vp^3]\big(\kp\big) \Big\rangle_{(0)}\notag
&=& \frac{\lambda}{9}  \frac{1}{2 k^3} \int \frac{\d^3 p}{(2\pi)^3} \frac{1}{2 p^3}\notag \\[5pt]
&=& \frac{\lambda}{9}  \frac{H^2}{2 k^3} \frac{1}{4 \pi^2} \log \frac{[aH]}{K}\,.
\eea
Combing these results we have 
\begin{align}
H^2 \Big\langle \bvar(\k) \bvar(\kp) \Big\rangle'_{(1)} =   \frac{H^2}{3} \left( \log \frac{k}{[aH]} + \frac{1}{3}\right) \frac{\lambda}{4 k^{3}} \frac{\log [aH] / K}{2 \pi^2}+ H^2 \Big \langle \bvar(\k) \bvar(\kp) \Big\rangle'_{{\rm  IC}^{(1)}}\,.
\label{eq:EFT_comb_cutoff}
\end{align}
Comparing the UV expression given in \cref{eq:UVOneLoop2PtFn} with the combined EFT results, \cref{eq:EFT_comb_cutoff}, we see that we need
\beq
\delta \alpha^{(1)} =  \frac{\lambda}{24 \pi^2} \left( \gamma_E - \frac{7}{3}  + \log 2 \right) \ .
\eeq
This expression differs from our result with dynamical dim reg, \cref{eq:deltaAlpha}.  This is not entirely surprising as the precise definitions of the parameters in the UV are scheme dependent and thus this scheme dependence is also inherited through matching.

\subsubsection*{Trispectrum}

Next, we will perform the calculation to match the trispectrum taking $\alpha =0$ on all legs and regulating all integrals with a hard cutoff. As was argued above, ${\cal K}_4$ is the only term that can generate log divergences from the loop momentum integrals.  Taking $p \gg k_i$ we have 
\begin{align}
{\cal K}_4 &\simeq \frac{1}{(k_2 k_3 k_4)^3} \int \frac{\d^3 p}{(2\pi)^{3}}   \int^\tau \frac{\d\tau_1}{(-\tau_1)^4}  G(\k_1;\tau,\tau_1) \notag \\[5pt] 
&\hspace{22pt} \times 2{\rm Im} \int^{\tau_1} \frac{\d\tau_2}{(-\tau_2)^4}   \frac{ G(\p,\tau_1,\tau_2)}{p^3}(1+i p\tau_1) (1-i p \tau_2) e^{-i p (\tau_1-\tau_2)} +{\rm permutations} \nonumber
\\[8pt]
 &\simeq \frac{H^4 \lambda^2}{16(k_2 k_3 k_4)^3}  \int \frac{\d^3 p}{(2\pi)^d} \frac{1}{p^3} \bigg[ \frac{10}{81} - \frac{1}{27} \gamma_E(2+ 3 \gamma_E) - \frac{5}{36} \pi^2 \notag \\[5pt]
&\hspace{50pt}+ \frac{1}{9} \left(\log \frac{2p}{[aH]}\right)^2 + (1+ 3\gamma_E) \log \frac{2p}{[aH]} + \frac{4}{9} \log \frac{k}{[aH]}
\bigg] + {\rm permutations}\notag\\[8pt]
&=\frac{H^4 \lambda^2}{8 (k_2 k_3 k_4)^3} \frac{1}{2 \pi^2} \log \frac{[aH]}{K}\bigg[ \frac{10}{81} - \frac{1}{27} \gamma_E(2+ 3 \gamma_E+6 \log 2) - \frac{5}{36} \pi^2 + \frac{2 \log 2 + 3 \log^2 2}{27} \bigg]\notag \\[5pt]
&\hspace{5cm}+ {O}\Big((\log[aH])^2\Big) + {\rm permutations}\,.
\end{align}
As we did in the main text, we are focused on the single $\log[aH]/K$ term because it cannot be absorbed into the initial conditions and the RG implies that higher powers of log should be products of logs already present in lower order diagrams.  

In order to compare this to the EFT, we need to keep track of our field redefinition to order $\lambda^2$.  Specifically, we need
\beq
 \bvar \equiv H \left( [aH]^{-\alpha} \varphi_+ + [aH]^{-\beta} \varphi_- + \frac{\lambda}{9}\frac{1}{ 3!} [aH]^{-3\alpha} \varphi_+^{3}  +  \frac{\lambda^2}{81 (3!)}  [aH]^{3- 5\alpha}  \varphi_+^{5}\right) \ .
\eeq
to remove the $\vp^6$ operator in the EFT Lagrangian.  This additional term contributes to the trispectrum at one loop: 
\begin{align}
\Big\langle \bvar\big(\k_1\big) \,...\, \bvar\big(\k_4\big)\Big\rangle' &\supset \frac{\sum_i k_i^3 }{(k_1 k_2 k_3 k_4)^3}  \frac{\lambda^2}{81} \frac{5!}{3! 2} \int \frac{\d^3 p}{(2\pi)^3} \frac{1}{p^3}  \notag\\[7pt]
&= \frac{\sum_i k_i^3 }{(k_1 k_2 k_3 k_4)^3}  \frac{\lambda^2}{81} \frac{5!}{3! 2} \frac{1}{2\pi^2} \log \frac{[aH]}{K}\,,
\end{align}
which matches the leading term in the UV expression, namely the factor of $10/81$.  After matching this term, we see a fairly complicated expression remains for the linear log.  This can be absorbed into the $c_{3,1}$ using \cref{eq:eft_tri}, such that 
\beq
c_{3,1}\to \lambda - \frac{\lambda^2}{2 \pi^2} \bigg( \frac{1}{9} \gamma_E(2+ 3 \gamma_E+6 \log 2) + \frac{5}{12} \pi^2 - \frac{2 \log 2 + 3 \log^2 2}{9} \bigg)\,.
\eeq
This agrees with \cref{eq:shiftc31}, which was computing using dynamical dim reg.

\subsection{Composite Operator Mixing}
We continue to demonstrate how the calculations proceed using a hard cutoff regulator.  In this section, we will compute the correlators that yield composite operator mixing, thereby determining the dynamical RG equations.  
\subsubsection*{One Loop}

The first (and simplest) non-trivial calculation to do is the anomalous dimension of $\vp^2$, which we derive from
\begin{align}
\Big\langle \vp^2[0] \vp\big(\k_1\big) \vp\big(\k_2\big)\Big\rangle &= \int \frac{\d^3 p_1 \d^3 p_2}{(2\pi)^6} \Big\langle \vp\big(\p_1\big) \vp\big(\p_2\big) \vp\big(\k_1\big) \vp\big(\k_2\big) \Big\rangle \notag \\[5pt]
&= \int \frac{\d^3 p}{(2\pi)^3} \Big\langle \vp\big(\p\s\big) \vp\big(-\p- \k_1-\k_2\big) \vp\big(\k_1\big) \vp\big(\k_2\big) \Big\rangle'\,.
\end{align}
Since our goal is to reproduce the $O(\lambda)$ log divergence we found above in the main text, we can compute the correlator in terms of $\vp^2$ as opposed to using $\bvar$.  Using $\vp$, there is already of a term proportional to $\log k/[aH]$ from the tree-level time evolution that would only give a $\log^2$ term after integrating over $p$.  Instead, we are interested in the contribution from the initial conditions:
\begin{align}
\Big\langle \vp\big(\k_1) \,...\, \vp\big(\k_4\big) \Big\rangle'_{{\rm IC}^{(1)}}  &= \frac{\lambda}{8(k_1 k_2 k_3 k_4)^3} \Bigg[ \frac{(\gamma_E-2) }{3} \sum_i k_i^3 -\frac{k_1 k_2 k_3 k_4}{k_t} \nonumber \\
&\hspace{14pt}- \frac{1}{9} k_t^3 + 2 \sum_{h<i<j} k_h k_i k_j +\frac{1}{3} k_t \Bigg(\sum k_i^2 - \sum_{i<j} k_i k_j\Bigg) \Bigg] \,.
\end{align}
Taking $k_1 \simeq k_2 = p$ and expanding in $k_3, k_4 \ll p$, we find
\begin{align}
\Big\langle \vp^2[0] \vp\big(\k_1\big) \vp\big(\k_2\big)\Big\rangle &\simeq  P_+(k_1) P_+(k_2) \int \frac{\d^3 p}{(2\pi)^3} \frac{1}{2 p^3} \frac{2}{3} (\gamma_E-2)   \notag\\[5pt]
&= \frac{2}{3} (\gamma_E-2) \frac{1}{2\pi^2}\log \frac{[aH]}{K} P_+(k_1) P_+(k_2)\,.
\end{align}
We can then apply the same steps to evaluate a correlator with an arbitrary composite operator to find
\begin{align}
\Big\langle \phi^n[0] \phi\big(\k_1\big) \,...\, \phi\big(\k_n\big) \Big\rangle &\supset \frac{n!}{2^n (k_1 \,...\, k_n)^3}  \lambda \int \frac{\d^3  p}{(2\pi)^3} \frac{n(n-1)}{4 p^6} \notag\\[5pt]
&\hspace{14pt} \times 2\s {\rm Im} \int \frac{\d\tau_1}{(-\tau_1)^4} (1-i p \tau)^2 (1+i p \tau_1)^2 e^{i 2 p (\tau-\tau_1)} \notag\\[7pt]
&\supset \frac{n!}{2^n (k_1 \,...\, k_n)^3}  \lambda \binom{n}{2} \frac{1}{3} \left( \log \frac{2 k}{aH}+\gamma_E - 2  \right) \int \frac{\d^3  p}{(2\pi)^3} \frac{1}{2 p^3}\,.\notag\\[7pt] 
&\supset \frac{n!}{2^n (k_1 \,...\, k_n)^3}  \lambda\binom{n}{2} \frac{1}{6\pi^2} \left( \log \frac{2k}{aH}+\gamma_E - 2  \right) \log \frac{[aH]}{K}\,.
\end{align}
The term of interest in this expression is the log divergence, which is multiplied $\binom{n}{2}$. We see this coefficient is scheme dependent as this result differs slightly from the result using dynamical dim reg in \cref{eq:b1_dim_reg}.

We can extend this to order $\lambda^2$ using 
\begin{align}
 \Big\langle \vp^2[0] \vp\big(\k_1\big) \,...\, \vp\big(\k_4\big) \Big\rangle
&= \int \frac{\d^3 p}{(2\pi)^3} \Big\langle \vp\big(\s\p\,\big) \vp\big(-\p\,\big) \vp\big(\k_1\big) \,...\, \vp\big(\k_4\big) \Big\rangle \,.
\end{align}
We will evaluate this expression for $k_i \ll p$, which is a different limit of the 6-point function as compared to the previous calculation.  We also want to isolate the piece proportional to $P_+(k_1)\,...\,P_+(k_4)$, so we can simply isolate a subgraph that looks just like two mass insertions:
\beq
\Big\langle \vp\big(\s\p\,\big) \vp\big(-\p\,\big) \vp\big(\k_1\big) \,...\, \vp\big(\k_4\big) \Big\rangle \supset \Gamma_{2,4}(p) P_+(k_1)\,...\, P_+(k_4)\,,
\eeq
with
\begin{align}
\Gamma_{2,4}(p) &= \int^\tau \frac{\d\tau_1}{(-\tau_1)^{4}} \int^{\tau} \frac{\d\tau_2}{(-\tau_2)^{4}} \Big\langle  \phi^2(p,\tau_1)\phi^2(p,\tau) \phi^2(p,\tau_2) \Big\rangle \notag\\[5pt]
&\hspace{13pt}- 2{\rm Re}\int^\tau \frac{\d\tau_1}{(-\tau_1)^{4}} \int^{\tau_1} \frac{\d\tau_2}{(-\tau_2)^{4}}\Big\langle  \phi^2(p,\tau) \phi^2(p,\tau_1) \phi^2(p,\tau_2) \Big\rangle\,.
\end{align}
By direct calculation we find that  
\begin{align}
\Gamma_{2,4}(p) =  \, \frac{1}{216 p^3}\bigg[& 16 + 4 \gamma_E (-11 + 3 \gamma_E) + 3\pi^2 +4 (-11 + 6\gamma_E+3 \log 2)\log 2 \notag \\
&+ {O}\bigg(\log\frac{p}{[aH]}\bigg) \bigg]\,,
\end{align}
which agrees with~\cref{eq:b2Final} above.
This calculation is illustrated in terms of Witten and Feynman diagrams as shown in \cref{fig:gamma24_massive}.  Performing the momentum integral, we find
\begin{align}
 \int \frac{d^3 p}{(2\pi)^3} \Gamma_{2,4}(p) =    \frac{1}{2\pi^2} \log \frac{aH}{K} \left[ b_{2,4}  + { O}\left(\log \frac{aH}{K}\right)  \right]+ {\rm finite}\,,
\end{align}
where   
\beq
b_{2,4} =  \frac{\lambda^2}{216 }\Big[ 16 + 4 \gamma_E (3 \gamma_E-11 ) + 3\pi^2 +4\log 2 ( 6\gamma_E+3 \log 2-11) \Big] \ .
\eeq

\subsubsection*{Two Loops}
The two-loop mixing of $\vp^3$ and $\vp$ is calculated using
\beq
\Big\langle \vp^3[0] \vp\big(\k\s\big)\Big\rangle =  \int \frac{\d^3 p_1 \d^3 p_2 \d^3 p_3}{(2 \pi)^9} \Big\langle \vp\big(\p_1\big) \vp \big(\p_2\big) \vp\big(\p_3\big) \vp\big(\k\big) \Big\rangle \,,
\eeq
see \cref{eq:twoLoopStart}.
When we use a hard cutoff as the regulator, we may use the tree-level four-point function with all massless fields:
\begin{align}
\Big\langle \vp\big(\p_1\big) \vp \big(\p_2\big) \vp\big(\p_3\big) \vp\big(\k\big) \Big\rangle'_{{\rm IC}^{(1)}} & \simeq \frac{\lambda}{8 (p_1 p_2 p_3 k)^3} \Bigg[ \Big(\sum_i p_i^3\Big)\frac{1}{3} \left(\log \frac{p_t}{p_i} + \gamma_E + \frac{1}{3}-2\right) \nonumber \\[5pt]
&\hspace{14pt} - \frac{1}{9} p_t^3 + 2 \sum_{i<j<\ell}  p_i p_j p_\ell \nonumber  +\frac{1}{3} p_t \Big(\sum p_i^2 - \sum_{i<j} p_i p_j\Bigg) \Bigg]\,  .
\end{align}
where $p_t \equiv p_1 +p_2+p_3$.  Here we have assumed $k / p_i \ll 1$ and kept only the leading terms in this expansion since higher orders will not contribute to the mixing. Expanding this out and (implicitly) imposing the momentum conserving $\delta$-function so that $\p_3 = -\p_1-\p_2$ we get
\beq
\Big\langle \vp^3[0] \vp\big(\k\s\big)\Big\rangle 
\simeq \lambda P_+(k)\int \frac{\d^3 p_1 \d^3 p_2}{(2\pi)^6}  \frac{1}{12 (p_1 p_2 p_3)^3 } \left( \sum_i \kappa_i p_i^3  - p_1 p_2 p_3 +\sum_{i\neq j} p_i^2 p_j  \right)\,,
\label{eq:2LoopIntermediate}
\eeq
where $\kappa_i = \log p_t/p_i + \gamma_E + 1/3-2$. The first term (proportional to $\kappa$) factorizes into two logarithmically divergent integrals. The only single log comes from the second term, which can be evaluated using a change of variables
\beq
\int \frac{\d^3 p_1 \d^3 p_2}{(2\pi)^6}  = \int \frac{\d^3 p_1}{(2\pi)^3} p_1^3  \frac{3!}{(2\pi)^2} \int_{1/2}^1 \d x_2 \int_{1-x_2}^{x_2} \d x_3 x_2 x_3\,,
\eeq
where $x_2 = p_2/p_1$ and $x_3=p_3/p_1$.  We then get
\begin{align}
\Big\langle \vp^3[0] \vp\big(\k\s\big)\Big\rangle &\supset -\frac{\lambda}{12} P_+(k) \int \frac{\d^3 p_1}{(2\pi)^3} \frac{1}{p_1^3}  \frac{3!}{(2\pi)^2} \int_{1/2}^1 \d x_2 \int_{1-x_2}^{x_2} \d x_3 \frac{1}{x_2 x_3} \notag\\[5pt]
&= -\frac{\lambda}{12} \frac{3!}{(2\pi)^2} \frac{\pi^2}{12} P_+(k)\int \frac{\d^3 p_1}{(2\pi)^3} p_1^3  \notag\\[5pt]
&= -\frac{\lambda}{192 \pi^2} P_+(k) \bigg(\log \frac{[aH]}{K}\bigg) \,.
\end{align}
We see this result matches the result using dynamical dim reg from the main text given in \cref{eq:2loopResult}.  This is a further confirmation that the $d_0$ coefficient in \cref{eq:d0} is scheme independent.

Finally, we will argue that the third term in \cref{eq:2LoopIntermediate} will produce a $\log^2$ in this description, and therefore does not contribute to the RG.  If we define 
\beq
\rho(a,b,c)=  \int \frac{\d^3 p_1 \d^3 p_2}{(2\pi)^6}  \frac{1}{p_1^{2a} p_2^{2b} p_3^{2c}}\,,
\eeq
then the final term corresponds to taking $a=3/2$, $b=1$ and $c=1/2$, plus permutations thereof.  We can use the  methods discussed in the main text to calculate $\rho(a,b,c)$ as
\beq
\int \frac{\d^3 p_1 \d^3 p_2}{(2\pi)^6}  = \int \frac{\d^3 p_1}{(2\pi)^3} p_1^3  \frac{3!}{(2\pi)^2} \int_{1/2}^{1-\epsilon} \d x_2 \int_{1-x_2}^{x_2} \d x_3 x_2 x_3 \frac{1}{x_2^{2b} x_3^{2c} }\,,
\eeq
where we have included an additional regulator $\epsilon$ to address additional divergences that do not appear in the $p_1$ integral.  
\beq
\rho(3/2,1,1/2)+ {\rm permutations} = \frac{1}{2 \pi^2 }\log\big([aH]/K\big) \times \log \epsilon \propto  (\log aH)^2\,.
\eeq
The sum over permutations is essential in this calculation as the split into $p_1$, $x_2$ and $x_3$ breaks the manifest permutation invariance of the measure of integration, which is only valid if the integrand itself is permutation invariant.  This calculation reproduces the result from the main text where this contribution is $\log^2$, although the need for two separate regulators makes this less transparent.

%\clearpage
\phantomsection
\addcontentsline{toc}{section}{References}
\small
\bibliographystyle{utphys}
\bibliography{dSRefs}

\end{document}